%% file: main.tex
\documentclass[acmsmall]{acmart}
\usepackage{algorithm}
\usepackage{algpseudocode}
\usepackage{subfig}
\AtBeginDocument{%
  }

\setcopyright{acmlicensed}
\copyrightyear{2025}
\acmYear{2025}
\acmDOI{XXXXXXX.XXXXXXX}

\acmJournal{JACM}
\acmVolume{37}
\acmNumber{4}
\acmArticle{111}
\acmMonth{8}

\begin{document}

\title{From Profiling to Optimization: Unveiling the Profile Guided Optimization}

% TODO: complement the info
\author{Bingxin Liu}
\email{liubingxin@mail.bnu.edu.cn}
\orcid{0009-0005-4361-009X}
\affiliation{%
  \institution{Beijing Normal University}
  \city{Beijing}
  \country{China}}
  
\author{Yinghui Huang}
\email{huangyinghui@mail.bnu.edu.cn}
\orcid{0009-0004-1430-7211}
\affiliation{%
  \institution{Beijing Normal University}
  \city{Beijing}
  \country{China}}
  
\author{Jianhua Gao}
\email{gaojh@bnu.edu.cn}
\orcid{0000-0002-3828-0015}
\affiliation{%
  \institution{Beijing Normal University}
  \city{Beijing}
  \country{China}}

\author{Jianjun Shi}
\email{shijianjun@bnu.edu.cn}
\orcid{0009-0007-5641-6354}
\affiliation{%
  \institution{Beijing Normal University}
  \city{Beijing}
  \country{China}}

\author{Yongpeng Liu}
\email{liuyongpeng@phytium.com.cn}
\orcid{0009-0001-5560-0340}
\affiliation{%
  \institution{Phytium Technology Co., Ltd.}
  \country{China}}
  
\author{Yipin Sun}
\email{sunyipin@phytium.com.cn}
\orcid{0009-0001-0345-8127}
\affiliation{%
  \institution{Phytium Technology Co., Ltd.}
  \country{China}}

\author{Weixing Ji}
\email{jwx@bnu.edu.cn}
\orcid{0000-0002-3250-0435}
\affiliation{%
  \institution{Beijing Normal University}
  \city{Beijing}
  \country{China}}

\thanks{Weixing Ji is the corresponding author. This work was supported by the National Natural Science Foundation of China under Grant 62372046, co-sponsored by CCF-Phytium Fund.}

% \renewcommand{\shortauthors}{Liu et al.}

\begin{abstract}
  \input{contents/sections/sec0_abstract.tex}
\end{abstract}

%% TODO: put our own definitions here
\begin{CCSXML}
<ccs2012>
   <concept>
       <concept_id>10011007.10011006.10011041</concept_id>
       <concept_desc>Software and its engineering~Compilers</concept_desc>
       <concept_significance>500</concept_significance>
       </concept>
   <concept>
       <concept_id>10011007.10010940.10011003.10011002</concept_id>
       <concept_desc>Software and its engineering~Software performance</concept_desc>
       <concept_significance>500</concept_significance>
       </concept>
   <concept>
       <concept_id>10003752.10010124.10010138.10010143</concept_id>
       <concept_desc>Theory of computation~Program analysis</concept_desc>
       <concept_significance>500</concept_significance>
       </concept>
 </ccs2012>
\end{CCSXML}

\ccsdesc[500]{Software and its engineering~Compilers}
\ccsdesc[500]{Software and its engineering~Software performance}
\ccsdesc[500]{Theory of computation~Program analysis}

\keywords{compilers, profile guided optimization, instrumentation, hardware sampling}

\received{20 February 2007}
\received[revised]{12 March 2009}
\received[accepted]{5 June 2009}

\maketitle

% \section{Introduction}
\input{contents/sections/sec1_introduction.tex}
\input{contents/sections/sec2_background.tex}

% \section{Instrumentation}
\input{contents/sections/sec3_instrument.tex}
\input{contents/sections/sec4_compiler.tex}
\input{contents/sections/sec5_experiments.tex}

% \section{Discussion}
\input{contents/sections/sec6_future.tex}
\input{contents/sections/sec7_conclusion.tex}

% \begin{acks}
%     This work was supported by the National Natural Science Foundation of China under Grant 62372046, co-sponsored by CCF-Phytium Fund.
% \end{acks}

\bibliographystyle{ACM-Reference-Format}
\bibliography{contents/refs}

\end{document}

%% file: contents/sections/sec0_abstract.tex
Profile Guided Optimization (PGO) uses runtime profiling to inform compiler decisions, effectively combining static analysis with actual execution behavior to enhance performance. Runtime profiles --- acquired through instrumentation or hardware- and software-assisted sampling --- provide detailed insights into control flow, branch predictions, and memory access patterns. This survey systematically categorizes PGO research by profiling method (instrumentation vs. sampling), optimization stages (compile time and link/post-link time), compiler integration (GCC, LLVM), and target architectures. Key algorithms and frameworks are shown in terms of design principles. Performance experiments on representative examples demonstrates PGO's speedups, overheads, and integration maturity. Finally, we identify open challenges, such as reducing sampling overhead, supporting dynamic input workloads, and enabling cross-architecture portability, and propose future research directions to advance adaptive, low-overhead optimizing compilers.

%% file: contents/sections/sec1_introduction.tex
\section{Introduction}

Optimization of program performance has been a key research direction, particularly for compute-intensive and latency-sensitive applications\cite{10.1145/93548.93550}. As the post-Moore's Law era sets in, ``clock frequencies plateau and core counts climb'', software increasingly rely on advanced compiler optimizations to extract instruction-level parallelism and minimize memory stalls. In this context, compilers play a critical role as the bridge that translates high-level language programs into efficient machine code for the target platform. Traditional compiler optimizations primarily employ static analysis techniques to optimize the program's intermediate representation before actual execution, improving the performance of final executable. These static optimizations are widely adopted due to their ease of use and low runtime overhead\cite{10.1145/351403.351408}.

However, with the evolution of computer architectures, such as the prevalence of out-of-order execution and multi-stage pipelined processors, there is an increasing demand for more efficient micro-instruction scheduling within the compiler. To fully exploit the processor's instruction-level parallelism, optimizations often align branch targets based on predetermined heuristics. If the aligned (predicted) branch is taken at runtime, the CPU avoids pipeline stalls, yielding substantial performance gains\cite{10.1145/13310.13312}. Conversely, a misprediction forces the CPU to flush the pipeline and restore registers, resulting in significant penalties. Consequently, compiler designers strive to maximize branch-prediction accuracy.

In the early days, programs were simple enough that static analysis provided sufficiently accurate guidance for optimization. But as applications have grown in size and complexity, static techniques struggle to quickly, comprehensively, and accurately characterize runtime behavior or identify hot code regions. The inability to accurately predict dynamic execution paths marks a significant limitation of purely static optimization\cite{1675827}. Meanwhile, software tracing or hardware-based sampling on modern CPUs provided elaborate hardware support (Performance Monitoring Units, Last Branch Records, instruction-based sampling) that can reveal precise runtime statistics with low overhead\cite{10.1145/1772954.1772963}.

Profile Guided Optimization (PGO), also known as feedback-directed optimization (FDO), bridges the gap between static analysis and dynamic behavior, enabling compilers to reorder, inline, and lay out code based on actual execution counts, branch-taken probabilities, and memory access profiles. By driving optimizations with real-world execution traces, PGO can yield speedups often in the range of 5\%-30\% on real applications \cite{10.1145/384285.379246,10.1145/2851502}, far surpassing what purely static heuristics can deliver.
% However, PGO faces several challenges:
% \begin{itemize}
%     \item Data Collection Overhead vs. Precision: Instrumentation-based profiling offers exact basic-block and edge counts but can unacceptable slowdown during profiling runs. Hardware-based sampling minimizes overhead (< 5 \%) but must correct for imprecise event-to-IR mappings.
%     \item Mapping Raw Events to IR: Turning PMU-sampled addresses into frequency-annotated CFGs requires careful binary-to-source mapping, especially in the presence of address randomization, code alignment, and dynamic linking.
%     \item Integration into Compiler Pipelines: Different compilers (GCC, LLVM, commercial toolchains) consume profile data at compile time, link time (BOLT, LTO hooks), or even post-link (dynamic binary instrumentation), each with its own tradeoffs.
% \end{itemize}

As a result, the body of PGO research now spans instrumentation algorithms, software- and hardware-based sampling methods, mapping algorithms, compiler-internal optimizations, and full end-to-end toolchains\cite{10.1145/2813885.2737990, 10.1145/3585341.3585359, 10.1145/3338502.3359763}.
%This survey gathers all those threads into one place so that compiler researchers, tool developers, and HPC practitioners can see the big picture and identify open challenges.
In this article, we provide a comprehensive overview of PGO by first examining the spectrum of profiling techniques, ranging from software instrumentation-based edge and path profiling, software sampling-based dynamic profiling techniques to hardware-sampling mechanisms. Building on this foundation, we show how major compiler frameworks integrate feedback data at various stages of the build process. To illustrate PGO's practical impact, we present published empirical results on overhead versus speedup across representative benchmarks and contrast instrumentation-based and sampling-based approaches on both x86 and ARM platforms. Finally, we identify open research directions --- such as zero-overhead sampling, dynamic PGO integration in JIT environments, machine-learning-driven heuristic tuning, and cross-architecture portability --- and propose a roadmap for advancing profile guided optimization. Through this taxonomy of techniques, comparative analysis of mapping strategies, detailed survey of compiler-level PGO passes, and overview of performance data, our work aims to serve as a one-stop reference for compiler researchers, tool developers, and performance engineers seeking to understand and extend the state of the art in profile guided optimization.

Our survey draws on 61 primary references spanning 1971 through 2024, covering five decades of research. The earliest work dates to 1971 and the most recent to 2024 study on matching stale profiles to evolving binaries. Roughly 15 references originate in the 1970s-80s, chiefly foundational algorithmic and static-optimization papers; 20 come from the 1990s-2000s, where classic PGO instrumentation and hardware-sampling techniques were introduced; and 26 are from 2010 onward, reflecting modern sampling-based PGO, compiler integrations, and binary-rewriting toolchains.

These references appear in top-tier outlets --- ACM's PLDI, CGO, ASPLOS, SIGPLAN, ISCA, IEEE MICRO, and journals like J. ACM and IEEE TPDS. Their collective span reflects the evolution of PGO from textbook static heuristics through heavy-weight instrumentation to today's light-weight, hardware-accelerated, and deployable feedback systems. By organizing them thematically and chronologically, readers can trace how advances in profiling (both software and hardware), compiler frameworks, and mathematical inference have steadily improved PGO's efficiency, accuracy, and practicality.

The remainder of this article is organized as follows. In Section 2, we review essential background on compiler internals and hardware support, including control-flow graphs, intermediate representations, and basic profiling concepts. Section 3 then focuses on profiling techniques, first describing software-based approaches that achieve fully correct profiling through instrumentation, followed by software-based sampling methods that trade some precision for lower overhead, and finally examining hardware-based sampling mechanisms such as PEBS, IBS, and Last Branch Recording. Building on these profiling foundations, Section 4 explores how collected profiles are consumed at different stages of the build process: we discuss compile-time optimizations, link-time and post-link-time strategies, and runtime optimizations. In Section 5, we present our empirical evaluation of PGO's impact on both ARM and AMD64 platforms, reporting speedups and overheads for representative benchmarks. Section 6 identifies open research directions --- such as zero-overhead sampling, dynamic PGO in JIT systems, machine-learning-driven heuristic tuning, and cross-architecture portability --- and outlines a roadmap for future work. Finally, Section 7 concludes with a summary of key insights and recommendations for advancing PGO research.

%% file: contents/sections/sec2_background.tex
\section{Background}

Modern CPUs deploy deep, multi-stage pipelines to extract instruction-level parallelism, speculating on branch outcomes to keep the pipeline utilization. When a branch is mispredicted, the cost of flushing and refilling these stages can exceed the cost of executing straight-line code. According to a study by Su and Zhou \cite{acabps} on the SPECint95-beta and SPECfp92 benchmarks, branch instructions account for 10\%–20\% of operations in integer programs and about 5\% on average in floating-point programs. Therefore, the extra overhead caused by branch misprediction becomes more significant as the application scale increases.

To minimize this penalty, McFarling and Hennesey  \cite{10.1145/17356.17402} proposed performing branch prediction at compile time. They introduced a series of static (compile-time) and dynamic (hardware-assisted) prediction strategies. Dynamic branch prediction attaches a few bits to each branch instruction and updates them at runtime to show the execution bias. Static branch prediction, by contrast, marks one branch direction at compile time, and the branch is always predicted to follow that direction.
%  As modern applications grow in scale, the extra overhead of pipeline flushing and re-execution due to mispredicted branches becomes unacceptable.

% 随着计算机硬件的发展，二进制指令层面的并行技术在上世纪得到了广泛的研究和发展。多级流水线的出现大大增加了单位时间内中央处理器执行指令的密度。为了充分利用硬件资源，程序中的分支语句常常被随机执行或按特定分支执行，即在运行时确定应该执行某个分支之前就开始执行指令，从而尽可能占用机器的闲置资源。但一旦分支预测失败就会造成严重的资源浪费，在指令级并行架构中，随机执行分支语句失败之后的流水线清理与重新执行所带来的开销可能远远大于执行程序原本指令的开销。根据 Su 和 Zhou \textcite{} 两人的一项针对 SPECint95-beta 以及 SPECfp92 基准测试的调查研究显示，分支语句在整形数程序中的占比可达 10\% ~ 20\%，而在浮点数程序中也占到平均 5\% 的比重。随着现代程序规模的膨胀，分支执行失败带来的流水线清空和程序重新执行的额外开销变得不可接受。为了尽可能减少分支重新执行所带来的额外开销，斯坦福大学的 Scott 和 John 提出在程序编译阶段就对程序进行分支预测 \cite{10.1145/17356.17402} 。该研究提出了一系列基于静态（编译时）和动态（硬件辅助）的分支预测策略。动态分支预测通常在分支语句上的附着几个比特位，并在程序运行时设置这些比特位从而反应分支语句概率上的偏向。而静态分支预测则在编译时标记某一个分支方向，然后分支总是被预测为朝该方向发展。
% A Comparative Analysis of Branch Prediction Schemes
% Reducing the cost of branches
%% A comparative analysis of schemes for correlated branch prediction

Early static branch-prediction optimizations relied on manual tuning based on experience—such as the `likely' annotations in the Linux kernel or heuristic algorithms to guide prediction. Manual tuning and heuristics heavily depend on expert knowledge and can be time-consuming. With the advent of kernel-level profiling tools like GNU perf in the late 1980s and early 1990s, programmers are enhanced with more efficient and systematic tools to analyze execution bottlenecks. As profiling tools matured, compiler developers began to collect runtime profile data and use it to guide compile-time optimizations. Joseph and Stefan \cite{10.1145/143365.143493} were among the first to propose improved static branch prediction by using runtime profile information rather than purely algorithmic or manual prediction. Chang et al. \cite{10.1002/spe.4380211204} further using runtime statistics for instruction-layout optimization. They described a compiler with two new components: a performance profiler that inserts probes into the program to collect runtime data, and a profile-guided optimizer that uses the accumulated profiles to optimize the final executable. This instrumentation, collection, and then, optimization workflow laid the foundation for modern PGO techniques.
% 早期静态分支预测优化技术主要依靠经验对程序进行手动调优，如 Linux 内核中的 Likely 标记，或是根据启发式算法优化分支预测。手动调优与启发式算法严重依赖于经验与算法的准确性，且优化过程漫长。自上世纪七、八十年代内核性能分析工具如 Gnu Perf 的出现，程序员得以更加高效系统的分析程序的执行瓶颈。九十年代后，随着性能分析工具的成熟，编译器开发者开始尝试使用性能分析工具获得运行时程序剖析数据，并在编译阶段使用运行时描述信息进行程序优化。Joseph 和 Stefan \cite{10.1145/143365.143493} 率先提出通过观察程序实际运行时的状态优化静态分支预测的准确性。Pohua P. Chang 等人使用程序运行时统计数据进行指令布局优化 \cite{10.1002/spe.4380211204}。该研究提出了一种可一在运行时自动生成剖析信息以辅助代码优化的编译器。该编译器由两个新组件组成：执行剖析器和基于剖析文件的优化器。执行剖析器在输入程序中插入探针来统计运行时信息。通过执行程序的不同输入，累积运行信息，并将这些信息提供给优化器以供优化器优化最终程序。这种插桩--收集——优化的方式奠定了 PGO 技术的主体框架。在这一阶段，区别于依赖对于程序执行阶段的算法预测或人工预测，优化器采用程序运行时的实际的运行剖析信息进行优化。
% Predicting Conditional Branch Directions From Previous Runs of a Program
% Using profile information to assist classic code optimizations
% TODO:
% 启发式算法补充

PGO gathers profiling data by inserting instrumentation at compile time, recording execution counts for basic blocks, branches, and function calls. This approach, however, imposed high runtime overhead. Fortunately, driven by hardware evolution and the urgent need for low-overhead analysis, CPU vendors incorporated Performance Monitoring Units (PMUs) into processors. PMUs allow software to collect runtime information, such as cache hit rates, branch-prediction accuracy, and event counts, with minimal overhead. Jeffrey et al. introduced ProfileMe \cite{645821}, an instruction-based sampling method that records detailed pipeline-stage information for sampled instructions rather than sampling at the basic-block level. AMD's Instruction Based Sampling (IBS) \cite{amd64vol2} and Intel's Processor Event-Based Sampling (PEBS) and Last Branch Record (LBR) \cite{intel_sdm} exemplify such hardware-sampling features. These mechanisms ensure that the instruction causing a counter overflow interrupt is uniquely associated with the event that incremented the counter, enabling accurate mapping back to source code. Hardware-based sampling avoids the prohibitive overhead and extra compile–run steps of software instrumentation. Subsequently, compiler researchers seek to exploit this data to improve optimizations. AutoFDO proposed by Chen et al. \cite{45290} and BOLT proposed by Panchenko et al. \cite{10.5555/3314872.3314876} both implement sampling-based PGO using PMU data, targeting compile-time and link-time optimizations respectively. Wenlei et al. \cite{10.1145/3498714} provided the theoretical foundation for these methods.
% TODO: 加不加采样开销
% sometimes up to two orders of magnitude in large applications
%, as shown in Figure \ref{}.
% 早期的反馈优化技术是通过编译时插入指令来采集程序剖析信息，即在编译应用时在程序中插入记录基本块、跳转、调用函数执行次数的指令集合。这种模式的弊端在与为获取运行时程序描述信息而进行的额外记录步骤，因此会产生巨大的运行开销。有时，这种额外的开销在大规模程序中可能达到百倍之多如图 \ref{} 所示。
% 此外，该模式对训练时的输入极为敏感，若不幸使用了有偏差的输入集，则有可能反而降低程序的执行效率。
    % TODO: 补充数据
    % TODO: 运行效率图改进

% 幸运的是，随着计算机硬件的发展以及软硬件分析优化的迫切需要，硬件厂商在处理器内部加入性能监控单元（Performance Monitor Unit，PMU）以辅助软件人员以更小的开销获取软件运行时的信息，其中包括处理器的利用率、指令执行的效率、缓存利用率、分支预测正确率、事件计数器等有用信息。根据硬件设计，Jeffrey 等人提出了一种名为 ProfileMe \cite{645821} 的采样分析方法，使采样工具能够以指令为单位进行采样而非传统上对基本块等单位进行采样。随着被采样的指令在多级流水线上移动时，该采样技术可以收集值得关注的事件和指令在流水线上各个阶段的详细记录，从而提供程序在运行时的关键性能信息。该项技术在超微半导体（Advanced Micro Device, Inc. AMD）公司所开发的中央处理器中的基于指令的采样技术（Instruction based sampling, IBS）\cite{} 中有所体现。英特尔（Intel ）公司则在 PMU 中提供了基于事件的采样方法（Processor event-based sampling, PEBS）和最近分支记录（Last branch record）两种硬件采样技术\cite{}。该技术可以确保造成计数寄存器溢出中断的指令与使计数寄存器递增的动态指令唯一对应。从而使构造指令到源代码的反射结构成为可能。这类基于硬件的采样技术可以避免基于软件插针技术的采样方法所带来的不可避免的运行时开销以及额外的编译运行步骤。自然的，优化工作者希望从这类信息中发掘可以用来优化程序性能的部分。Chen 等人在所提出的 AutoFDO 技术 \cite{45290} 以及 Panchenko 等人提出的 BOLT 优化技术就是基于 PMU 硬件的采样反馈优化技术。前者主要针对编译阶段的优化而后者则关注链接阶段的优化。而 Wenlei 等人的工作则为这种优化方式提供了理论基础\cite{10.1145/3498714}。
    %ProfileMe: hardware support for instruction-level profiling on out-of-order processors
    %AMD manual
    %INTEL manual
    %AutoFDO 2016
    %Profile Inference Revisited
    
In summary, hardware-based sampling PGO offers a low-overhead, source-insensitive direction for compile-time optimization. Sampling is performed over the optimized executable file using PMU events. After sampling, specialized algorithms correct and correlate machine instructions with the compiler's intermediate representation. By annotating the control-flow graph (CFG) with vertex and path weights derived from the profiles, the optimizer can globally reorder code to improve layout. Decoupling the profiling runs from actual program runs allows iterative integration of sampling-based feedback into the compiler's optimization pipeline.

\input{contents/figs/PGO-pipeline.tex}

Figure \ref{fig:pgo-pipeline} encapsulates the PGO pipeline at a glance.  We begin by gathering runtime counts, either through lightweight sampling hardware or fine-grained instrumentation, and then translate these binary-addressed events into source-level basic-block and edge frequencies.  With a CFG-consistent profile, the compiler applies targeted optimizations (e.g.\ hot-edge inlining, branch reordering, cache-friendly layout) before emitting an optimized binary.  Finally, this binary can itself be re-profiled to refine the feedback loop.

% TODO: 加不加章节简介
% The remainder of this article details the implementation of feedback-directed techniques. Chapter 3 examines software and hardware sampling methods. Chapter 4 discusses algorithms for analyzing sampling results. Chapter 5 presents concrete optimization implementations and strategies. Chapter 6 focuses on practical applications of feedback-directed optimization.
% 综上所述，基于硬件采样的反馈优化技术为我们提供了一种运行时开销极小，对原代码改变不敏感的编译优化方向。其采样程序直接针对程序经过优化的发布版本二进制程序采集 PMU 收集性能信息。为了编码效率，该信息往往以二进制形式呈现。在对目标程序片段的采样完成后，由专门的算法进行矫正，并建立机器指令与源代码的中间语言（Intermediate Representation）的反射。通过解析统计信息，编译器驱动可以生成针对源代码控制流图（Control Flow Graph，CFG）上的顶点和路径的权重标注。最终优化器根据流图权重进行全局代码布局优化。由于成功的将训练程序执行和实际的程序执行解耦，基于采样的反馈优化技术可以轻松的集成在编译优化器中实现迭代优化。

%% file: contents/figs/PGO-pipeline.tex
\begin{figure}[t]
  \centering
  \includegraphics[width=0.9\linewidth]{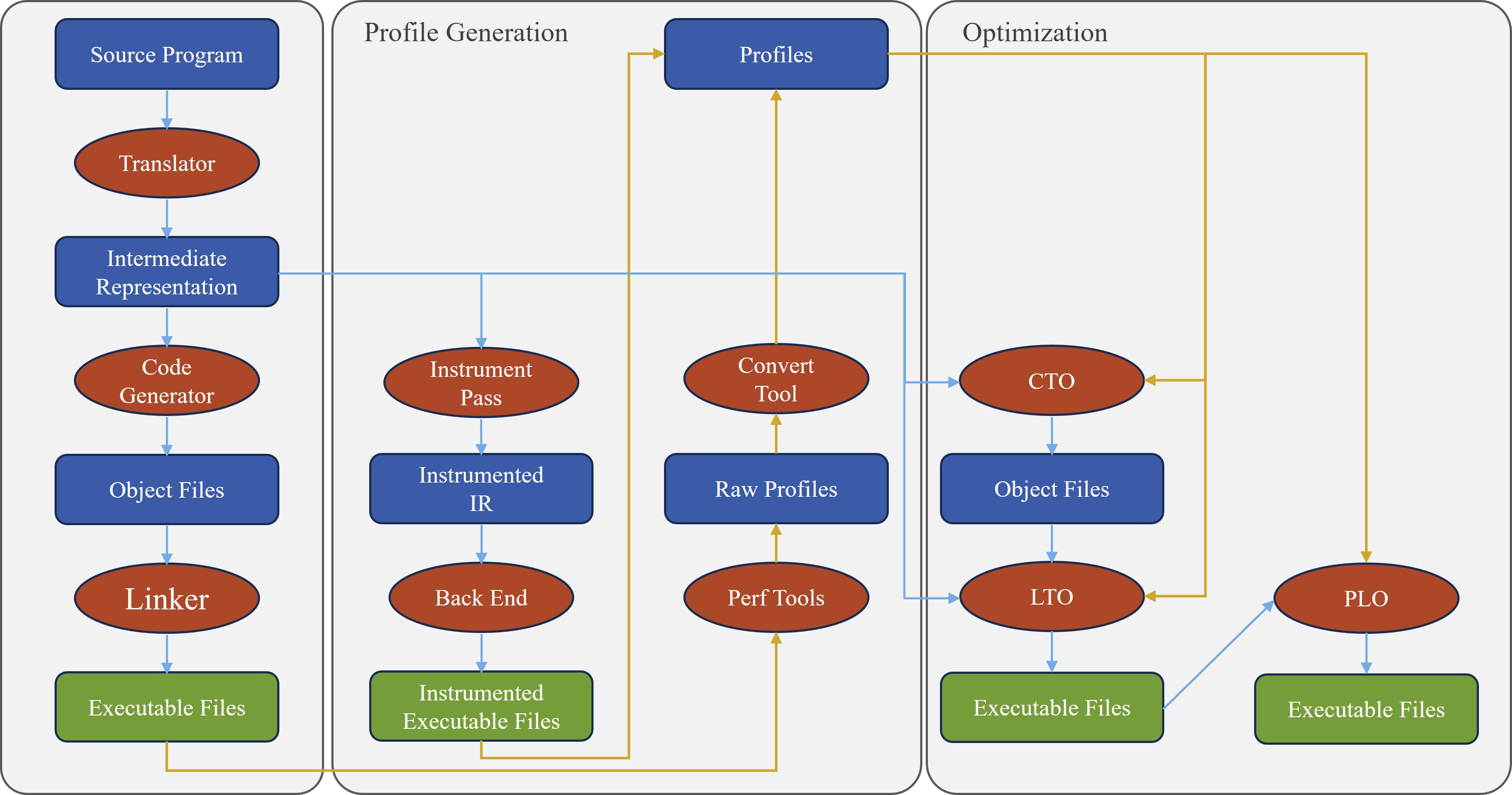}
  \caption{End-to-end PGO workflow: the left most sub-figure shows normal compile stages; The middle one shows profiling steps via sampling or instrumentation; The right-most sub-figure shows profile guided optimization passes in different compile stages.}
  \label{fig:pgo-pipeline}
\end{figure}

%% file: contents/sections/sec3_instrument.tex
\section{Program Profiling}
% \section{获取程序描述信息}
The profile guided optimization pipeline use runtime profile to drive optimisations such as inlining, basic-block layout, and indirect-call promotion. To make the optimization reasonable, the compiler must first collects a high-quality profile whose accuracy is related to its collection cost. This section therefore focuses on the process of profile: it classified the profiling techniques into full correct profiling (instrumentation) versus partial correct tracing (sampling-based) approaches, contrasts their overhead trade-offs, and reviews reconstruction techniques that lift raw addresses back to IR-level frequencies.

Table~\ref{tab:sec3-profile-collection-overview} summarizes three broad classes: edge and path instrumentation, software sampling, and hardware sampling. Instrumentation-based methods yield exact execution counts but incur high runtime overhead due to inserted probes at every branch or path. Software sampling reduces overhead by collecting only partial traces at fixed intervals or events, reconstructing full profiles statistically at moderate cost. Hardware sampling leverages built-in PMU support to provide low-overhead, approximate and remapped profiles that often suffice for guiding global optimizations.

% Pogram profiling enables the compiler to obtain a comprehensive and faithful view of the runtime state and make optimization decisions. Runtime information includes not only software and hardware status but also program-specific metadata. Besides, correctly mapping execution behavior to source line of code and providing this data to the compiler is a key step for realizing profile-guided optimization. Typically, profiling tools collect runtime information via software instrumentation or hardware sampling and bind this data to code regions for compiler.
% 获取程序描述信息可以使得编译器获得全面真实的运行时状态，并进行针对性的优化。程序运行时信息不仅包括运行时软硬件状态，也包括程序本身的信息，将运行状态与程序本身局部性信息进行正确映射并提供给编译器进行优化，是反馈优化获得优化收益的前提条件。通常，信息收集工具将通过软件插入或硬件采样等方式收集程序运行时信息，并将信息与代码片段绑定以提供给编译器进行优化。

\begin{table}[t]
  \caption{Overview of profile-collection techniques}
  \label{tab:sec3-profile-collection-overview}
  \small
  \centering
  % \begin{tabular}{p{2.8cm}p{1.5cm}p{1.5cm}p{1.2cm}p{5cm}}
  \begin{tabular}{lccl}
    \toprule
    \textbf{Techniques} & \textbf{Overhead} & \textbf{Correction} & \textbf{References}\\ 
    \midrule
    Instrumentation & $\ast\ast\ast$ & Exact & \cite{10.1145/183432.183527, 566449, 10.1145/301618.301678, 10.1145/301618.301678, 10.5555/935616}\\
    Software sampling & $\ast\ast$ & Statistical & \cite{traub2000ephemeral, 10.1145/1250734.1250746, 10.1145/1064978.1065034, bruening2020dynamorio, 10.1145/378795.378832, 6494982, 4145115, 10.1145/337449.337483, Hirzel2001BurstyTA}\\
    Hardware sampling & $\ast$ & Statistical & \cite{717402, 903267, 645821, 5452049, 10.1145/258915.258924, 36576, DBLP:journals/corr/WichtVCL14, 10.1007/978-3-540-77560-7_20, 10.1145/1772954.1772963, 10.1145/258916.258924, 10.1145/3498714, 10.1007/978-3-642-39038-8_27, 10.1145/2851502, 10.1145/3640537.3641573, 10444807}\\
    \bottomrule
  \end{tabular}
\end{table}

\subsection{Instrument-Based Profiling}
% \subsection{程序剖析技术}
% TODO： 改例子描述

Edge profiling and path profiling are two core analysis and popular profiling techniques. Edge profiling records the execution frequency of each edge (i.e., jump between basic blocks) in the control-flow graph, revealing branching behaviors. For example, it can quickly identify hot branches and guides the compiler to reorder basic blocks to improve instruction cache locality or tune the branch-prediction heuristics. Path profiling goes further by counting the number of times each complete path from entry to exit is executed, capturing contextual execution correlations, such as the distribution of different loop-exit paths or specific nested-condition invocation patterns. Although path profiling incurs higher complexity and overhead than edge profiling, it provides a more complete characterization of program behaviors. Together, these two techniques supply the compiler with temporal and frequency-based profile information, enabling optimizations such as code-layout adjustment and informed inlining decisions.
% 在众多信息收集方法中，边缘剖析（Edge Profiling）与路径剖析（Path Profiling）是两种核心的分析技术。前者通过统计控制流图中每条边（即基本块间的跳转）的执行频率，呈现程序分支的局部特征。例如，它可快速定位高频跳转的分支，指导编译器调整基本块排列顺序以提升缓存利用率，或优化分支预测策略。后者则更进一步，追踪从程序入口到出口的完整路径执行次数，揭示代码执行的上下文关联性。例如，某段循环中不同退出路径的执行分布，或嵌套条件语句的特定组合调用模式，均可通过路径剖析捕捉。尽管路径剖析的实现复杂度与开销相对边缘剖析较高，但其对程序行为能够进行更完整的刻画。这两种技术分别从时间与频率维度，为编译器提供描述程序运行的状态信息，使代码布局调整、函数内联决策等优化手段成为可能。

% 边缘剖析和路径剖析在信息范围、粒度、实现复杂性和开销方面存在关键差异。边缘剖析侧重于基本块之间的局部转移，提供更细粒度的单个控制转移信息，实现相对简单且侵入性较低，适合生产环境。而路径剖析则捕获整个转移序列，提供更全局的执行流信息，实现更为复杂，通常用于需要详细分析的剖析阶段。总的来说，这两种技术都用于指导编译器优化，但在细节层次和开销上有所不同。边缘剖析适用于快速识别代码中频繁使用的转移，而路径剖析则提供了执行序列的全面视图，能够推动更复杂的优化，但代价是额外的复杂性和运行时开销。如需进一步了解技术细节，可以参考Ball-Larus路径剖析的相关文献或讨论这些插桩技术的标准编译器优化教材。

% \input{contents/figs/cfg-profile-instrument-example}
% \begin{figure}[htbp]
%   \centering
%   \subfloat[Original CFG with edge frequencies]{
%     \includegraphics[width=0.32\linewidth]{contents/figs/cfg-original.jpg}
%     \label{fig:cfg-original}}
%   \hfill
%   \subfloat[MST (hot edges in orange)]{
%     \includegraphics[width=0.32\linewidth]{contents/figs/cfg-mst.jpg}
%     \label{fig:cfg-mst}}
%   \hfill
%   \subfloat[MST complement (cold edges in green)]{
%     \includegraphics[width=0.29\linewidth]{contents/figs/cfg-profiled-edges.jpg}
%     \label{fig:cfg-mst-complement}}
%   \caption{Select suitable profiling edges.}
%   \label{fig:spanning-tree-layout}
% \end{figure}

\subsubsection{Edge Profiling}
% \subsubsection{边缘剖析}

% TODO:
% 更改 VF EF 的定义，补充流量等式 flow equation
Edge profiling only collecting execution-frequency data for each edge or basic block in the control flow graph. The compiler inserts instrumentation so that whenever a branch or jump executes, the counter associated with that edge is incremented. This approach yields detailed local information about the most frequently executed basic blocks and the most possible taken branches. Edge profiling's advantages lie in its simplicity and relatively low overhead, while still providing fine-grained data for local optimizations such as branch-prediction tuning, basic-block reordering, and hotspot identification.
% 边缘剖析（Edge Profiling）是一种收集程序控制流图（CFG）中每条边或每个基本块执行频率数据的技术。控制流图中的每条边代表了两个基本块之间可能的跳转。编译器会在代码中进行插桩，使得每当执行一个分支或跳转时，与该特定边关联的计数器就会增加。这种方法提供了关于最频繁跳转分支执行频率的局部信息。边缘剖析的优势在于其简单性和较低的实现开销，同时能够提供关于单个控制转移频率的详细描述。它适用于指导局部优化，如分支预测改进、基本块重排序以及确定哪些代码路径是“热点”（频繁执行）。

% The CFG is a rooted directed graph defined as $G = (V, E)$, with a special exit vertex \texttt{EXIT} (distinct from the root). Each vertex corresponds to a basic block; each edge represents a control transfer between blocks; the root vertex is the first block executed, and the exit vertex is executed last; there is a directed path from the root to every vertex and from every vertex to the exit.
% 控制流图（Control Flow Graph, CFG）是一个有根的有向图，我们定义这个图为 G = (V, E)。且该图有一个特殊的顶点 EXIT （不同于根顶点）。控制流图可以与程序源代码做如下对应：每一个顶点代表一个代码的基本块；每一条边代表从一个基本代码块到另一个基本代码块的分支走向（控制转移）；根顶点代表程序中第一个执行的代码块，相应的，结束顶点最后执行；从根顶点到每一个顶点都有一条有向的路径连接两者，同样的，从各个顶点到结束顶点也有一条有向路径连接两者。

% check pl and epl
To obtain basic-block execution counts, one could simply assign a counter to each block and increment it on execution. However, this brute-force approach incurs unnecessary overhead. We can collect part of the information and completely rebuild a profile which includes both vertex frequencies and edge frequencies for the whole program.
%Let $VF(G,pl)$ denote the frequency of a vertex, given a set of profile log points, denoted as $pl$, of vertices and edges in $G$.
The challenge is to place counters within CFG such that the execution frequency of every vertex can be exactly derived from the recorded counts. To minimize overhead, counters should be placed in regions of low execution frequency.
% 为了获取程序基本块的执行次数，最简单的办法即为每一个基本块非配一个计数器。每次执行某个基本块时，增加该基本块对应的计数器的值。但是这种暴力统计常常产生不必要的统计开销。对于控制流图，获得完整的统计信息需要两方面的信息，即基本块（顶点）的执行频率以及分支语句（边）的执行频率。对于基本块的执行频率(记为 $VF(G,pl)$)，关键问题在于决定计数器在pl(程序流图中顶点和边组成的集合)中的摆放位置从而使每个顶点在任意一次程序执行中出现的频率都可以完全的从计数器记录下的数值中推导还原。为了降低额外的开销，这些计数器应该被放置在执行频率较低的区域，即程序中执行频率较低的区域。

Thomas and James \cite{10.1145/183432.183527} first observed the similarity between optimal counter placement in a CFG and the well-known maximum spanning-tree problem \cite{10.1137/1.9781611970265}. They proposed an algorithm that selects counter positions based on predicted or measured frequencies of basic blocks and branches. The solution corresponds to finding a minimum-cost subset edges with profile log points, denoted as $epl$, whose removal yields a spanning tree on which all other edge frequencies can be inferred via flow-equations. Formally, if $E - epl$ contains no cycles, then $epl$ is an optimal solution for edge frequence $EF(G,epl)$ \cite{10.5555/578398}, and the minimum number of counters required is $|E| - (|V| - 1)$.
% Thomas 和 James \cite{10.1145/183432.183527} 率先意识到程序控制流图中分支计数器摆放位置和著名最大生成树问题 \cite{10.1137/1.9781611970265} 的相似之处。基于最大生成树问题提出了一种用于检测和记录分支执行情况的算法。算法可以根据程序中每个基本块和分支的预测或执行频率选择最优的优化计数器位置。并且可以获得最频繁执行基本块的子序列，该子序列的长度根据程序的执行情况进行优化，从而可以高效的重新生成追踪序列。我们称使成本函数 $cost(G,pl,W)$ 在给定权重矩阵的控制流图 $G$ 中最小的顶点与边的集合 $pl$ 为 $VF(G,pl)$ 的最优解。很明显获取分支语句执行频率的最小成本的解与以最小成本插入计数器获取基本块执行频率的解为同解。为了求出最优解，可以利用网络编程领域中的一个著名结论，即对于放置在分支语句上的一组计数器 $epl$，当且仅当 $(E-epl)$ 所构成的图中不存在环形结构时，有对于 $EF(G,pl)$ 的最优解\cite{10.5555/578398}。同时因为控制流图的生成树代表着无环结构下的最大子集，可以得到 $epl$ 是 $EF(G,epl)$ 的最小解的充要条件是 $(E-epl)$ 是控制流图 $G$ 的一个生成树，即求解 $EF(G,epl)$ 的最少计数器数量为 $|E| - (|V| - 1)$。
%另一个类似的是分支语句的执行频率，定义为 $EF(G,pl)$ 。
    % optimally profiling and tracing programs
    % R.E. Tarjan, Data Slrur.lures and Network Algorithms, Society for Industrial and Applied Mathematics, Philadelphia, PA (1983).
    %Algorithms for Network Programming
Thus, placing counters on the edges in $epl$ (initialized to zero) and incrementing them on each taken branch suffices to reconstruct all edge frequencies. If a vertex $v$ is a leaf in the spanning tree $(V, E - epl)$, its entry edges' frequencies can be uniquely determined by the flow-equation at $v$ and the known frequencies of other entry edges. Figure~\ref{fig:cfg2-original} illustrates this process: solid edges belong to $epl$, and the remaining dashed line edges form the spanning tree. For vertex $k$, the flow equation is $(K \rightarrow B) = (I \rightarrow K) + (J \rightarrow K)$.
Knowing $(I \rightarrow K)$ and $(J \rightarrow K)$ yields $(K \rightarrow B)$, and so on, until all edge frequencies are reconstructed.

% 因此，我们仅需要在最优解 $epl$ 的路径上放置计数器，并将计数器数值初始化为0，在每次执行该路径所对应的分支语句时增加该语句上的计数器的值即可获得整个程序的执行情况。若顶点$v$(基本块)是生成树的一个叶子节点，那么一定存在一个连通该顶点的路径存在与 $epl$ 中。该路径的权重信息（频率）已知，则未测量边的频率可以由 $v$ 的流量方程以及其他已知频率的输入输出边所唯一确定。若 $(E - epl)$ 不是控制流图 $G$ 的一个生成图，则未测量边的频率无法唯一确定。图 \ref{fig:cfg-profile-instrument-example} 展示了这一过程。控制流图中的打黑色原点的边表示 包含在 $epl$ 中的边。其他边则属于 $(E-epl)$ 并且构成了控制流图的一个生成树。对于顶点 $Q$ 有流量等式 $(P \rightarrow Q = Q leftarrow A + Q leftarrow B)$ 。已知 $P \rightarrow Q$ 与 $Q \rightarrow A$,易知 $Q \rightarrow B$。已知 $Q \rightarrow B$ ，则 $B \rightarrow R$ 也可推出。以此类推，可以得到控制流图中所有边的频率。

% TODO:
% 介绍算法1 
Using this reconstruction method, a post-order depth-first search on the spanning tree computes each edge's frequency. Once all edges frequency to a vertex $v$ except $e$ are known, the final edge $e$ is deduced by the flow-equation.

\subsubsection{Path Profiling}
% \subsubsection{路径剖析}

% TODO:
% Bursty Tracing 捕捉因果关系
% 补充图

\begin{figure}[t]
  \centering
  \subfloat[Original CFG with edge frequencies]{
    \includegraphics[width=0.32\linewidth]{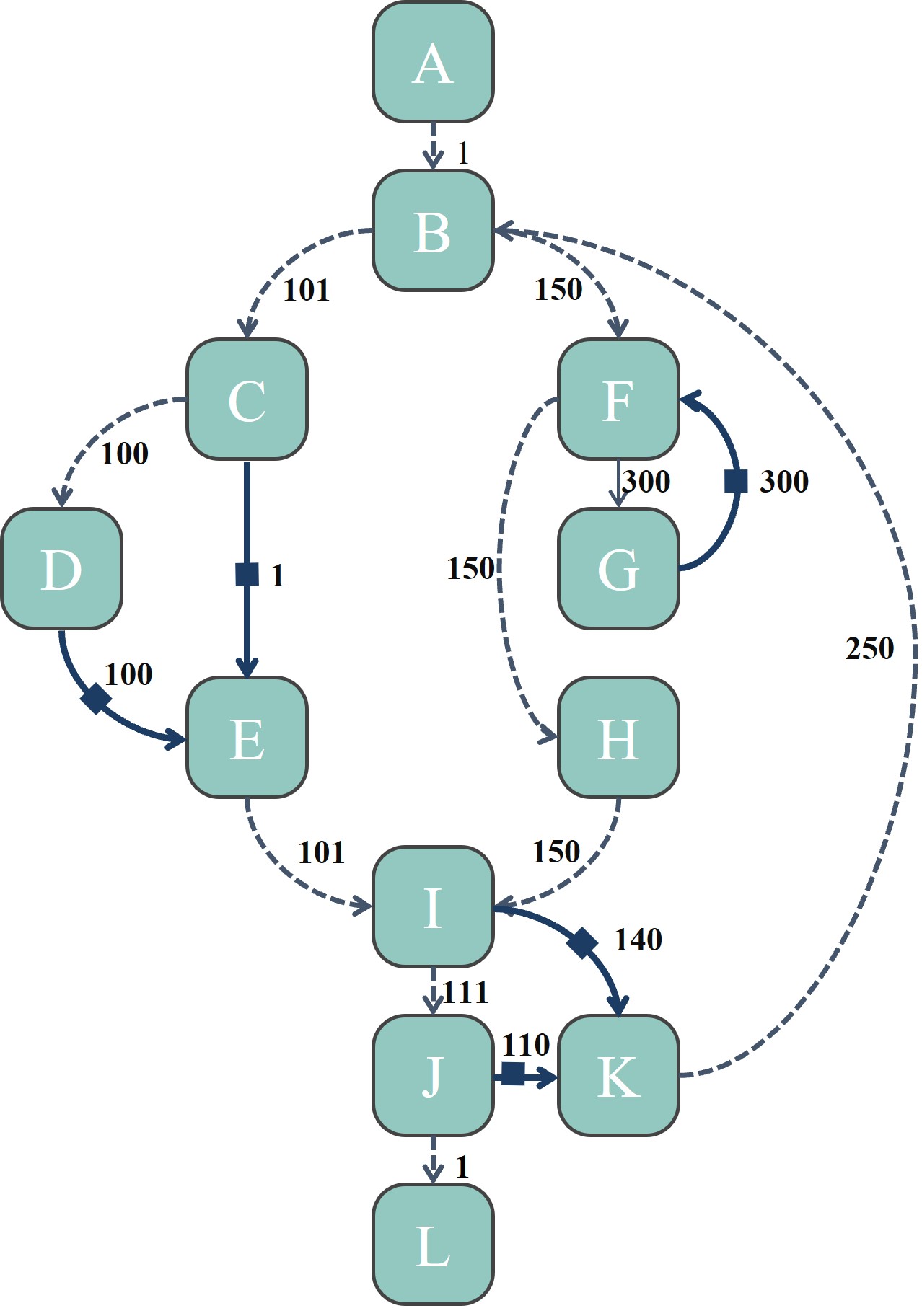}
    \label{fig:cfg2-original}}
  \hfill
  \subfloat[Hot-path candidate I]{
    \includegraphics[width=0.32\linewidth]{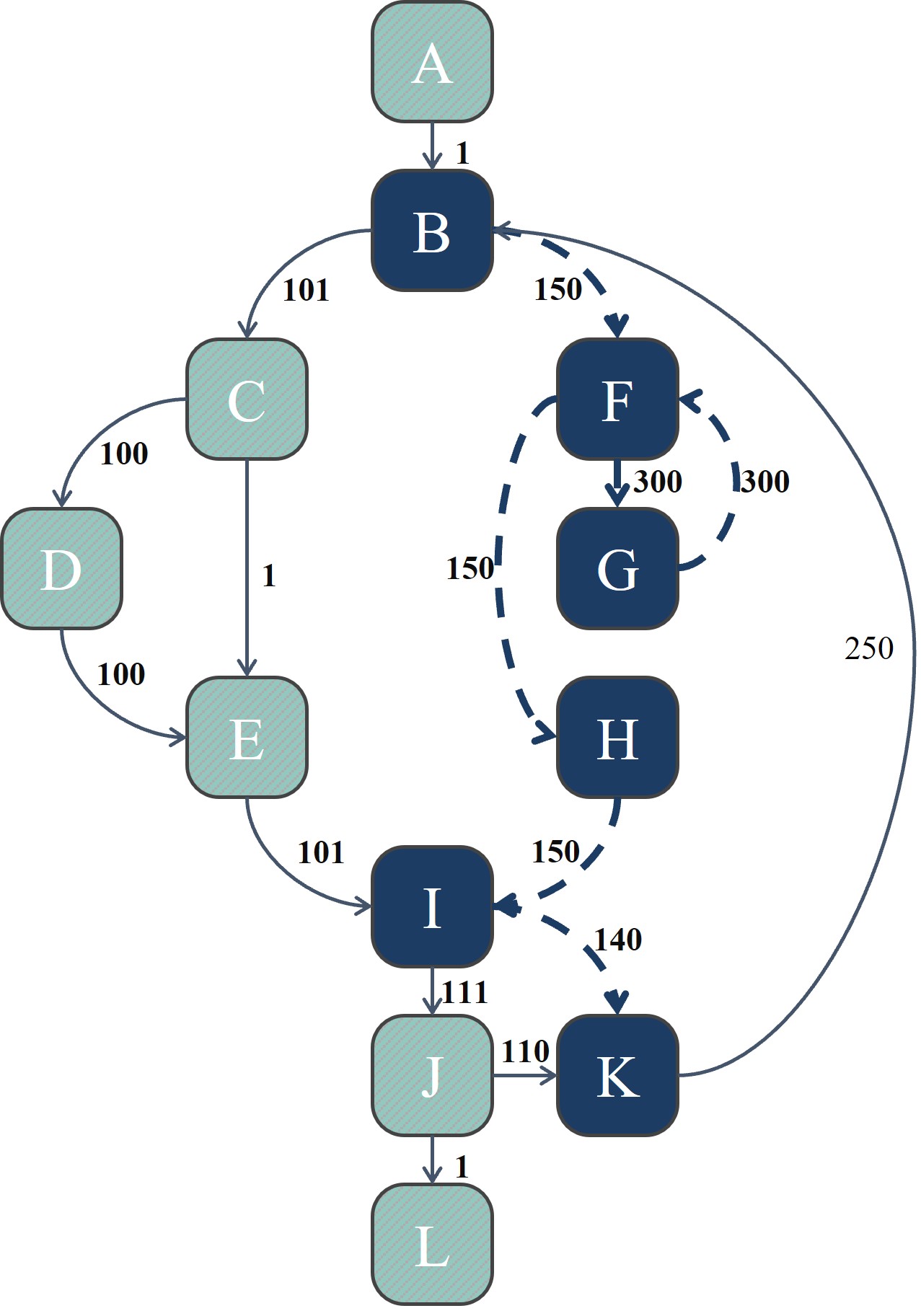}
    \label{fig:mess-cfg-1}}
  \hfill
  \subfloat[Hot-path candidate II]{
    \includegraphics[width=0.32\linewidth]{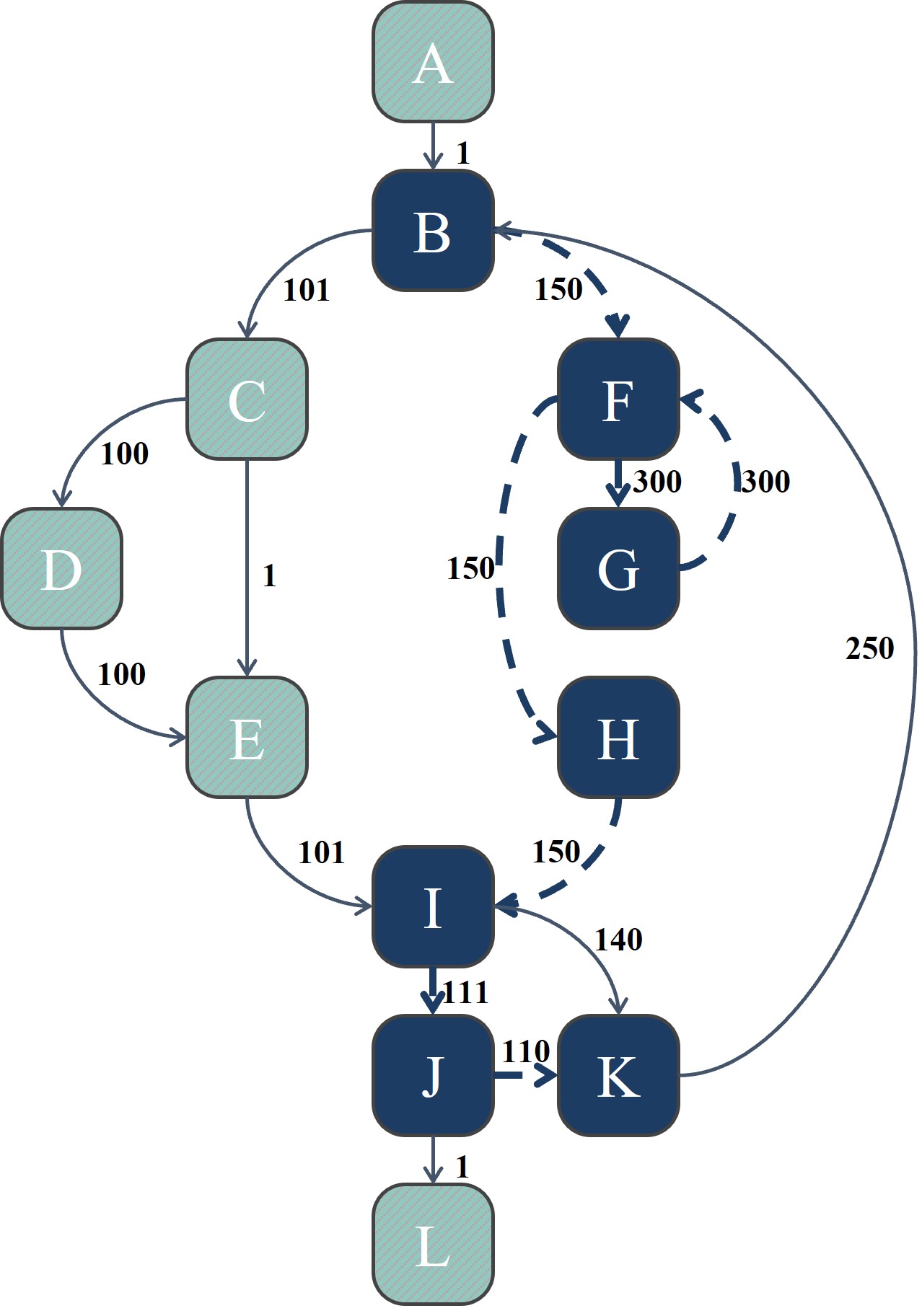}
    \label{fig:mess-cfg-2}}
  \caption{Different hot-path extractions with the same execution count. Orange-colored blocks and edges are hot.}
  \label{fig:mess-cfg}
\end{figure}

\input{contents/figs/instrument-edge-algorithm}

Traditional compiler optimizations rely on edge profiling, which infers program hotspots by only recording branch-jump frequencies. Its advantages are low overhead (only branch instructions are instrumented) and ease of implementation. However, edge profiling provides only local control-flow statistics and cannot capture the dynamic behavior of full paths—for example, differences in loop-nested paths or correlations across function-call chains. This limitation motivates \emph{path profiling}, which records the execution count of each acyclic path. A “path” is a sequence of edges from a function or procedure entry to its exit. Path profiling's strength lies in capturing the execution order of blocks, delivering a more faithful description of program behavior, and helping to identify frequently executed paths for aggressive optimizations (such as code-layout improvements or inlining decisions). Yet, path profiling is more complex to implement and can incur higher runtime overhead and memory usage, because the number of potential paths usually far exceeds the number of edges.
% 传统编译器优化依赖边剖析，通过记录分支跳转频率推断程序热点，其优势在于低开销（仅需对分支指令插桩）和易实现。然而，边剖析仅能提供局部控制流统计，无法捕捉完整路径的动态行为，例如循环嵌套中的路径差异或跨函数调用链的关联性。这种局限性催生了路径剖析(Path Profiling)的需求，即记录每条无环路径的执行次数。一个“路径”是从函数或过程的入口点到出口点的边序列。路径剖析的优势在于它能够捕获块的执行顺序，提供更真实的程序行为描述，并有助于识别频繁执行的路径，从而进行更积极的优化（如代码布局优化或内联决策）。然而，路径剖析的实现更为复杂，且由于潜在路径数量通常远大于边的数量，可能会导致更高的运行时开销和内存使用。

In the CFG shown in Figure~\ref{fig:mess-cfg}, acyclic paths (\textit{ABFHIK}, \textit{ABFHIJK}) could yield identical edge frequencies under traditional edge profiling, yet their actual execution counts differ greatly (for example, path \textit{HIK} executes 140 times in Figure~\ref{fig:mess-cfg-1}, whereas another possible hot path could be path \textit{HIJK} executed 110 times in Figure~\ref{fig:mess-cfg-2}).

Ball and Larus's path-profiling algorithm\cite{566449} , shown in Algorithm~\ref{alg:edge-values-dag}, introduces an incremental encoding scheme to avoid explicit path enumeration while still obtaining complete runtime path information. This method assigns a unique identifier (or index) to every feasible path in the code region. During execution, the profiler computes the path index incrementally and updates the counter for that complete path. Experiments show that its instrumentation overhead is only $1.3\times$ that of contemporaneous edge-profiling techniques.
% Ball-Larus \cite{566449}路径剖析算法提出了一种基于增量编码的路径标识方法避免了显示路径枚举，并可以获取程序运行的完整路径信息。该方法为代码区域中每个可行的路径分配一个唯一的标识符（或编号）。在执行过程中，剖析器会计算这个路径编号，并更新与该完整路径对应的计数器。实验表明其插桩开销仅为同时期边剖析技术的1.3倍。

% \input{contents/figs/edge-order-example}
\begin{figure}[t]
  \centering
  \subfloat[Simplified control-flow graph]{
    \includegraphics[width=0.275\linewidth]{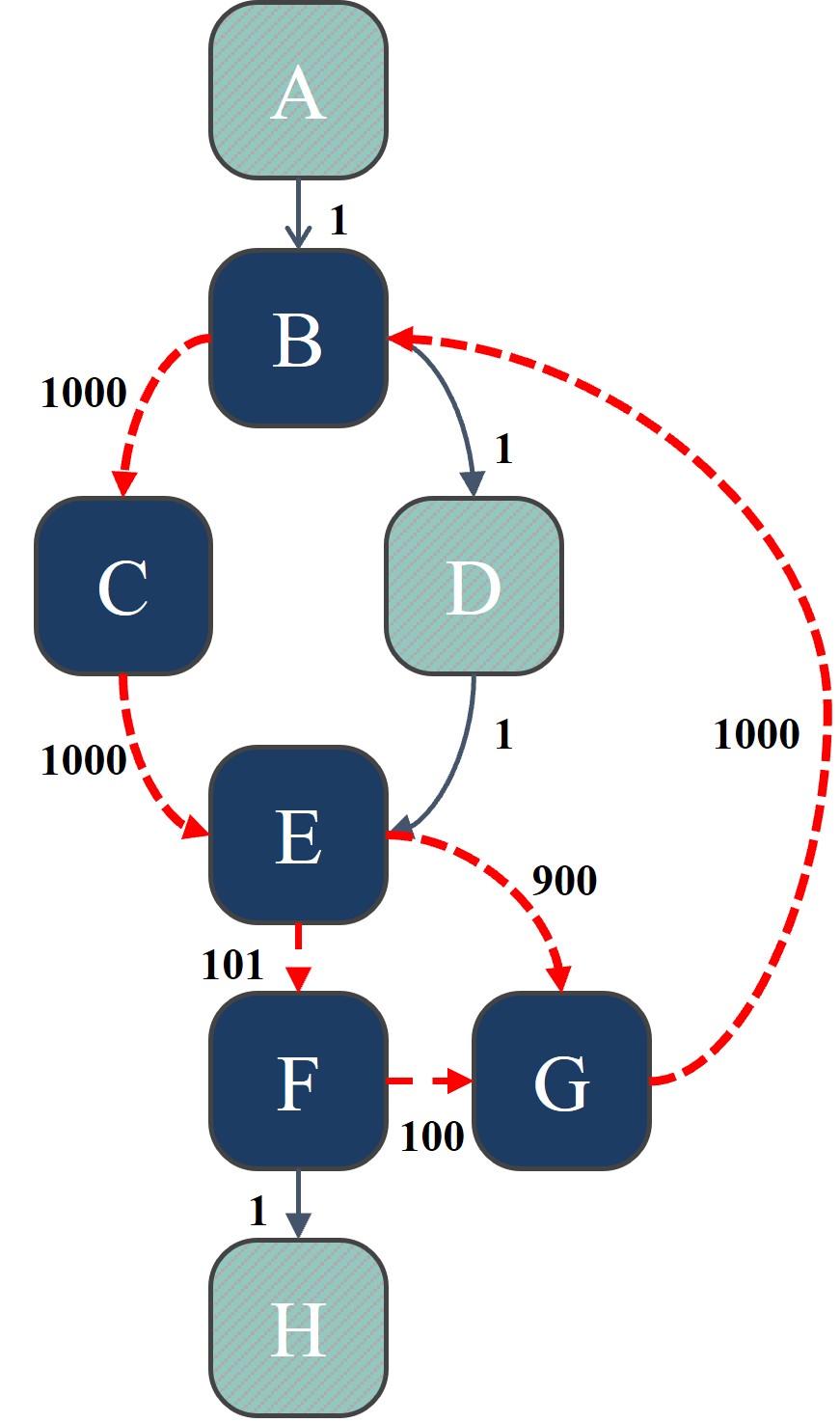}
    \label{fig:cfg3-original}}
  \hfill
  \subfloat[Edge-value assignment via Ball-Larus algorithm]{
    \includegraphics[width=0.24\linewidth]{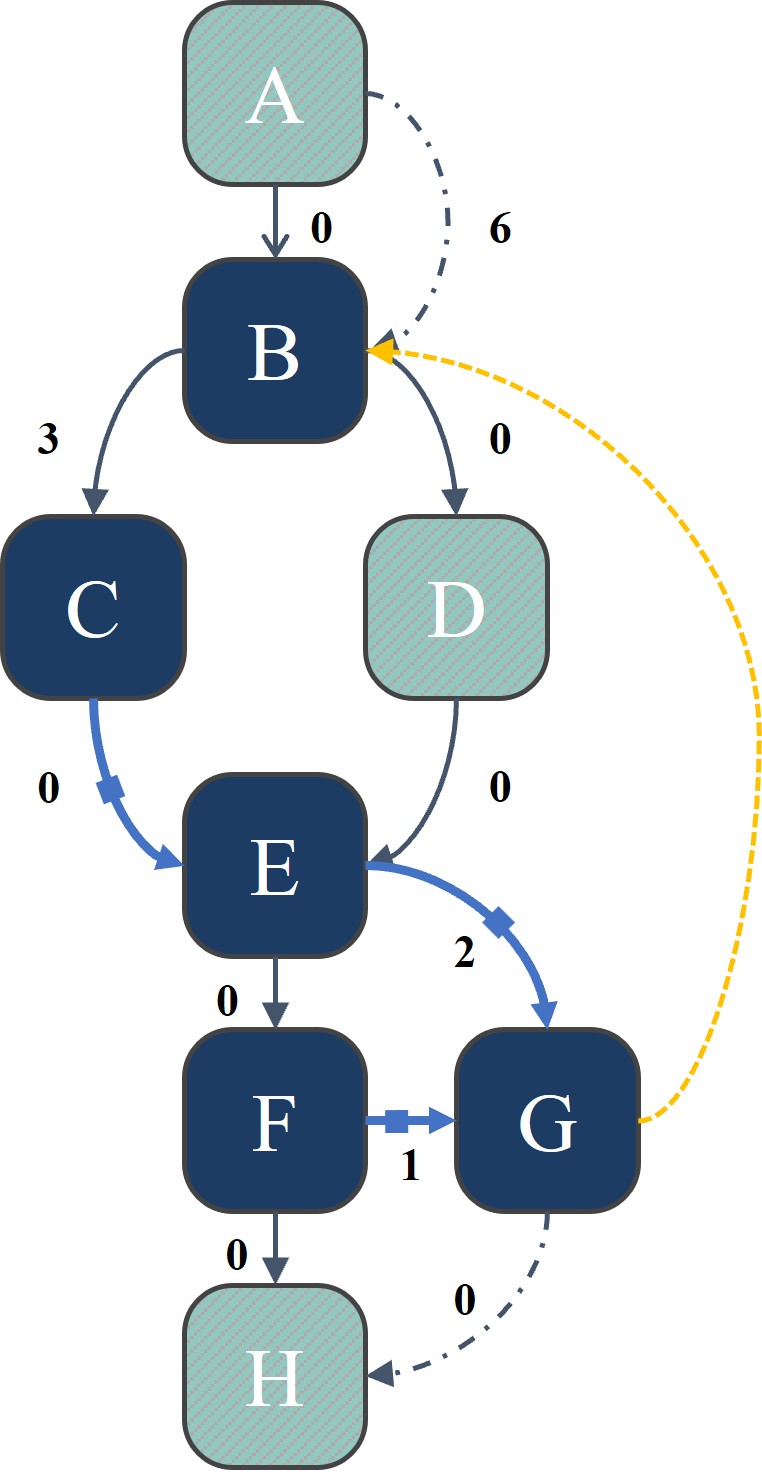}
    \label{fig:cfg-assgin-values}}
  \hfill
  \subfloat[Instrumented CFG with initial state and recording points]{
    \includegraphics[width=0.29\linewidth]{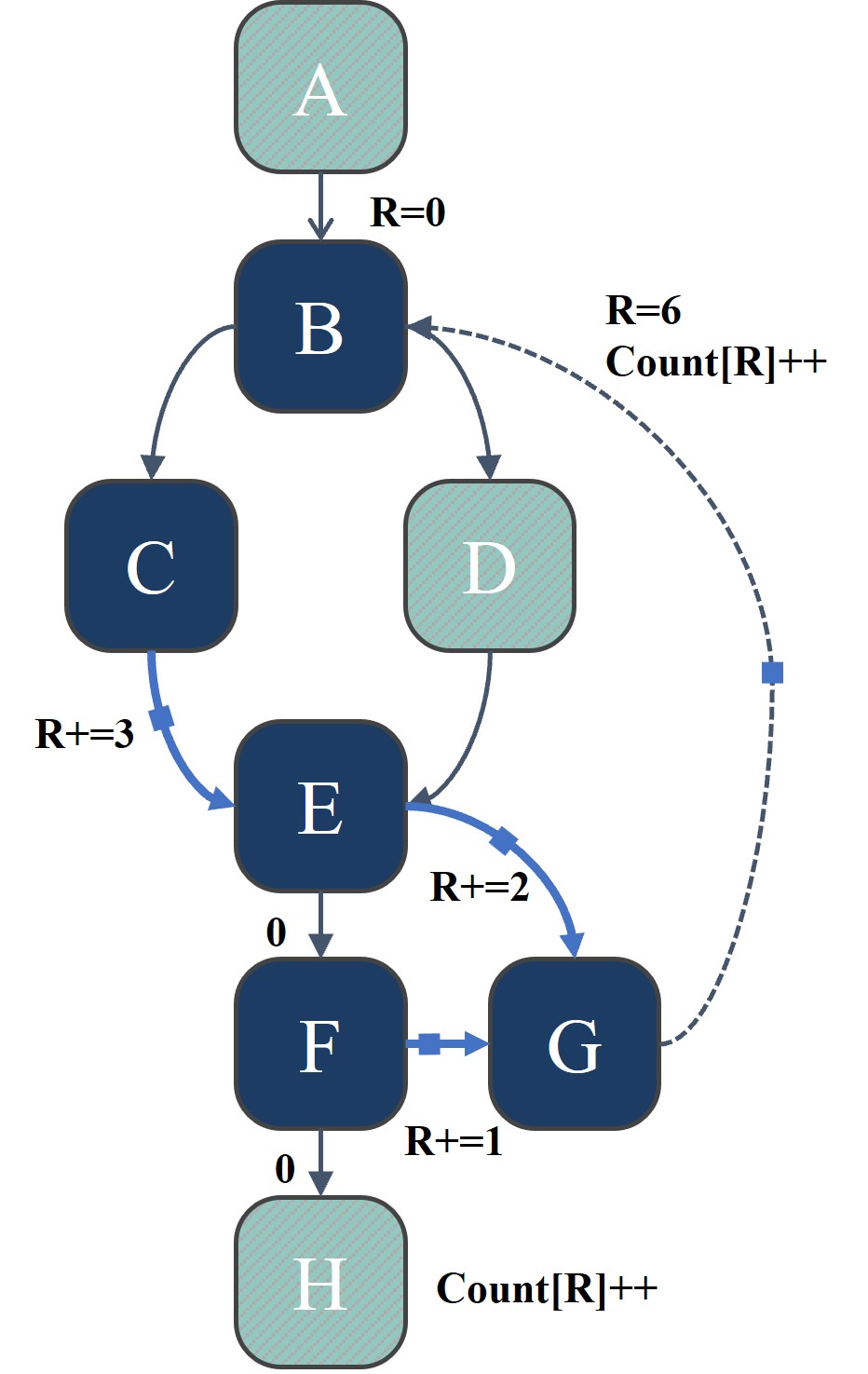}
    \label{fig:cfg-assign-counters}}
  \caption{From a raw CFG to a Ball-Larus-instrumented graph: (a) raw graph, (b) value calculation, (c) final instrumentation layout. Orange-colored (shallow-colored) blocks are hot blocks; Blued (dotted) edges are the complement edges of MST.}
  \label{fig:path-profiling-example}
\end{figure}

% check The Ball-Larus algorithm
As shown in Figure~\ref{fig:path-profiling-example}, the Ball-Larus algorithm instruments only the chords (edges not in the spanning tree, dotted edges in Figure~\ref{fig:cfg-assgin-values} and Figure~\ref{fig:cfg-assign-counters}) with small increment updates to give each acyclic path a unique index.
% 在图~\ref{fig:cfg-profile-example}所示的控制流图中，有六条无环路径（ACDF、ACDEF、ABCDF、ABCDEF、ABDF、ABDEF）在传统的边剖析（Prof1 与 Prof2）下呈现相同的边频率，然而这些路径的真实执行次数却大相径庭（例如路径 ABCDEF 在 Prof1 与 Prof2 中都执行了 100 次，而基于最高频边的启发式会选中 ACDEF）。Ball-Larus 路径剖析算法通过在最小生成树之外的弦边上插装微小的增量更新，实现对每条无环路径的唯一编号。
The instrumentation proceeds in four main phases:  

% TODO: 斜体？
\begin{itemize}
% 该算法分为四个主要阶段：  
\item \emph{DAG conversion}: Add a dummy edge \texttt{EXIT(H)\(\to\)ENTRY(A)} and remove all back-edges ($G \rightarrow B$) to transform the CFG into an acyclic directed graph (DAG) (dashed edges in Figure~\ref{fig:cfg-assgin-values}).

\item \emph{Edge-value assignment}: Traverse the DAG in reverse topological order, maintaining an array \texttt{NumPaths} at each vertex. For each edge \(e: v\to w\), set \(\mathit{Val}(e)=\mathit{NumPaths}[v]\) and then update \(\mathit{NumPaths}[v]\mathrel{+}= \mathit{NumPaths}[w]\), ensuring that the sum of edge values along each \texttt{ENTRY(A)\(\to\)EXIT(H)} path is unique and lies in \([0,n-1]\), where \(n\) is the number of acyclic paths.

\item \emph{Spanning-tree selection}: Choose a maximum-weight spanning tree based on static frequency estimates to minimize the number of chords to instrument. Each chord edge, carrying value \(\mathit{Val}(e)\), incurs a single register-add instruction at runtime.

\item \emph{Register and counter updates}: Initialize a register \(r=0\) at program entry. Upon traversing each instrumented(chord) edge, execute \texttt{r += Val(e)}; at \texttt{EXIT(H)}, perform \texttt{count[r]++} and reset \(r\) in loop scenarios. The final \(r\) directly indexes the counter array, precisely recording each acyclic path's count.
% 1. DAG 转换：通过添加伪边 \texttt{EXIT\(\to\)ENTRY} 并移除所有循环返回边，将 CFG 转化为无环有向图（DAG）。2. 边值分配：按照逆拓扑顺序遍历 DAG，对每个顶点维护 \texttt{NumPaths}，对于每条边 \(e: v\to w\)，先令 \(\mathit{Val}(e)=\mathit{NumPaths}[v]\)，再累加 \(\mathit{NumPaths}[v]+= \mathit{NumPaths}[w]\)，确保每条 ENTRY→EXIT 路径的边值之和唯一且分布在 \([0,n-1]\) 之间，其中 \(n\) 是无环路径数。3.生成树选择：基于静态频率估计选取最大权生成树，以最小化需插装的弦边；所有弦边根据其 \(\mathit{Val}(e)\) 在运行时通过一条简单的寄存器加法插装。4.寄存器与计数器更新：在程序入口初始化 \(r=0\)，遍历每条弦边时执行 \texttt{r += Val(e)}；到达 EXIT 时执行 \texttt{count[r]++}，并在循环情形中重置 \(r\)。最终 \(r\) 的值直接索引计数数组，精确记录无环路径次数。
\end{itemize}

For example, the six paths in Figure~\ref{fig:path-profiling-example} can be indexed from 0 to 5 (\textit{ACDF}$\rightarrow$0, \textit{ACDEF}$\rightarrow$1, \textit{ABCDF}$\rightarrow$2, \textit{ABCDEF}$\rightarrow$3, \textit{ABDF}$\rightarrow$4, \textit{ABDEF}$\rightarrow$5). During execution, the register \(r\) accumulates the increments on chord edges to yield the unique path identifier at \texttt{EXIT}. In loops, each back-edge is instrumented to “end the current path, update the counter, and reset \(R\),” and virtual edges at loop entry/exit separate and accurately count intra- and inter-loop paths.
% 例如，图~\ref{fig:cfg-edge-order-instrument-example} 的六条路径可被编号为 0 到 5（如 ACDF→0、ACDEF→1、ABCDF→2、ABCDEF→3、ABDF→4、ABDEF→5） ；程序执行过程中，每次经过弦边时累加相应增量，即可在 EXIT 处获得该路径的唯一标识。在处理循环时，每条回边均插装“结束当前路径并增量计数、重置 \(r\)”的代码，并在循环入口和出口引入虚拟边以确保循环内外的路径分离与统计准确 。

% 尽管收集的信息比边剖析更为详尽，路径剖析的运行时开销与边剖析相当：在 SPEC95 基准上，PP（路径剖析）平均开销 30.9\%，而高效边剖析为 16.1\%。在反馈导向优化中，精确的路径分布可为代码布局、分支预测和内联决策提供更细粒度的依据，从而提升优化效果。 

To address the limitation that traditional path profiling cannot capture control flows across procedures and loops, Larus et al. \cite{10.1145/301618.301678} extended path profiling by dynamically merging repeated path segments and statically precomputing common prefixes, reducing overhead to linear time and enabling global, cross-loop, and cross-procedure path methods for large programs. With the advent of dynamic compilation and language virtual machines, Bond et al. \cite{1402089} combined global flow information to distinguish hot and cold edges, thereby reducing instrumentation in hot paths. David \cite{10.5555/935616} studied context-sensitive path identifiers in multi-procedure settings, enabling full-path tracking across function boundaries.
% and supporting system-level optimizations.
% 为了进一步解决路径分析无法完整记录跨过程、跨循环的全局控制流造成的优化空间受限的问题。Larus 等人 \cite{10.1145/301618.301678} 对路径分析的工作进行扩展，通过动态合并重复路径片段和静态预计算公共前缀将路径分析的粗处开销降低至线性级别，从而使跨循环与跨过程的全局路径方法成为可能，为大型程序的全路径分析优化提供方法。随着动态编译以及语言虚拟机的兴起， Bond 等人 \cite{1402089} 则在此基础上结合全局程序流对程序中的冷热边进行识别，从而减少高频路径上的采样程序片段。而 David \cite{10.5555/935616} 则针对多过程交互场景下的路径分析进行研究，提出对调用上下文敏感的路径标识方法，实现了跨函数库的完整路径追踪方法，为系统级优化提供支持。
% efficient_path_profiling_1996
% Whole program paths
% Practical path profiling for dynamic optimizers
% Inter-procedural path profiling and the inter-procedural express-lane transformation
% Abstract execution: A technique for efficiently tracing programs

\subsection{Sampling-Based Profiling}
% \subsection{采样技术}

Efficient observation of a program's behavior is fundamental to profile-guided optimizations and performance tuning. However, software instrumentation-based profiling instruments monitoring instructions into the target code to capture execution traces (such as call paths), but its runtime overhead often reaches several to tens of times compared with the original running cost, making it impractical. This trade-off motivated the rapid development of software- and hardware-assisted sampling. Instead of fully recording the whole runtime behaviours, software sampling works in a shared library and let user select or automatically insert trace entries into the program to sampling the execution sequences.  Modern CPUs embed dedicated PMUs and programmable counters that trigger sampling on hardware events with minimal interference to program performance, thereby collecting profiling data at low overhead.
%However, while hardware sampling eliminates most runtime cost, it introduces a new challenge: accurately mapping sampled data back to source of program to reflect true execution behavior and places and support feedback-driven optimizations.
% 程序动态行为的高效观测是编译器优化、性能调优的核心基础。传统软件采样技术通过在目标代码中注入监控指令，能够有效捕获程序的执行轨迹（如函数调用路径等），但其运行时开销常达到数倍甚至数十倍，导致难以在生产环境中使用。这一矛盾催生了硬件辅助采样技术的快速发展：现代处理器往往内置独立的性能监控单元（PMU）以及可编程计数器，以在不显著干扰程序运行时性能的前提下，通过硬件事件触发采样行为，从而以极低的运行时开销收集程序运行剖析数据。然而，硬件采样在摆脱运行时开销的同时引入了新的问题，即如何将采样数据精准映射到程序逻辑片段中，以反映程序的运行时行为，从而支撑反馈驱动优化。

\subsubsection{Software Sampling}
% \subsubsection{软件采样}

Depending on when the sampling probes are injected, software sampling can be classified as \emph{static} or \emph{runtime} sampling. Static sampling inserts instrumentation during compilation or linking, fixing the sampling behavior before execution. This approach is deterministic, sampling points are known at compile time, so they can be optimized alongside the source, avoiding dynamic checks and switches at runtime and resulting in relatively low overhead.
% 根据插入采样指令片段的时间，可以将软件采样分为静态采样和运行时采样两种常见的软件采样方式。静态采样是指在程序编译或链接阶段插入采样逻辑，采样行为以及采样目标在程序运行前就已确定。这种采样方式因其采样点在编译时确定，具有确定性，可以在编译时与源程序一同被优化，且避免运行时的动态判断和切换而具有较低的运行时开销的特点。运行时采样即在程序运行的过程中动态的插入采样指令片段和触发采样行为，这种采样方式更加灵活，其行为可以根据程序运行的实际状态进行调整。
Runtime sampling, by contrast, dynamically injects sampling code and triggers during program execution, offering flexibility: sampling actions and targets can be adjusted at runtime. Typical runtime-sampling frameworks are dynamic, highly flexible, but incur significant overhead. Based on runtime events (e.g., branch-misprediction, cache misses) and intermediate results, runtime sampling employs feedback mechanisms (such as hotspot detection or event triggers) to adjust its scope and frequency. Being dynamic, it can capture context information, like call stacks or data flows, at various granularities (instruction-, basic-block-, function-, or module-level), but must pay the cost of runtime branchings, context switches, and data processing.
% 典型的运行时采样程序具有动态性、高灵活性以及较高运行时开销等特点。根据程序运行时行为（如分支预测失败、缓存未命中等）及采样结果，运行时采样可以通过反馈机制（如热点检测、事件触发）优化采样的范围及频率。对于可采集的数据，运行时采样因其动态性，可以针对性的捕获程序运行过程中的上下文信息（如调用栈、数据流等）并进行多种粒度的采样（如指令级、基本块级、函数级、模块级等多层次采样）。由于运行时采样的动态性和灵活性，程序将不可避免的引入多种采样逻辑判断、切换和数据处理引入的开销。

With the rise of dynamic languages like Java, \emph{Dynamic Binary Instrumentation} (DBI) emerged to analyze runtime behavior without recompiling programs. Traditional methods, separating instrumentation build and optimization build, and, therefore, cannot switch between sampling and native execution. To address this problem, Omri et al.\ \cite{traub2000ephemeral} introduced \emph{Ephemeral Instrumentation}, allowing probes to be dynamically inserted or removed without altering code layout. By modeling CFG nodes and edges with an irreducible finite-state Markov chain, they inferred execution frequencies with reduced overhead. Industry soon produced several DBI frameworks, such as Valgrind, Pin, and DynamoRIO. Valgrind \cite{10.1145/1250734.1250746} uses dynamic binary translation (DBT) to lift the binary to an IR, insert probes, and tracks state with shadow registers and memory modeling—trading performance for deep introspection. Pin \cite{10.1145/1064978.1065034} employs JIT insertion of instrumentation into the binary, with code-cache warmup to compile frequently executed probes into native code, boosting sampling performance. It offers layered APIs at instruction, basic-block, and routine granularity, and can enable or disable specific probes at runtime. DynamoRIO \cite{bruening2020dynamorio} emphasizes extensibility and speed, letting users freely insert probes and modify program behavior while maintaining high performance.
% 早期随着 Java 等动态语言的流行，催生了动态二进制插桩技术(Dynamic Binary Instrumentation)用以分析程序运行时的行为及状态。动态二进制插桩技术允许开发者在不重新编译程序的情况下分析和改变程序的行为。为了解决传统方法（如完全插桩）需构建插桩版本程序，并无法切换采样与原生执行的问题， omri 等人 \cite{traub2000ephemeral} 提出短暂插桩(Ephemeral Instrumentation)技术,其可以在不改变程序代码布局的情况下动态插入或移除插桩代码，并通过不可约有限状态马尔可夫链推导控制流图中的节点与边的执行频率，从而降低采样开销。随后，业界根据不同优化方向提出了不同动态采样框架。其中以 Valgrind \cite{10.1145/1250734.1250746}, Pin \cite{10.1145/1064978.1065034}，以及 DynamoRIO \cite{bruening2020dynamorio} 等框架最为经典。Valgrind 通过动态二进制翻译(Dynamic Binary Translation, DBT)技术，将原始二进制程序转换为中间表示(IR)后再插入采样逻辑。并使用了影子寄存器(shadow registers)和内存建模技术，跟踪程序运行的状态。这种技术虽然会极大降低程序的运行效率，但可以允许使用者获取近乎任何数据，从而进行深度分析。Pin 则使用 JIT (Just-in-time) 技术，允许采样代码动态的插入二进制执行文件中，并采用代码缓存预热技术将频繁执行的插桩代码编译为原生指令从而大大提升了采样本身的性能。并且，Pin提供了分层插桩API,提供指令级(INS)、基本块级(BBL)、以及函数级(RTN)等多粒度接口且支持运行时动态的启动或禁用特定的采样检测点。而 DynamoRIO 则强调其采样行为的多样性与性能，即允许使用者自由的插入所需的采样逻辑以及改变程序本身的行为方式，并保持采样的性能。
% Ephemeral Instrumentation for Lightweight Program Profiling

As hardware became ubiquitous and performance demands grew, researchers focus on reducing sampling overhead. Arnold and Ryder \cite{10.1145/378795.378832} designed a low-overhead instrumentation framework using fine-grained sampling. They maintain two code versions, a fully instrumented version and a near-native “checkpoint” version, with a global counter to trigger sampling checks inserted by the compiler. Hyoun et al.\ \cite{6494982} enhanced DynamoRIO to periodically switch between native and instrumented execution, tuning sampling frequency and window size to balance overhead and data volume, and employed hardware pre-sampling to cache probe fragments, combining low overhead with dynamic flexibility. Moseley et al.\ \cite{4145115} parallelized instrumentation by spawning a shadow process on a separate core to run probes, while the main process runs natively; copy-on-write and system-call virtualization minimize memory and context-switch costs, achieving >90\% accuracy at <1\% overhead. To handle verbose context data in large applications, John \cite{10.1145/337449.337483} sampled on OS clock interrupts or context switches, tagging stack frames to distinguish call chains, and built a partial call-context tree (PCCT) with dynamic pruning and merging to control storage growth. More recently, for data-center-scale workloads, Martin and Shiraz \cite{Hirzel2001BurstyTA} proposed \emph{Bursty Tracing}, extending the Arnold-Ryder framework to support long-burst sampling across procedure boundaries, making sampling sensitive to sudden load surges and capturing long-range causal relationships to accelerate feedback to the optimizer.
% 随着计算机硬件的快速普及，生产及消费环境对性能的敏感度的提升，研究者着力解决采样造成的性能损耗问题。Arnold 以及 Ryder \cite{10.1145/378795.378832} 设计了一种低开销的代码插桩框架，通过细粒度采样技术减少性能损耗，使动态优化系统能够在运行时高效收集分析数据。框架提出使用双工采样的策略，即维护一份包含完整的插桩逻辑的插桩版本代码以及一份与源代码几乎一致的仅插入检查点的检查版本代码，使用全局计数器（globalCounter）控制采样间隔，通过编译器插入的检查点代码实现低开销触发。Hyoun 等人 \cite{6494982} 则通过修改 DynamoRIO 框架周期性的切换原生执行与插桩执行，通过控制采样频率与采样长度等参数来平衡采样开销与收集的数据量。此外，其工作还使用硬件进行预采样，提前对插桩程序片段进行缓存，从而既规避了传统插桩的高开销，又突破了硬件剖析的灵活性限制。Moseley 等人 \cite{4145115} 则通过创建影子进程，在独立核心执行插桩代码，主进程继续原生运行的方式进行并行化插桩，并通过写时复制 (Copy on Write) 避免内存操作的冗余开销，结合系统调用虚拟化和动态参数调整，实现了高精度（>90\%）与低开销（<1\%）的平衡。针对大规模程序上下文信息冗长，传统工具对于上下文不敏感的问题，John \cite{10.1145/337449.337483} 通过操作系统时钟中断或上下文切换触发法采样，从而控制频率。同时在栈帧中设置标记位，区分新旧调用链并结合树状数据结构建立部分调用上下文树(PCCT)并结合动态剪枝与合并控制存储存储规模。近些年随着数据中心级应用的需求，研究者开始探索尝试解决采样在大规模并发程序中遇到的挑战。Martin 和 Trishul \cite{Hirzel2001BurstyTA} 根据现代大规模程序负载突增的情况提出了突发追踪 (Bursty Tracing) 技术。该技术扩展了 Arnold-Ryder 框架，通过在检查代码与插桩代码之间动态切换支持对跨越过程边界的执行轨迹进行长突发(Long Bursts)采样，对大规模程序负载如突然增加的情况更为敏感并可以捕捉长序列间的因果关系，从而可以加速对优化器的反馈。

% A framework for reducing the cost of instrumented code
% Instant profiling: Instrumentation sampling for profiling datacenter applications
% Shadow Profiling: Hiding Instrumentation Costs with Parallelism
% A portable sampling-based profiler for Java virtual machines
% Bursty Tracing: A Framework for Low-Overhead Temporal Profiling

% \paragraph{静态采样}

% 总结
% 当前软件插桩技术已进入"生产级可用"阶段，在Google、Meta等公司的数据中心实现常态化部署。未来发展方向将聚焦于零感知插桩（zero-overhead instrumentation），通过硬件辅助和算法创新，最终实现运行时优化的完全自动化闭环。

\subsubsection{Hardware Sampling}
% \subsubsection{硬件采样}

Although the software sampling mitigate the run time compared with instrumentation-based profiling, the cost of sampling is still non-negligible. Further, researchers began exploring hardware-assisted sampling methods. Early work focused on leveraging on-chip units to perform low-perturbation sampling, overcoming the inherent performance bottlenecks of software-based instrumentation and sampling.
% 为了解决传统软件插桩采样技术运行时开销难以满足生产环境的要求这一矛盾。研究者开始探索使用硬件采样替代软件插桩采样的方法。早期研究主要聚焦于利用硬件模块实现低干扰的采样，突破传统软件插桩带来的固有性能瓶颈。

With the widespread adoption of hardware branch-prediction structures (e.g., the BTB in Pentium and PowerPC), it became possible to sample hardware state to reflect runtime behavior. Thomas et al.\ \cite{717402} first used the branch-target buffer to count branch-direction frequencies for estimating basic-block weights, and combined a two-level predictor to record branch history, reducing sampling overhead to under 5\%. However, accurately sampling indirect jumps and low-frequency paths remained challenging. To mitigate these issues, the work replaces the BTB with programmable counters to avoid read/write contention and employed write-after-increment buffering to prevent interference with critical-path execution. They also investigated methods to index hardware-event samples. Craig et al.\ \cite{903267} proposed a design for a standalone programmable coprocessor with dynamic resource management (e.g., prioritizing hot instructions) as a direction for future CPU.
% 随着硬件分支预测机制（如 Pentium、PowerPC 中的 BTB）的普及，通过硬件采样以反应程序运行时状态成为可能。Thomas \cite{717402} 等人率先利用分支目标缓冲区统计分支方向频率估算基本块权重，并使用两级分支预测器（HRT + PT）记录分支历史。将采样开销降低到 5\% 以内。但是，硬件采样带来的对间接跳转、低频路径的准确采样造成的干扰仍需进一步研究。为了解决这一问题，Thomas 等人使用计数器代替缓冲区以避免大量读写冲突带来的精度损失，并采用写后更新侧绿避免采样影响关键路径的执行。此外，研究还探索了硬件采样事件的索引问题。Craig 等人 \cite{903267} 则提出使用定制的独立可编程协处理器与动态资源管理策略（如优先分析高频指令）为未来处理器提供了设计方向。

% Using branch handling hardware to support profile-driven optimization
% A programmable co-processor for profiling

As multicore processors and hardware monitoring evolved, mainstream architectures (such as x86-64 and ARM64) incorporated dedicated PMUs to record hardware events during program execution. Users can collect PMU data via tools like GNU Perf on Linux. This event-based sampling is commonly called Event-Based Sampling (EBS).
% 随着多核处理器的普及与硬件技术的发展，主流架构处(如 AMD64 以及 Arm)理器都在硬件层面上设置了专门监控和处理软件在硬件上运行时产生的事件与状态的硬件单元 -- 性能监控单元。用户可以通过专门的硬件事件采样软件（如 Linux 中的 GNU Perf）对软件运行时产生的硬件信息进行记录。这种基于硬件事件的采样技术被称为基于事件采样(EBS)。

Precise Event Sampling (PES) is a profiling feature in commercial CPUs that precisely samples hardware events and attributes them to the instructions causing the events. PES is widely used for application performance analysis, identifying issues such as false sharing in multithreaded code, long-latency remote memory accesses, poor data locality, bandwidth bottlenecks, and cache-miss behavior. By sampling the instruction pointer and effective addresses, PES helps pinpoint the source code locations and data objects responsible for performance problems. Different CPU architectures and even different implementations within the same ISA (e.g., Intel vs. AMD on x86-64) provide their own solutions.
%In the following paragraphs, we outline and compare these architectures' PES implementations.
% 精确事件采样 （Precise Event Sampling, PES）是一种在商业处理器中采样硬件事件以及精准定位引起事件发生的指令的一种剖析功能。该功能大量的被应用与对应用程序的性能分析，以识别应用程序运行中所遇到的性能平静。如检测多线程代码间的错误共享、长延时远程内存访问瓶颈、数据局部性问题、带宽浪费所带来的性能降低问题、以及缓存冲突所造成的未命中问题。此外，PES 还可以通过对指令指针以及操作的有效地址进行采样，以帮助标定造成瓶颈的源代码以及数据对象。不同架构的处理器如 AMD64 、ARM 、RSICV ，甚至相同架构的不同实现如 INTEL 和 AMD 公司的产品都有自己相对应的解决方案。本节将阐述各个硬件架构 PES 相关技术的异同。

While the basic implementation of sampling --- periodic sampling when overflow of hardware counters --- are common across platforms, each CPU family exposes a unique set of event types, registers, and overflow behaviors. In the following, we first describe how two mainstream ISAs on x86-64 support PES, and then briefly turn to the problem on ARM64 architecture.
% In order to leverage low-overhead, high-precision profiles in Profile-Guided Optimization, compilers and tooling must be tuned to the precise PMU interfaces of the target architecture.  

\paragraph{x86-64 Architecture}
% \paragraph{AMD64 架构}

% TODO: 描述图标
% TODO: PEBs PES 注意

Dean et al.\ \cite{645821} introduced \textsc{ProfileMe}, a hardware-software co-design for out-of-order CPUs that randomly samples instructions, tagging them on pipeline entry and recording events such as cache misses, branch outcomes, and stage latencies during execution. They also implemented “paired sampling” to capture concurrency and resource usage of potentially simultaneous instructions. AMD's Instruction-Based Sampling (IBS) \cite{5452049, amd64vol2} adopts this instruction-level sampling approach. Each core's IBS unit comprises two control registers, two internal counters, and several Model-Specific Registers (MSRs) for event data. IBS supports two sample modes: \texttt{IBS\_fetch} and \texttt{IBS\_op}. Users program the control registers for desired events; the internal counters then track occurrences and log associated information in MSRs. In \texttt{IBS\_op} mode, IBS captures instruction-emission and loop-iteration data.
% Dean 等人 \cite{645821} 提出了 Profile 一种在乱序执行处理器上通过对指令进行随机采样来收集细粒度性能信息的硬件-软件协同系统。它在指令进入流水线时附加标签，并在指令执行过程中记录缓存缺失、分支预测、各流水线阶段延迟等丰富事件，以实现对单条指令的精确归因和分析。还引入了“配对采样”技术，可以同时采样两条可能并发执行的指令，揭示指令间的并发度和资源利用情况。AMD 公司的 PES (即 IBS) 采用了这种基于硬件的指令采样技术 \cite{5452049} 。每个 CPU 内核中的 IBS 设备都由以下几个部分组成：两个控制寄存器，两个 IBS 内部的计数器，以及一些用于采样的特定型号寄存器（Model Specific Registers, MSRs）。AMD IBS 技术只允许两种采样方式，即取指令采样（IBS fetch）以及微码采样（IBS op） \cite{amd64vol2}。用户可以针对根据感兴趣的事件对相应的控制寄存器进行编程，之后相应的内部计数器将对被监控的事件进行计数，并且在 MSRs 中记录与该事件相关的信息。对于微码采样来说，IBS 则可以采集始终循环信息或微码发射信息。%图 \ref{} 展示了 AMD IBS 模块的结构及设置和采样的流程。
% ProfileMe
% Incorporating Instruction-Based Sampling into AMD CodeAnalyst
% AMD64 Architecture Programmer's Manual Volume 2: System Programming.
% \begin{figure}[h]
%     \centering
%     \includegraphics[width=0.4\textwidth]{AMD-IBS.png}
%     \caption{AMD-IBS 工作流程}
%     \label{fig4}
% \end{figure}

Intel's PES technique, called precise event based sampling (PEBS), is implemented entirely within the PMU module \cite{intel_sdm}. The PMU includes global control registers to enable event counting, status registers indicating which events are supported and which counters have overflowed, event-select registers for choosing hardware/software events (e.g., instruction retire, page faults), and performance counters. Counters can be programmed to overflow after a fixed interval; on overflow, PEBS captures the next occurrence of the monitored event, snapshots CPU state into a PEBS buffer, and then stops sampling. When the buffer fills to a threshold, a hardware interrupt is raised; the OS handler reads and clears the buffer, then signals the user process. The process can then analyze the data to identify performance bottlenecks.
% Intel 公司的 PES 实现则全部集中在 Intel CPU 的 PMUs 模块中 \cite{intel_sdm}。该模块含有数个寄存器和计数器。其中全局寄存器被用来控制是否开启全局的事件计数器，以及各个计数器的功能。状态寄存器则包含了 PMUs 能够支持的事件以及各个事件计数器是否溢出。事件选择寄存器则用来存储被选择的软硬件事件，如指令完成、指令加载、页错误等错误。性能监控计数器则记录了这些被选择事件发生的数量。这些计数器可以被配置为每隔一定次数的事件（即采样间隔）就产生一次溢出。当计数器溢出时， PEBS 将会捕获该被监控事件的下一次出现，当该事件出现时，机器状态将会被 PEBS 的辅助机制拷贝到 PEBS 缓冲区。此时，一次完整的采样过程结束。PEBS 辅助机制将收集到的所有数据归类到名为 PEBS record 的数据结构中。当缓冲区的记录达到一定阈值则会产生一个硬件中断，此时操作系统将根据注册好的中断处理程序读取缓冲区的数据并清空缓冲区，随后向用户进程发送信号。用户进程收到相应的信号后则可以根据情况进行数据分析，找出降低程序性能的瓶颈。%图 \ref{} 展现了采样中的一次硬件中断的产生过程。

%Intel Manual
% \begin{figure}[h]
%     \centering
%     \includegraphics[width=0.4\textwidth]{AMD-IBS.png}
%     \caption{Intel-PEBS 工作流程}
%     \label{fig5}
% \end{figure}

\paragraph{ARM64 Architecture}
% \paragraph{ARM 架构}

% TODO: 拓展

% ARM64 implements a Statistical Profiling Extension (SPE) similar to PES. However, studies have shown that ARM64's SPE support for precise event sampling is not yet as mature as PEBS or IBS: SPE may exhibit sampling bias on branch instructions and under-sample instructions immediately following branches. Since hardware-sampling-based feedback optimizations are less common on ARM, we limit our discussion here.
% ARM 架构则采用了与 IBS 类似的 Statistical Profiling Extension (SPE) 采样方法，即 PES 模块。但是研究表明架构对精确事件采样的支持尚不成熟，无法与 PEBS 与 IBS 技术进行比较。如 SPE 可能会对采样的指令产生统计数据上的偏移，对分支指令的额外采样数量和对紧随分支目标指令的某些指令低采样技术。考虑到反馈优化技术没有在 ARM 架构上普及，因此本文章不过多的介绍 ARM 架构中的硬件采样技术。

Arm's Statistical Profiling Extension (SPE) is a hardwar-assisted, sampling-based profiling mechanism introduced in the Armv8.2-A architecture to provide precise, low-overhead visibility into instruction-level performance events. Unlike traditional PMU sampling, SPE employs a down-counter, which is initialized with a programmable interval plus a small random skew, to select individualinstructions or micro-ops for detailed tracing. When the counter reaches zero, the core buffers a “trace packet” for the sampled operation into an memory buffer; only upon buffer mark or end-of-buffer does the hardware generate an interrupt to offload the collected data. This design combines the buffering benefits of PEBS (no per-sample interrupt) with the rich per-instruction detail of IBS.

Each SPE trace packet records the sampled instruction's program counter, execution latency, and microarchitectural events encountered by that instruction. For memory operations, SPE logs the virtual data address, cach-hierarchy source (L1/L2/LLC/DRAM), and TLB miss indicators; for branches, it can indicate taken/not-taken and in some implementations misprediction. Programmable filters allow tuning SPE to record only loads, stores, branches, or operations exceeding a latency threshold, thus focusing profile collection on the most performance-critical events. Because logging is handled entirely in hardware and buffered, SPE imposes very little runtime distortion—even under moderate sampling rates, overhead remains on the order of a few percent, and when overloaded it simply drops excess samples rather than stalling execution.

%  TODO references
The rich, statically sampled profiles produced by SPE can directly feed into a hardwar-sampling PGO workflow on ARM64, analogous to AutoFDO on x86-64. A representative workload is run with SPE enabled, producing an aux-buffer trace that is post-processed into a sequence of sampled PCs and associated event attributes. With these PC-frequency counts mapped to basic blocks or edges, after symbolizing and aggregating, the compiler uses them as weighted profiles to guide optimizations. Miksits et al.\ \cite{10.1109/SCW63240.2024.00139} confirm that SPE-driven PGO achieves high fidelity (more than 99\% hotspot identification) with under 5\% overhead when properly tuned, delivering performance gains close to instrumentation-based PGO while eliminating the dual-build burden. By providing a balanced blend of precise per-instruction detail and hardwar-buffered low-overhead sampling, Arm SPE could fill the role of PEBS/IBS on x86-64 and enable practical, production-scale PGO on ARM64 platforms.

Table \ref{tab:hw-sampling-compare} presents a comparison of the three hardware sampling facilities: AMD IBS, Intel PEBS, and Arm SPE. The ``sampling trigger'' row shows that while PEBS and SPE both rely on programmable counter overflows, IBS further allows sampling at both fetch and execution stages with adjustable depth. Under data recorded, all three capture the program counter, but only IBS and SPE include rich microarchitectural details (e.g.\ cache/TLB source and branch outcomes), whereas PEBS is limited to architecturally visible events.  For buffering, SPE's AUX buffer most closely mirrors PEBS's MSR region approach but is augmented by watermark-driven offload, ensuring minimal skid without stalling execution. Notably, in precision, SPE achieves zero skid—every sample maps exactly to the retired instruction, whereas PEBS may skid by a few instructions and IBS requires more dedicated logic to maintain precision. The programmable filters row highlights SPE's unique ability to apply latency thresholds in addition to event-type filtering, which focuses profiles on the most performance-critical operations. Finally, SPE overhead, lower than both PEBS and IBS, making it especially attractive for production-scale profiling.

\input{contents/tables/intel-PEBs-vs-amd-IBS}

\subsubsection{Mapping Profile Data Back to Source}
% \subsection{程序剖析信息与源程序的映射}
Raw profiling data collected via hardware counters associates performance events with binary addresses and low-level counters. To make this information actionable in the compiler, we must map it back into source-level constructs such as functions, basic blocks, and loop headers.  Accurate mapping enables profile-driven optimizations (e.g., inlining hot call sites, unrolling frequently executed loops, and reordering branches) to operate on the program's actual structure and semantics. Moreover, reconstructing the full runtime behavior ensures a CFG-consistent profile, by resolving sparse or noisy samples into complete edge and path frequencies under flow-conservation constraints, which drives code-layout, branch-prediction, and other downstream passes with both precision and reliability. 

\paragraph{Challenges of Hardware Tracing}
% \subsubsection{硬件追踪造成的问题}
A key challenge in hardware sampling is extracting insights from vast, real-time event streams. Traditional tools aggregate data too coarsely to distinguish between different call contexts, information critical for advanced optimizations. Ammons et al.\ \cite{10.1145/258915.258924} addressed this by embedding lightweight probes at path entry and exit, combining flow-sensitive and context-sensitive analysis to bind hardware samples to exact execution paths and calling contexts. They initialize counters at path entry, read them at exit, and record calling contexts in a Calling Context Tree (CCT). This approach differentiates a function's performance under recursion and normal calls and accurately identifies hot segments along complex paths. With SPEC95, they found that a few hot paths accounted for most L1 cache misses, guiding targeted optimizations. While improving precision and reducing probe overhead, this method still faces storage and compute challenges due to exponential path growth.
% 在硬件采样数据的利用上，如何从大量、实时的硬件事件中提取出真正有助于优化的信息也成为一个关键问题。传统工具由于数据过度聚合，往往无法区分同一代码单元在不同调用上下文中的行为差异，而这正是一些高级优化策略所迫切需要的信息。对此，Ammons 等人 \cite{10.1145/258915.258924} 提出了一种全新的分析方法，该方法通过在程序执行路径中嵌入轻量级插桩代码，并结合流敏感（flow-sensitive）与上下文敏感（context-sensitive）的分析策略，将硬件采样数据与具体执行路径和调用上下文精细绑定。具体来说，该方法在路径入口处初始化硬件计数器，在出口处读取采样数据，然后利用构造调用上下文树（Calling Context Tree, CCT）的方式记录动态调用链。通过这种方法，不仅能够区分同一函数在递归调用与普通调用中的性能表现，还可以准确捕捉到复杂执行路径中的热点部分。在 SPEC95 基准测试中，研究表明少数热点路径贡献了大部分 L1 缓存未命中情况，为后续的性能调优提供了明确方向。虽然该方法在降低插桩干扰和提升数据精度上取得了显著成果，但面对路径数量呈指数级增长的问题，其在存储和计算上的开销仍需要进一步优化。
% Exploiting Hardware Performance Counters with Flow and Context Sensitive Profiling

Unlike software-instrumentation profiles, hardware samples record only instruction addresses and event counts. To leverage this data in the compiler, instruction-level samples must be mapped onto the compiler's IR. This requires mapping each sampled address back to the corresponding source location (line or IR block) so that the compiler can annotate the CFG with frequency or event weights. In theory, all instructions within a basic block share the same frequency, but hardware sampling bias can distort measurements. Below, we describe these problems and their mitigations in the following three aspects.
% 相比于基于插针技术产生的统计信息，基于硬件采样产生的统计信息仅仅是一种对于指令地址及事件的记录。为了使编译器能够使用这些信息，需要将指令级别的统计信息转化为基于编译器中间语言（IR）的标注统计信息。为此，需要先将针对每一条指令的统计信息进行指令地址至源码位置的反射，从而体现在指令对应的源代码（行）的频率或是值的信息。在优化过程中，编译器读取统计数据中的频率数据来对控制流图进行标注。理论上来说，每个基本块内的指令出现的频率应当相等。但由于硬件产生的统计信息也有自身的局限性，一些信息不能够被准确的记录下来，导致统计数据失真。本节将介绍这些局限的表现形式和解决办法。

\begin{itemize}
% \paragraph{Sampling Synchronization}
% \paragraph{采样同步问题}
    \item \textbf{Sampling Synchronization} Sampling period is critical. If the sampling interval synchronizes with a program loop period of $k$ instructions, only one instruction per iteration will be sampled, while others are omitted. Randomizing the sampling period by adding a fresh, per-sample random offset to the PMU interval ensures more uniform coverage without fixed alignment.
    % 对于一个采样行为，采样频率是其重要的属性。如果选择的采样的周期与程序运行中的对部分子程序的调用周期同步，那么只有少数本分的代码会被统计。例如如果一次循环中有 $k$ 条指令，而采样周期被选为 $k$ 的倍数。则循环中仅有一条指令会被统计。随机化采样周期可以有效避免这种情况的出现。通过将 PMU 的采样周期设置为一个固定的值加上一个随机的偏移量可以使采样的样本分布更加均匀。注意随机的偏移量不是固定的，在每次采样后都会重新选择一个新的偏移量。因此不会出现偏移问题。

% \paragraph{Sampling Skid}
% \paragraph{采样滑移问题}
    \item \textbf{Sampling Skid:} Hardware overflow interrupts occur some cycles after the actual event-causing instruction, shifting the reported $PC$. Jeffrey et al.\ \cite{645821} observed a six-cycle skid: the sampled $PC$ corresponds to the instruction six retirements after the true one. While timing samples are unaffected, frequency counts bias toward long-latency instructions (the “aggregation effect”) and can shadow nearby instructions (the “shadow effect”), inflating their counts.
    % 硬件通过计数器的溢出产生的中断信号通知软件进行一次采样。理想情境下，造成计数器溢出的指令就是当前程序计数器 $PC$ 的记录值，即指令的地址。遗憾的是实际情况下，采样报告的程序计数器的值往往是在处理器中许多周期之后执行的指令的地址。这种现象被称为“滑移”。在Jeffrey等人 \cite{645821} 的工作中展示了这一现象，在其工作中滑移现象显示报告的指令地址与六个指令周期后执行的指令的地址相对应。这种采样的偏移不会对针对时间的采样方式有影响，但对频率信息有较大影响。这是因为触发长时间迟滞的指令会累积同一基本块中的其他指令出现的频次最终导致统计次数异常增加。这种现象被成为聚集效应（aggregation effect）。同时这些长延时指令积累会遮盖住本身正确的采样记录，造成阴影效应（shadow effect）。%图 \ref{} 展示了该效应的一种影响。 可以观察到随机数指令造成的指令执行的延迟使指令采样频率异常升高。
    % ProfileMe: hardware support for instruction-level profiling on out-of-order processors
    % \begin{figure}[H]
    %     \centering
    %     \includegraphics[width=0.4\textwidth]{skid-problem.png}
    %     \caption{采样滑移}
    %     \label{skid-problem}
    % \end{figure}

% \paragraph{Simultaneous Retirements}
% \paragraph{多指令同时结束问题}
    \item \textbf{Simultaneous Retirements:} Modern superscalar cores retire multiple instructions per cycle. When one retire causes overflow, interrupts from other retirees in the same cycle are masked, so only one instruction is recorded. If a fixed group of instructions always retires together, their combined count is attached to a single representative instruction.
    % 在大多数现代超标量处理器上，在一个给定的时钟周期内允许有多个指令结束执行。但是结束指令的并发性会造成一定程度上的混乱。这是因为当一条指令结束执行并且造成某个计数器溢出，进而产生中断信号时，其他计数器溢出产生的中断信号则会被屏蔽。因此若存在总是一起结束的一组指令，那么同一时间内可能只有一条指令被记录，且这条指令的采样数将是同组指令的采样总和。
\end{itemize}

\input{contents/figs/min-cost-flow-algorithm}

Fortunately, Levin et al.\ \cite{10.1007/978-3-540-77560-7_20} demonstrated that the aforementioned minimum-cost flow (MCF) algorithm (Algorithm~\ref{alg:min-cost-flow}) can effectively recover inaccurate profiling data, with the recovery accuracy depending on the choice of the cost function. Typically, if a basic block's sampled data is deemed reliable, its corresponding edges in the residual plot of CFG are assigned high cost weights; conversely, edges with less trustworthy samples receive low cost weights. The key challenge lies in correctly assessing the cost weight of each block's sampled information. Building on this idea, He et al.\ \cite{10.1145/3498714} proposed an enhanced model that extends the MCF algorithm for post-processing sampled data. Their method resolves inconsistencies between block counts and edge frequencies and, under multiplicity and connectivity constraints, derives a consistent profile, improving the accuracy of sampling-based profiles. This mathematical-model approach allows sparse hardware samples to be reconstructed into full CFG-consistent execution paths, providing the compiler with more reliable data for optimization decisions.

Algorithm~\ref{alg:min-cost-flow} implements a method to solve the MCF problem on a directed graph \(G=(V,E)\) with capacities \(\mathit{cap}(e)\) and costs \(c(e)\) for each edge \(e\in E\), a designated source \(s\), sink \(t\), and required flow \(F\). In the context of sample-based FDO, raw hardware samples yield only sparse counts for a subset of CFG edges (or instructions), which the algorithm treats as ``flow supplies''.  By constructing the residual graph where sampled edges carry initial supply and unsampled edges carry zero, and by defining each cost \(c(e)\) inversely proportional to sample confidence (e.g.\ based on sample counts and estimated error), the successive shortest-path solver computes augmenting paths with minimum reduced cost. Enforcing flow-conservation through primal-dual updates yields a full, CFG-consistent profile—assigning inferred execution frequencies to every edge that best explain the observed samples under a global cost minimum. These reconstructed basic-block and edge counts can then be fed into the compiler's PGO pipeline to guide optimizations (code layout, inlining, branch hints, etc.), achieving near-instrumentation accuracy with only hardware-sampling overhead.  

Ramasamy et al.\ \cite{36576} proposed mapping sparse hardware samples to estimated basic-block and edge frequencies via heuristics, smoothing data over the CFG under flow-conservation constraints and mapping addresses back to source via debug symbols. This approach reduced runtime overhead to under 2\%, making real-time profiling feasible in production despite sparsity and lost high-level semantics. nehao et al.\ \cite{10.1145/1772954.1772963} addressed this by using multiple hardware counters to estimate, for each basic block, the confidence and adequacy of its sampling counts. This shifts the problem to determining which hardware events most faithfully represent actual program execution. The authors employed Support Vector Regression (SVR) to compute regression weights for each event type, finding that the \texttt{INST\_RETIRED} event correlates most strongly with true block execution frequency.

Both the aggregation effect and the shadow effect typically occur within the same basic block, so one must decide which effect dominates. In the aggregation effect, the number of samples is proportional to instruction latency, so one can mitigate it by estimating each instruction's delay and discounting its effect. Since long-latency instructions overwhelmingly cause aggregation and are less affected by sampling skid, events tied to these instructions serve to quantify aggregation. To gauge shadowing, compare the total cycles sampled for a block against the number of retired-instruction samples; the relative magnitudes indicate which effect prevails. With this information, one can adjust the MCF cost function to reconstruct more accurate frequency profiles.
% 对于前文提到的聚集效应和阴影效应，他们通常出现在同一个基本块中。因此需要决定哪种效应占主导地位。对于聚集效应，指令采样的数量与指令延迟的数量呈正比，因此可以通过估算每条指令的延迟时间减少其造成的影响。由于造成聚集效应的指令绝大多数都是停滞事件较长的指令，通常不会受到指令滑移的影响。因此可以使用对这些导致长延迟的指令采样的事件对聚集效应进行量化。对于阴影效应，可以将基本块中的周期数（由"CPU\_CLK\_UNHALTED"事件采样得到）与周期中指令结束的事件数相比较。通过比较两种效应的影响因子，可以得到一个程序基本块中的主导因素。根据上述信息，可以对 MCF 中代价函数进行调整从而估算出相对正确的统计信息。

On modern Intel x86-64 processors, PEBS ensures that the program counter reported on a counter-overflow interrupt exactly matches the dynamic instruction that incremented the counter. Newer Intel CPUs also include LBR registers to log the most recent branch instructions; these LBR entries extend the BTB and, in PMU-based sampling, reorder interrupts to improve profile fidelity and avoid multi-instruction interrupt ambiguities. Wicht et al.\ \cite{DBLP:journals/corr/WichtVCL14} incorporated last branch record to capture source and target addresses of branches, reconstructing edges and basic-block frequencies with high fidelity. Their end-to-end toolchain --- Perf, Gooda, and AutoFDO --- maps LBR records through debug symbols directly into the compiler's optimization pipeline. On SPEC C++, LBR-based profiling incurred only 1.06\% overhead and achieved performance gains comparable to instrumentation-based PGO, sometimes exceeding it, while eliminating the dual-build constraint and enabling more flexible, real-time feedback.
% 随着硬件采样技术的不断成熟， Wicht 等人 \cite{DBLP:journals/corr/WichtVCL14} 则进一步引入了 Last Branch Record（LBR）机制。LBR 能够捕捉分支指令的源地址和目标地址，通过重建实际的执行路径来实现对基本块及边缘频率的精细统计。该方法不仅借助调试符号将二进制地址精确映射到源代码，同时构建了包括 Perf、Gooda 与 AutoFDO 在内的完整工具链，实现了硬件采样数据与编译器优化流程的直接耦合。值得一提的是，实验结果表明，在 SPEC C++ 基准测试中，采用 LBR 模式的采样方法平均开销仅为 1.06\%，而其优化收益接近于传统基于插桩的 PGO 方法，甚至在某些场景下超出传统方法。这种基于 LBR 的采样方法不仅显著降低了额外开销，而且通过消除传统双编译模式的约束，提升了数据采集与优化反馈的灵活性和实时性。
% Hardware Counted Profile-Guided Optimization
% 在上文提到的现代的 Intel x86 处理器中 PEBS （即 precise event based sampling）可以确保计数器溢出中断所报告的程序计数器的地址与导致计数器递增的动态指令完全对应。在最新的英特尔处理器中，存在一组名为 LBR(Last Branch Record) 的寄存器专门用来记录最后几个分支命令的执行情况。这些寄存器是分支追踪缓存 BTB (Branch Trace Buffer) 的扩展，在基于 PMU 的硬件采样中可以被用来整理中断的顺序，从而提升采样信息的准确性，避免多指令同时结束时产生中断造成的影响。
% TODO check the algorithm
Clearly, every branch logged by LBR corresponds to an edge path in the CFG. For edges not recorded by LBR, their frequencies can be reconstructed via MCF. Because one overflow interrupt can record up to $N$ branches, one can compute instruction frequencies directly from LBR data without inferring via edge frequencies. The MCF algorithm accumulates the counters for all recorded branches accordingly.
% 很明显，对于 LBR 记录的每一条分支语句，在程序控制流图中都有一条边与之对应。对于 LBR 没有记录到的边的频率信息，可以使用相应算法（如图 \ref{} 所示）进行重建。在算法中首先对基本块按拓扑顺序进行标注。每个基本块的频率为其输入边的频率的加和。则没有执行分支语句的部分的频率为该基本块的采样频率减去该基本块的所有输出边的频率。根据LBR的特性，即一旦计数器产生一个溢出中断，则有N个分支的执行被记录下来，可以直接计算出指令的频率信息而无需经边的频率信息计算得到。算法如图 \ref{} 所示。该算法会增加所记录的所有分支信息对应的指令计数器的值。

% \begin{figure}[H]
%     \centering
%     \includegraphics[width=0.5\textwidth]{LBR-BB.png}
%     \caption{通过LBR计算基本块的频率信息}
%     \label{LBR-BB}
% \end{figure}

% \begin{figure}[H]
%     \centering
%     \includegraphics[width=0.5\textwidth]{LBR-instruction-profile.png}
%     \caption{通过LBR计算指令的频率信息}
%     \label{LBR-BB}
% \end{figure}

These two methods together yield instruction-level statistics. LBR-based sampling still faces the three hardware sampling challenges discussed earlier, but randomizing the sampling interval mitigates synchronization issues. On most modern processors, only one branch can retire per cycle, so LBR avoids the simultaneous-retirement problem. Moreover, LBR's skid is only ten cycles, one-third that of traditional sampling, and the branch instruction taken event occurs far less frequently (and at much larger intervals) than instruction retired event, reducing aggregation and shadow effects. Finally, each LBR sample captures multiple branch records in a single interrupt “pulse”, further smoothing the sampling distribution. 
% 通过以上两个算法都可以得到针对指令的统计信息。对于基于LBR的采样方式依然会受到 PMU 的影响，即会出现上文提到的三种问题。但是通过采用随机采样周期的方法可以缓解采样同步的问题。对于当前的绝大部分处理器而言，一个周期内只能结束一个分支语句的执行，因此 LBR 不会受到多指令同时结束造成的问题。对于采样位移造成的聚集和影子效应，基于 LBR 的采样模式也有明显的优势：首先LBR采样的位移只有10个指令周期，是传统的采样方式的位移指令周期的 $1/3$；其次分支指令结束事件（$BR\_INST\_RETIRED_{taken}$）的发生频率远远小于指令执行结束事件（$INST\_RETIRED$）的发生频率，并且事件发生的间隔也远大于指令偏移周期本身；最后，一次LBR采样本身就包含多个记录条目，因此采样是以脉冲的形式进行的，采样更加平滑。

Even so, hardware sampling accuracy remains affected by various sampling errors. Understanding and correcting these errors to maintain optimization effectiveness is another critical research area. Wu et al.\ \cite{10.1007/978-3-642-39038-8_27} systematically classified sampling errors, such as zero-counter errors, inconsistency errors, and branch-bias errors, through extensive experiments and proposed a simple statistical correction scheme. By using training inputs to generate error-pattern statistics, they adjust key errors in sampled profiles, enabling corrected profile files to deliver performance improvements that approach those of fully instrumented data. This lightweight post-processing approach improves sample quality without complex new sampling techniques, offering a practical path to efficient optimization at low sampling rates. Zhou et al.\ \cite{10.1145/2851502} further studied the impact of sampling errors on FDO at finer granularity and devised targeted correction strategies. Their work systematically analyzes how different error types at various sampling rates affect final optimization outcomes and uses training-input patterns to correct zero-count and branch-bias errors. Experiments show that this method not only boosts FDO speedups in typical scenarios but also exhibits high robustness across diverse inputs, demonstrating the effectiveness of simple statistical corrections in real-world compiler optimizations.
% 但即使如此，硬件采样数据的准确性依然受到各种采样误差的影响。如何理解并修正采样误差对优化效果的影响，成为了学界关注的另一重点。Wu 等人 \cite{10.1007/978-3-642-39038-8_27} 通过系统性实验揭示了采样误差的类型——例如零计数器错误、不一致性误差和分支偏置错误——并提出了一种基于统计模式的简单校正方法。研究者利用训练输入生成的统计模式，对采样数据中的关键误差进行调整，使得经过校正后的配置文件在反馈驱动优化中能够显著提升性能，进一步接近全插桩数据的效果。这种方法不依赖于额外的复杂采样技术，而是通过简单的后处理步骤改善了采样数据的质量，为低采样率下实现高效优化提供了切实的路径。Zhou 等人 \cite{10.1145/2851502} 则进一步从更细粒度上探讨了采样误差对反馈驱动优化的影响，并设计出相应的纠正策略。其研究系统地分析了不同采样率下，配置文件中各种误差对最终优化效果的影响，并提出利用训练输入模式来纠正零计数器和分支偏置错误的方法。实验结果表明，通过这种方式，不仅在常规应用场景中可以大幅提升 FDO 的加速比，而且在跨输入稳定性方面也表现出较高的鲁棒性，从而证明了基于训练输入的简单统计校正在实际编译器优化中的有效性。
% Simple Profile Rectifications Go a Long Way - Statistically Exploring and Alleviating the Effects of Sampling Errors for Program Optimizations
% Examining and Reducing the Influence of Sampling Errors on Feedback-Driven Optimizations

\paragraph{Reconstruction of Execution Profiles}
% \subsubsection{程序执行复原}

% TODO: 描述算法

% Early profilers relied primarily on static instrumentation --- for example, the traditional \texttt{gprof} tool—which inserts additional code into the program to record basic-block or function execution counts. However, this approach exhibits multiple drawbacks in practice: the instrumentation code can increase runtime overhead dramatically, sometimes degrading performance by 50\%-200\%, and requires a dual-build process (first compile an instrumented version, then use the collected data to optimize), severely limiting its applicability in production environments. To overcome this bottleneck, researchers began exploring the use of on-chip hardware performance counters in modern processors to collect runtime data via low-overhead sampling, thereby avoiding the high costs of instrumentation.
% 在早期，性能剖析主要依赖静态插桩技术，例如传统的 gprof 工具，其核心思想是通过在程序中插入额外的代码来记录基本块或过程的执行频率。然而，这种方法在实际应用中暴露出多方面的缺陷：插桩代码会大幅增加程序运行时的开销，甚至在一些场景下可能导致性能下降 50\% 到 200\% 以上，同时还需要通过双编译流程（先生成带插桩的版本，再利用采集数据进行优化）来获得反馈数据，这使得该方法在生产环境中的应用受到了极大限制。为了突破这一瓶颈，研究者开始探索利用现代处理器中内置的硬件性能计数器，通过低开销采样来获取运行时数据，从而尽可能地避免插桩带来的高额代价。

As hardware performance-counter capabilities matured, both academia and industry sought to correlate hardware-sampled data with program behavior. Traditional tools like \texttt{gprof} map simple aggregate metrics to source code units (e.g., functions or lines) but cannot capture the relationship between hardware events (such as cache misses or pipeline stalls) and the program's actual execution paths and calling contexts. To address this, Ammons et al.\ \cite{10.1145/258916.258924} proposed combining flow-sensitive and context-sensitive analysis techniques to precisely bind PMU-collected data to execution paths and call chains. Specifically, they initialize hardware counters at a path's entry, read the counts at the path's exit, and merge these measurements into a Calling Context Tree (CCT), enabling fine-grained analysis of complex control flow.
% 随着硬件性能计数器功能的不断完善，学术界和工业界开始尝试将硬件采样数据与程序的动态行为进行关联。传统工具如 gprof 只能将采样得到的简单指标映射到静态代码单元（如过程或语句），而无法捕捉到硬件事件（例如缓存未命中、指令停顿等）与程序实际执行路径以及调用上下文之间的内在联系。对此，Ammons 等人 \cite{10.1145/258916.258924} 提出利用流敏感（flow-sensitive）和上下文敏感（context-sensitive）分析技术，将硬件性能计数器收集到的数据同程序运行时的执行路径和调用链精确关联起来。具体来说，作者在路径入口处初始化硬件计数器，在路径出口读取计数值，并将这些数据与构造出的调用上下文树（Calling Context Tree, CCT）结合，从而实现了对程序内部复杂控制流的细粒度分析。
% Exploiting hardware performance counters with flow and context sensitive profiling

Hardware-sampling techniques have evolved beyond static data collection and correction to address the challenge of maintaining sampling accuracy in dynamic, rapidly changing environments. With frequent code updates and the rise of continuous-integration systems, \emph{profile staleness}—the decay of profile relevance over time—has become a major obstacle for practical PGO. Ayupov et al.\ \cite{10.1145/3640537.3641573} tackle this by combining multi-level hash matching with flow-conservation inference: they use a two-stage hashing strategy (loose hash followed by strict hash) to efficiently align stale profile data with new code versions, preserving most optimization benefits even when the code changes slightly. This method offers a concrete solution for continuously evolving codebases and exemplifies how hardware sampling adapts within dynamic environments.
% 硬件采样技术的发展并不仅仅局限于静态环境下的数据采集和校正，现代的研究开始关注如何在动态、多变的运行环境中保证采样数据的准确性。随着代码频繁更新和持续集成系统的普及，陈旧的采样数据（Profile Staleness）成为制约 PGO 实际应用的一大障碍。Ayupov 等人 \cite{10.1145/3640537.3641573} 针对这一问题提出了一种多级哈希匹配与流守恒推导的方法，通过松散哈希和严格哈希相结合的策略，实现了对陈旧采样数据与新版本代码之间的高效匹配和推断，使得即便在代码发生细微变动的情况下，旧有的配置文件数据也能被有效利用，恢复大部分优化收益。这种方法为面对持续变化的代码基提供了一种切实可行的解决方案，同时也反映出硬件采样技术在动态环境下不断自我迭代和完善的趋势。
% Stale Profile Matching

Another key challenge is handling debug-info loss and CFG distortion caused by aggressive compiler optimizations (e.g., inlining, code folding, loop unrolling), which break traditional DWARF-based mappings and blur the correspondence between sampled data and source code. To address this, He et al.\ \cite{10444807} introduce the CSSPGO framework: they embed \emph{pseudo-instrumentation metadata} in the compiler's IR as anchors linking code and sample data, protecting against debug-info gaps. By synchronizing LBR data with stack-sampling, CSSPGO reconstructs execution paths in each calling context, enabling truly context-sensitive profiling. Experiments in large data-center workloads show that CSSPGO adds negligible runtime overhead while closing much of the performance gap between AutoFDO and fully instrumented PGO.
% 在降低采样误差和提高数据映射精度的过程中，如何应对优化过程中调试信息丢失和控制流图失真问题也成为重要挑战。传统依赖 DWARF 调试信息的映射方法，在面对编译器的各种高级优化（如内联、代码合并、循环展开）时往往会失效，从而使得采样数据与源代码之间的对应关系变得模糊。对此，He 等人 \cite{10444807} 提出了 CSSPGO 框架，通过在中间表示（IR）中嵌入伪插桩元数据，作为代码与采样数据之间的锚点，避免了因调试信息丢失而导致的映射错误。同时，该方法通过结合同步的 Last Branch Record（LBR）和堆栈采样，重建了各个调用上下文中的执行路径，进而实现了上下文敏感的采样剖析。实验结果表明，在大规模数据中心环境下，该方法不仅在运行时几乎不产生额外开销，而且在性能提升上能够填补 AutoFDO 与传统插桩 PGO 之间的较大差距。
% Revamping Sampling-Based PGO with Context-Sensitivity and Pseudo-instrumentation

In summary, hardware sampling has undergone a sequence of innovations: from simple counter-based sampling to mathematical-model recovery of block and edge frequencies, to machine-learning and multi-objective corrections, and finally to context-sensitive pseudo-instrumentation. This trajectory reflects both improved efficiency in using hardware-counter resources and increasing precision in reconstructing true execution behavior with minimal overhead. Each advance narrows the gap between sampled and fully instrumented profiles, providing compilers with ever more accurate feedback for optimization and driving the evolution of PGO technologies.

%% file: contents/figs/instrument-edge-algorithm.tex
\begin{algorithm}[t]
  \caption{Assigning Values to Edges in a DAG}
  \label{alg:edge-values-dag}
  \begin{algorithmic}[1]
    \Require A directed acyclic graph (DAG) $G = (V, E)$
    \Ensure An integer value $\mathit{Val}(e)$ assigned to each edge $e\in E$
    \State Compute a reverse topological ordering $\pi$ of the vertices in $V$
    \ForAll{$v$ in $\pi$}  
      \If{$\mathrm{Out}(v)=\varnothing$}  
        \Comment{a leaf in the DAG}  
        \State $\mathit{NumPaths}[v] \gets 1$  
      \Else  
        \State $\mathit{NumPaths}[v] \gets 0$  
        \ForAll{edges $e=(v\rightarrow w)$ in $\mathrm{Out}(v)$}  
          \State \Comment{record current \#paths before exploring child}  
          \State $\mathit{Val}[e] \gets \mathit{NumPaths}[v]$  
          \State \Comment{accumulate child paths up into current vertex}  
          \State $\mathit{NumPaths}[v] \gets \mathit{NumPaths}[v] + \mathit{NumPaths}[w]$
        \EndFor  
      \EndIf
    \EndFor  
  \end{algorithmic}
\end{algorithm}

%% file: contents/tables/intel-PEBs-vs-amd-IBS.tex
\begin{table}[t]
  \centering
  \caption{Comparison of key features in AMD IBS, Intel PEBS, and Arm SPE hardware sampling mechanisms}
  \label{tab:hw-sampling-compare}
  \small
  \begin{tabular}{p{2cm} p{3.3cm} p{3.3cm} p{3.3cm}}
    \toprule 
    \textbf{Feature} & \textbf{AMD IBS} & \textbf{Intel PEBS} & \textbf{Arm SPE} \\
    \midrule
    \emph{trigger} &
      per-instruction &
      counter overflow &
      programmable interval + random skew \\
    \addlinespace
    \emph{data recorded} &
      PC + metrics &
      PC + events &
      PC + metrics + events \\
    \addlinespace
    \emph{buffering} &
      On-core, drained by interrupt &
      In MSR region, drained on overflow &
      AUX memory, drained on mark \\
    \addlinespace
    \emph{Skid / Precision} &
      Precise, minimal skid &
      Typically low skid  &
      Zero skid: exact instruction \\
    \addlinespace
    \emph{filters} &
      Event types, Depth, and fetch or execution &
      Event types and threshold &
      Event types and thresholds \\
    \addlinespace
    \emph{Interrupt / overflow behavior} &
      Interrupt or record-only modes; can drop samples when full &
      Interrupt on buffer full; can lose samples if not drained &
      Drops excess samples automatically; no execution stall \\
    \addlinespace
    \emph{overhead} &
      Moderate &
      Low &
      Very low \\
    \addlinespace
    \emph{first release} &
      Zen (2017+) &
      Nehalem (2008+) &
      Armv8.2-A (2017+) \\
    \bottomrule
  \end{tabular}

  \vspace{0.5ex}
\end{table}

%% file: contents/figs/min-cost-flow-algorithm.tex
\begin{algorithm}[t]
  \caption{Primal-Dual (Successive Shortest-Path) Algorithm for Minimum Cost Flow}
  \label{alg:min-cost-flow}
  \begin{algorithmic}[1]
      \Require Directed graph $G$ with capacity $\mathit{cap}(e)$ and cost $c(e)$ on each $e\in E$, source $s$, sink $t$, required flow $F$
      \Ensure A feasible flow $f(e)$ of value $F$ of minimum total cost, or report infeasible
      \Procedure{MinCostFlow}{$G=(V,E)$, $s$, $t$, $F$}
      \State Initialize flow $f(e)\gets 0$ for all $e\in E$
      \State Initialize potential $\pi(v)\gets 0$ for all $v\in V$
      \State $\,\mathit{flow}\gets 0,\;\mathit{cost}\gets 0$
      \While{$\mathit{flow}<F$}
        \Comment{Compute shortest augmenting path in residual graph}
        \State Let $d[v]\gets +\infty$ for all $v\in V$, and $d[s]\gets 0$
        \State Use Dijkstra on the residual graph with \emph{reduced cost} $\hat c(u,v)$ to compute $d[\cdot]$ 
        \State and parent pointers $\mathit{prev}[v]$, where \[
          \hat c(u,v) \;=\; c(u,v)\;+\;\pi(u)\;-\;\pi(v)
        \]
        \If{$d[t]=+\infty$}
          \State \textbf{return} “No feasible flow of value $F$”
        \EndIf
        \Comment{Update potentials to maintain non-negativity}
        \ForAll{$v\in V$}
          \State $\pi(v)\gets \pi(v) + d[v]$
        \EndFor
        \Comment{Find bottleneck capacity along the path}
        \State $\Delta \gets \min\bigl\{\mathit{cap}_{\text{res}}(u,v)\mid (u,v)\text{ on path }s\!\to\!t\bigr\}$
        \State $\Delta \gets \min(\Delta,\;F-\mathit{flow})$
        \Comment{Augment flow and accumulate cost}
        \State $P \gets$ the sequence of edges on the path from $s$ to $t$
        \ForAll{$(u,v)\in P$}
          \State $f(u,v)\gets f(u,v)+\Delta$
          \State $f(v,u)\gets f(v,u)-\Delta$
          \State $\mathit{cost}\gets \mathit{cost} + \Delta\cdot c(u,v)$
        \EndFor
        \State $\mathit{flow}\gets \mathit{flow}+\Delta$
      \EndWhile
      \State \Return $(f(\cdot),\,\mathit{cost})$
    \EndProcedure
  \end{algorithmic}
\end{algorithm}

%% file: contents/sections/sec4_compiler.tex
\section{Profile Guided Optimization Techniques}
% \section{反馈优化技术}

% TODO: 介绍部分合并 Proifle 格式

PGO uses runtime profile data to guide compiler optimization decisions, aligning code generation closely with actual program behavior. This section classifies the PGO techniques into three stages: compile time, link/post-link time, and runtime. Table~\ref{tab:pgo-classification}  categorizes existing techniques and approaches within these stages.

\begin{table}
  \caption{Categorization of PGO Techniques}
  \label{tab:pgo-classification}
  \begin{tabular}{p{3cm} p{10cm}}
    \toprule
    \textbf{Optimization Stage} & \textbf{Techniques and Approaches} \\
    \midrule
    Compile Time & 
      Branch-prediction optimization \cite{10.1145/330249.330255,10.1145/258915.258932,lin1973effective,held1971traveling} \newline
      Basic-block and procedure-level layout optimization \cite{5389544,10.1145/93548.93550,10.1145/3575693.3575727} \\
    Link/Post-link Time &
      Cross-module optimization \cite{36355, 10.1145/3575693.3575727} \newline
      Binary-level optimization \cite{10.5555/3314872.3314876,10.1145/3446804.3446843} \\
    Runtime &
      Dynamic recompilation and region-based compilation \cite{10.1145/778559.778562,10.1145/1111596.1111600,10.1145/1988042.1988048} \newline
      Modular frameworks and selective compilation \cite{10.5555/776261.776290,10.1145/504282.504295} \\
    \bottomrule
  \end{tabular}
\end{table}
\input{contents/tables/gcc-fdo-pipeline.tex}
\input{contents/tables/llvm-pgo-pipeline.tex}

\input{contents/sections/sec4-1_format.tex}

\input{contents/sections/sec4-2_CTO.tex}

\input{contents/sections/sec4-3_LTO.tex}
\input{contents/sections/sec4-4_RTO.tex}

%% file: contents/tables/gcc-fdo-pipeline.tex
\begin{table}[ht]
\centering
\small
\begin{tabular}{ll}
\toprule
\textbf{Pass / Flag} & \textbf{Purpose (FDO Effect)} \\
\midrule
\multicolumn{2}{c}{\emph{Instrumentation Phase}}\\
\midrule
\texttt{-fbranch-probabilities}   & Insert counters on CFG arcs to record branch weights.\\
\texttt{-fprofile-values}         & Collect histograms of runtime values (calls, divisors, etc.).\\
\midrule
\multicolumn{2}{c}{\emph{Compile Time Optimizations}}\\
\midrule
\texttt{-fbranch-probabilities}             & Insert counters on CFG arcs to record branch weights.\\
\texttt{-funroll-loops}                     & Aggressively unroll hot loops based on trip counts.\\
\texttt{-fpeel-loops}                       & Peel first iterations of hot loops to expose optimizations.\\
\texttt{-fsplit-loops}                      & Split hot/cold loop paths when profitable.\\
\texttt{-funswitch-loops}                   & Unswitch loop-invariant conditions on hot loops.\\
\texttt{-ftree-loop-distribute-patterns}    & Distribute hot loops by pattern to enable further transforms.\\
\texttt{-ftree-loop-vectorize}              & Vectorize hot loops using dynamic cost model.\\
\texttt{-ftree-slp-vectorize}               & Apply SLP vectorization aggressively on hot code.\\
\texttt{-fvect-cost-model=dynamic}          & Use profile-driven cost model for vectorization.\\
\texttt{-fvpt}                              & Record detailed value-distribution information.\\
\texttt{-ftracer}                           & Form hot traces (superblocks) by tail duplication.\\
\texttt{-fgcse-after-reload}                & Perform global CSE on hot code after register allocation.\\
\texttt{-finline-functions}                 & Inline hot call‐sites more aggressively.\\
\texttt{-fipa-cp}                           & Propagate constants across hot call graph edges.\\
\texttt{-fipa-cp-clone}                     & Clone functions for hot constant arguments.\\
\texttt{-fipa-bit-cp}                       & Bitwise constant propagation on hot paths.\\
\texttt{-fpredictive-commoning}             & Hoist and reuse computations in hot loops.\\
\midrule
\multicolumn{2}{c}{\emph{Link Time Optimizations}}\\
\midrule
\texttt{-fprofile-reorder-functions}        & Order functions into \texttt{.text.hot} / \texttt{.text.unlikely} by hotness.\\
\texttt{-finline-functions}                 & Inline hot call-sites more aggressively in link time.\\
\bottomrule
\end{tabular}
\caption{GCC 12 Optimization Passes Enabled or Tuned by FDO Profile Data, Organized by Compilation Stage}
\label{tab:fdo-passes-gcc12}
\end{table}
% & \textbf{Stage}
% & Compile (Instrument)   
% & Compile (Instrument)   
% & Loop Opt               
% & Loop Opt               
% & Loop Opt               
% & Loop Opt               
% & Loop Vectorization     
% & Loop Vectorization     
% & Loop Vectorization     
% & Value Profile Transform
% & Code Layout            
% & Codegen (Post‐RA)      
% & IPA                    
% & IPA                    
% & IPA                    
% & IPA                    
% & Memory Opt             
% & Link-Time           

%% file: contents/tables/llvm-pgo-pipeline.tex
\begin{table}[ht]
\centering
\small
\begin{tabular}{ll}
\toprule
\textbf{Pass} & \textbf{Purpose (PGO Effect)} \\
\midrule
\multicolumn{2}{c}{\emph{Instrumentation Phase}}\\
\midrule
InstrProfilingInsertion         & Instruments IR to collect execution counts at runtime. \\
\midrule
\multicolumn{2}{c}{\emph{Compile Time Optimizations}}\\
\midrule
PGOInstrumentationUse           & Applies profile data as IR metadata for optimizations.            \\
SampleProfileLoaderPass         & Loads sample-based profiles and annotates IR branches.            \\
ProfileSummaryInfoWrapperPass   & Generates global profile summaries for queries.                   \\
BranchProbabilityInfoWrapperPass& Computes branch probability information from profile metadata.    \\
BlockFrequencyInfoWrapperPass   & Derives basic-block execution frequencies foroptimizations.       \\
InlinePass                      & Adjusts inlining decisions based on function call frequencies.    \\
IndirectCallPromotionPass       & Promotes frequent indirect calls into direct calls with checks.   \\
HotColdSplitting                & Outlines cold basic blocks into separater sections for locality.  \\
BlockPlacementPass              & Reorders IR basic blocks to maximize fall-through on hot edges.   \\
LoopUnrollPass / FullUnrollPass & Unrolls hot loops extensively and cold loops conservatively.      \\
LoopUnswitchPass                & Hoists frequently taken branch conditions out of loops.           \\
\midrule
\multicolumn{2}{c}{\emph{Link Time and Post-link Time Optimizations}}\\
\midrule
FunctionImport (ThinLTO)        & Inlining hot functions across modules guided by profile.          \\
Symbol Ordering \& Sections     & Places hot functions contiguously in the binary.                  \\
MachineBlockPlacement           & Reorders machine-level basic blocks and aligns hot loops.         \\
MachineFunctionSplitter         & Splits cold machine blocks into separate sections for locality.   \\
BOLT      (PLTO, standalone)    & Rewrites the final binary to optimize layout and performance.     \\
Propeller (PLTO, standalone)    & Performs whole-program basic-block and function reordering.       \\
\bottomrule
\end{tabular}
\caption{PGO-Related Optimization Passes in LLVM 15, by Compilation Phase}
\label{tab:pgo-passes}
\end{table}

%% file: contents/sections/sec4-1_format.tex
\subsection{Instrumentation and Profile Formats}

\subsubsection{Instrumentation}

Instrumentation and profile formats form the foundational infrastructure for FDO in GCC PGO in LLVM. Instrumentation in GCC primarily involves augmenting the compiler's front end to insert probes for both control-flow arcs and selected value computations. Conditional jumps, calls, and switch-case edges receive counters recording precise execution frequencies during runtime. Complementary profile-values mechanisms also capture histograms of indirect calls, divisor operands, and value distributions for later specialization. GCC's runtime library aggregates these counts in memory, outputting them to disk in .gcda files upon program exit. Though hardware-based sampling alternatives like AutoFDO exist—collecting data from hardware counters to provide statistical coverage of hot code paths—instrumentation-based profiling remains highly accurate and detailed in GCC.
% % GCC instrumentation
% \paragraph{Instrumentation and Profile Collection}
% At the heart of FDO lies instrumentation, which augments the compiler's front end to emit probes for both control-flow arcs and selected value computations. When compiling with the branch-probabilities option, each conditional jump, call, and switch-case edge is equipped with a counter so that execution frequencies are recorded precisely at runtime. Complementing this, the profile-values mechanism installs histogram-style instrumentation around indirect calls, divisor operands, and other expressions whose value distributions can inform later specialization. During execution of the instrumented binary, GCC's runtime library transparently aggregates these counts in memory and writes them to disk upon program exit. Although hardware-based sampling alternatives exist, GCC's classic instrumentation approach remains the most accurate, producing .gcda files that faithfully reflect how often each path was taken and which values predominated in hot contexts. With this concrete profile in hand, the compiler is equipped to make informed decisions about code layout, inlining strategies, loop restructuring, and even low-level scheduling.

LLVM offers instrumentation-based PGO and sample-based PGO. Instrumentation in LLVM inserts lightweight counters either at the AST level or IR level, tallying function entries, basic-block visits, and branch edges. Instrumented programs produce .profraw files during execution, which are later converted into .profdata for use in subsequent compilations. LLVM also supports sample-based PGO, consuming externally collected sample data through the Sample Profile Loader pass, which maps sampled edge counts into IR metadata.
% LLVM instrumentation
% \paragraph{Compile-Time IR Instrumentation and Metadata Propagation}
% At the outset of PGO workflows, LLVM must gather or consume profile data. When a build is configured with -fprofile-generate, the instrumentation insertion phase weaves lightweight counters into the IR model of each function. These counters tally function entries, basic-block visits, and branch edges at runtime, constructing an execution profile that reflects true program behavior. Once the instrumented program runs, it produces a .profraw file that is later converted into .profdata.
% END

On subsequent compilation with -fprofile-use, LLVM's PGO instrumentation use pass reads this .profdata and attaches precise branch-weight and indirect-call target metadata directly to IR instructions. Branch instructions receive annotations indicating the relative likelihood of each successor edge, while indirect call sites are decorated with the most frequently observed targets and their frequencies. In parallel, the sample profile loader supports sample-based PGO (e.g., Linux perf data), mapping sampled edge counts into equivalent IR metadata. By embedding these runtime-derived hints into the IR, all downstream passes can reason about “hot” versus “cold” code paths with accuracy far beyond static heuristics.

Once metadata is in place, LLVM invokes its analysis passes—BranchProbabilityInfo and BlockFrequencyInfo—which synthesize the raw weights into probabilistic models of branch behavior and absolute basic-block execution frequencies. A complementary ProfileSummaryInfo pass constructs global summaries, identifying percentile thresholds that classify functions as hot or cold and computing cross-function call counts. These analyses underpin many of the subsequent IR-level optimizations, empowering LLVM to make nuanced decisions that balance performance gains against code-size costs.

\subsubsection{Profile Formats}
\input{contents/figs/profile-format.tex}
GCC's coverage-information format is built around two companion binary files (figure~\ref{fig:gcc-profile-format}): the “notes” file (.gcno), emitted at compile time, and the “data” file (.gcda), generated during program execution—each of which begins with a compact, fixed-size header that encodes a four-byte magic identifier (distinguishing file type and byte order), a four-byte version stamp (packed from a four-character ASCII string such as “B62r” to allow simple lexical comparisons), and a 32-bit “stamp” counter that ties together note and data generations. In the notes file only, an additional 32-bit word indicates whether unexecuted blocks should be supported, ensuring that downstream tools immediately know whether to annotate unreachable code. This versioning and stamping strategy ensures that mismatches in GCC version, target endianness, or compile/run cycle are detected early, allowing tools to reject or byte-swap files that cannot be safely interpreted.

Beneath each file's header lies a flat sequence of self-describing records, each beginning with a 32-bit tag word followed by a 32-bit length count (measured in 4-byte words) and then the payload items themselves. Primitive items are encoded in native machine endianness—32-bit integers as single words, 64-bit counters as two consecutive 32-bit words (low order followed by high), and strings as a length word plus raw bytes—so that high-performance I/O routines can stream data directly to and from memory with minimal overhead. The tag's 32 bits are conceptually split into four hierarchical levels (each one byte), where odd-valued level bytes mark active nesting; this bit-pattern convention allows tools to infer parent-child relationships among records without pre-knowing tag values, making the format robust to future extensions.

In the notes file, records describe program structure: each translation unit begins with a “unit” record carrying a source-file checksum and filename, followed by interleaved sequences of function announcements (identifiers, checksums, source ranges), basic-block flags, control-flow arcs (destination indexes and per-arc flags), and source-line mappings (line numbers interspersed with filename switches). At run time, the data file mirrors this structure but replaces skeletons with counter values: after a global program summary (runs merged and per-run maxima), each function is re-announced by identifier and checksum, a “present” marker signals which counters are available, and then a contiguous array of 64-bit execution counts is emitted in the same block/arc order defined earlier. By cleanly separating structural metadata from execution tallies—and by embedding sufficient version, checksum, and length information in every record—GCC's coverage format enables both lightweight streaming and flexible tooling (though modern users are encouraged to invoke gcov --json-format for higher-level language bindings).

LLVM's instrumentation-based profile format (figure~\ref{fig:llvm-profile-format}) begins with a compact, fixed-layout header that efficiently encodes all the metadata needed to locate and interpret the subsequent profiling records. At the very start of a raw profile file, a 64-bit magic number identifies the file as an LLVM PGO profile and signals its endianness, immediately followed by a version field that allows tools to detect and reject mismatched formats. Subsequent fields in the header specify the total number of instrumented functions, the aggregate count of 64-bit counters, the size of coverage bitmaps, and the length of the function-name blob. Crucially, the header also provides byte offsets to each major section—counters, bitmaps, and names—so that lightweight tools can seek directly to the data they need without parsing intervening records. This design balances the need for fast runtime writes (via compiler-rt's profiling runtime) with robust on-disk structure for llvm-profdata and other analysis tools.

Following the header, the profile file contains a contiguous array of per-function “ProfData” records, each of which describes the location and size of that function's runtime counters, coverage bitmap, and any associated value-profiling data. Each ProfData entry embeds a 64-bit hash of the function name and a secondary collision-checking hash, along with relative pointers into the counters and bitmap regions. A 32-bit counter count and a small array of per-value-kind site counts further describe how many 64-bit counters and value-profiling sites were collected. By co-locating these descriptors and then packing all raw counts and bitmaps into dense, byte-aligned blobs, LLVM minimizes both the runtime overhead of incrementing counters and the on-disk footprint of large profiles.

When value profiling is enabled, the format appends one or more “ValueProfData” blocks per function, each containing records for different profiling kinds—such as indirect-call targets or memory-access sizes. Within each block, a small header records the kind and the number of profiling sites of that kind, followed by per-site arrays indicating how many distinct values were seen. A sequence of fixed-size 16-byte entries then lists each profiled value (for example, a call target hash) alongside its execution count. This layered approach—header, ProfData, raw counters, bitmaps, and optional value tables—ensures that LLVM's instrumentation runtime, its merging and analysis tools, and downstream optimizers all share a precise, mutually agreed-upon binary layout for profile-guided optimization.

%% file: contents/figs/profile-format.tex
\begin{figure}[t]
  \centering
  \subfloat[GCC profile format]{
    \includegraphics[width=0.65\linewidth]{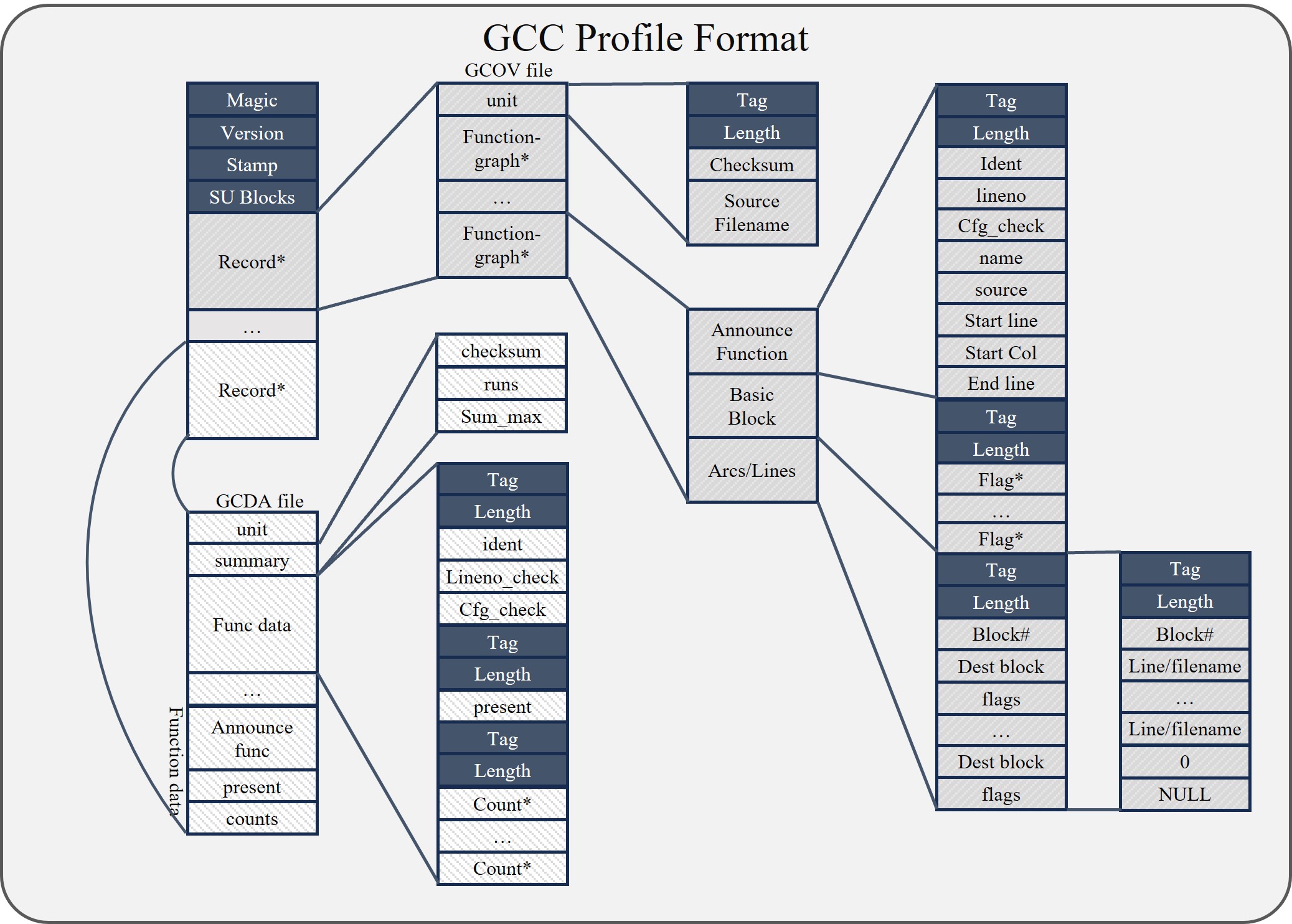}
    \label{fig:gcc-profile-format}}
  \hfill
  \subfloat[LLVM profile format]{
    \includegraphics[width=0.32\linewidth]{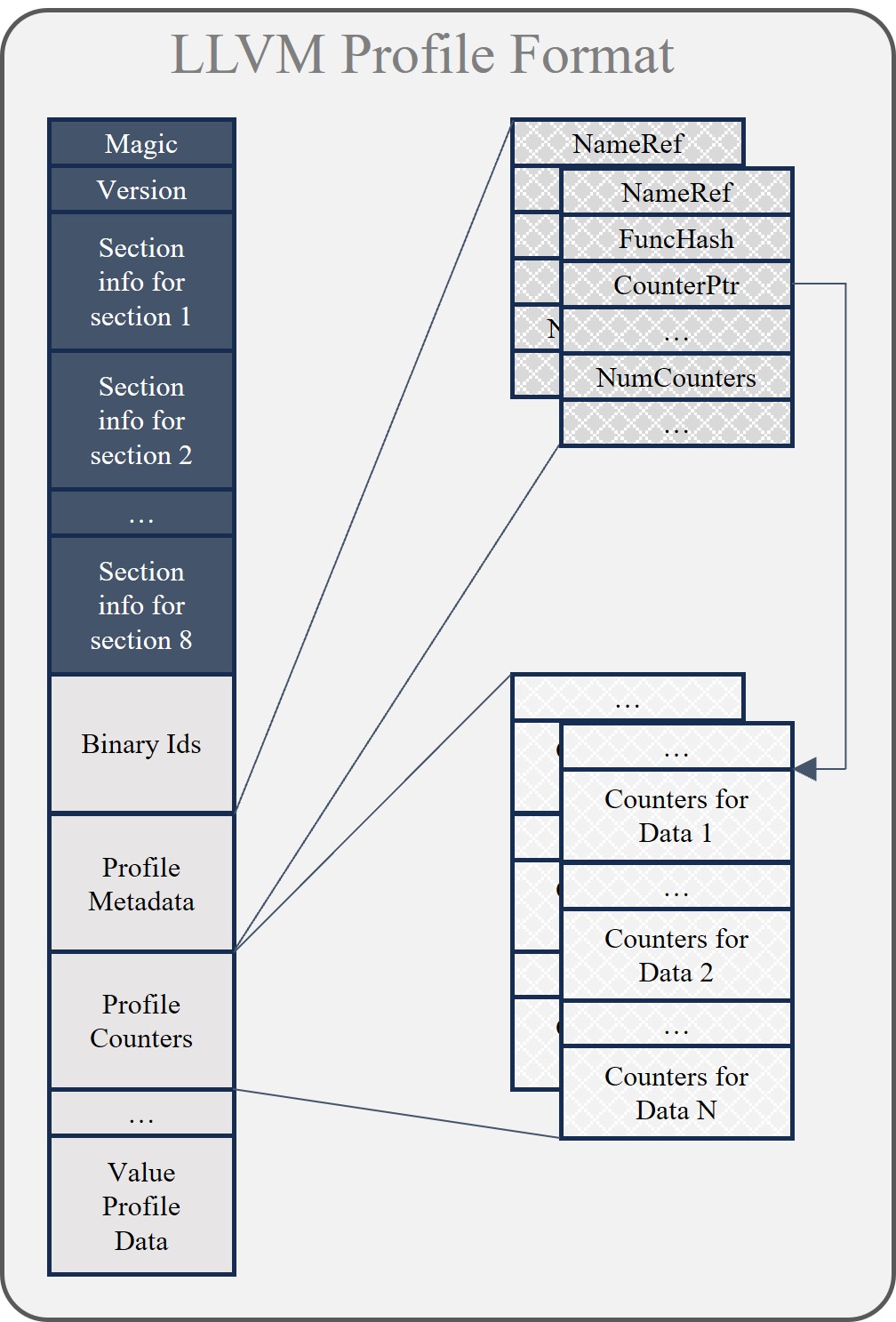}
    \label{fig:llvm-profile-format}}
  \caption{GCC and LLVM profile format in memory}
  \label{fig:profile-format}
\end{figure}

%% file: contents/sections/sec4-2_CTO.tex
% PGO offers a way to overcome static bottlenecks by using runtime profiling data—such as branch frequencies, cache hit rates, and call relationships—to guide more accurate code generation that matches actual execution characteristics. The core principles of profile-guided optimization can be summarized as three aspects:

% \begin{enumerate}
%   \item \textbf{Precise Hotspot Identification:} Use hardware counters (e.g., Intel LBR), lightweight instrumentation, or sampling to identify hot code regions (hot functions, hot branches) and inefficient patterns (cache thrashing, mispredicted branches), focusing optimization effort where it yields the greatest benefit;
%   \item \textbf{Dynamic-Static Synergy:} Combine static analyses at compile time with dynamic feedback at runtime to build multi-stage optimization strategies (e.g., staged compilation, incremental re-optimization), quickly producing usable code at startup and applying more aggressive optimizations during steady-state execution;
%   \item \textbf{Cost-Benefit Balance:} Control compilation time and memory overhead through lightweight profiling (e.g., LBR sampling), selective optimization (skipping cold functions), and distributed processing (parallel relinking), avoiding overall performance regressions from over-optimization.
% \end{enumerate}
% 精准热点定位：利用硬件计数器（如Intel LBR）、轻量级插桩或采样技术，识别高频代码段（如热函数、热分支）和低效模式（如缓存抖动、分支预测失败），将优化资源集中投入高收益区域；

% 动态-静态协同：结合编译时静态分析与运行时动态反馈，构建多层次优化策略（如多级编译、增量重优化），在启动阶段快速生成可用代码，在稳态阶段逐步应用更为激进优化的程序获取性能收益；

% 开销-收益平衡：通过轻量化剖析（如LBR采样）、选择性优化（如跳过冷函数）和分布式处理（如并行重链接技术），控制优化过程的时间与空间开销，避免因过度优化导致整体性能回退。

\subsection{Compile  Time Optimization}
During compile time optimization, the compiler uses profile data to make informed decisions early in the compilation process, optimizing code structure and generation directly from source or intermediate representations.

\subsubsection{Branch Prediction and Basic Block Reordering}
Branch-prediction and basic block reordering optimization aims to reduce the impact of conditional branches on pipeline efficiency by analyzing program runtime behavior. It focuses on two key dimensions: capturing branch-context correlations and optimizing instruction-stream continuity.
% 分支预测优化的核心在于通过程序行为分析与代码转换，降低条件分支指令对流水线效率的影响。其底层原理围绕两个关键维度展开：分支预测与上下文的相关性捕获与指令流连续性优化。

%分支相关性的静态捕获
Branches in modern programs often exhibit logical or statistical correlations. For example, one branch's outcome may depend on the path taken by a preceding branch (nested conditionals). Traditional static predictors use the majority-direction heuristic on a per-branch basis, ignoring cross-branch contextual information. Young et al.\ \cite{10.1145/330249.330255} propose constructing a \emph{history tree} from execution-trace branch sequences to identify high-frequency correlation patterns. This Static Context-Based Branch Prediction (SCBP) technique converts hardware branch-history information into compile time path analyses via code duplication, achieving context-sensitive prediction without extra runtime cost.
A basic block is the smallest unit of layout optimization. By analyzing jump frequencies between blocks, hot paths are laid out as contiguous code lines, reducing cache-line waste. Heisch \cite{5389544} performs global basic-block reordering based on execution traces, isolating cold code into separate pages to reduce working-set footprint.
% \paragraph{基本块级优化:} 基本块（无分支的指令序列）是优化的最小单元。通过分析基本块间的跳转频率，高频路径被排列为连续直线代码，减少条件分支误预测和缓存行浪费。例如，Heisch \cite{5389544} 利用执行轨迹全局重排基本块，将冷代码分离至独立内存页，减少实时内存占用。  
% Trace-directed program restructuring for AIX executables
Even with perfect prediction accuracy, control-flow transfers can disrupt instruction fetch. An unconditional jump costs one cycle, for instance, whereas a mispredicted branch costs five cycles. Young et al.\ \cite{10.1145/258915.258932} model basic-block reordering as a Directed Traveling-Salesman Problem (DTSP): nodes represent basic blocks, and edge weights capture the expected jump penalty (based on misprediction rate and target-address latency). Solving the DTSP via an iterative 3-OPT heuristic algorithm \cite{lin1973effective} with greedy augmentation for hot paths approximates the Held-Karp lower bound \cite{held1971traveling}, ensuring that blocks on the hottest paths are laid out contiguously to minimize fetch penalties.
% 即使分支预测准确，控制流跳转仍可能导致指令预取中断（Misfetch）。例如，Alpha 21164处理器中无条件跳转需1周期损失，误预测则需5周期损失。Young等人 \cite{10.1145/258915.258932} 将基本块重排序问题建模为有向旅行商问题（DTSP）：以基本块为节点，边权重为跳转损失函数（考虑误预测率与目标地址延迟），寻找使总损失最小的块排列顺序。该问题通过迭代3-Opt算法逼近理论下界（Held-Karp Bound），结合贪心策略优先连接高频路径，确保热路径上的基本块连续布局。
%例如，高频条件分支的“未跳转”目标块被排列为紧邻后继，减少跳转指令插入，同时提升指令缓存局部性。
% Near-optimal Intraprocedural Branch Alignment

\begin{figure}[t]
  \centering
  \subfloat[Original CFG with edge frequencies]{
    \includegraphics[width=0.25\linewidth]{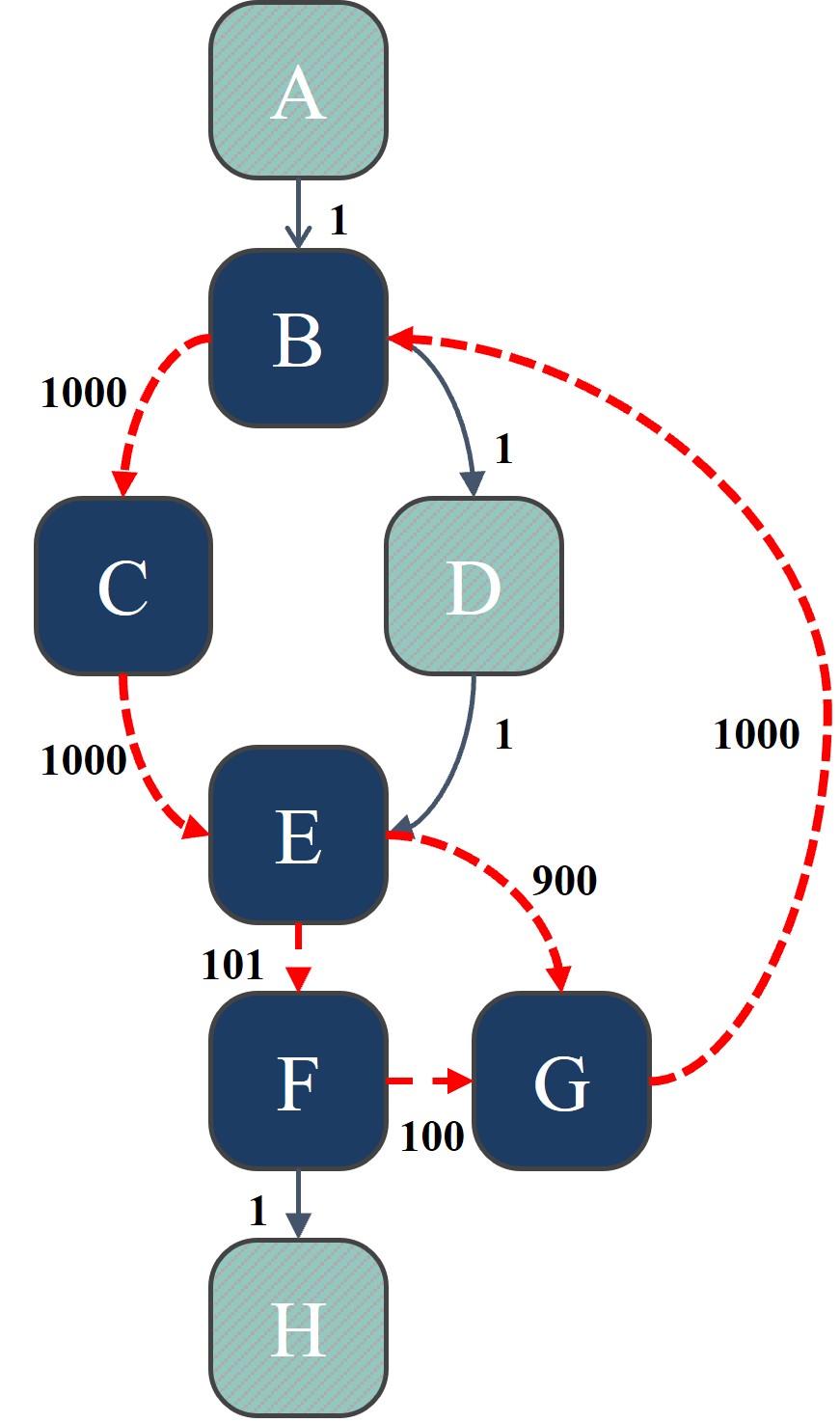}
    \label{fig:cfg4-original}}
  \hspace{0.02\linewidth}
  \subfloat[Before and after PGO reordering: hot blocks grouped for cache localit]{
    \includegraphics[width=0.48\linewidth]{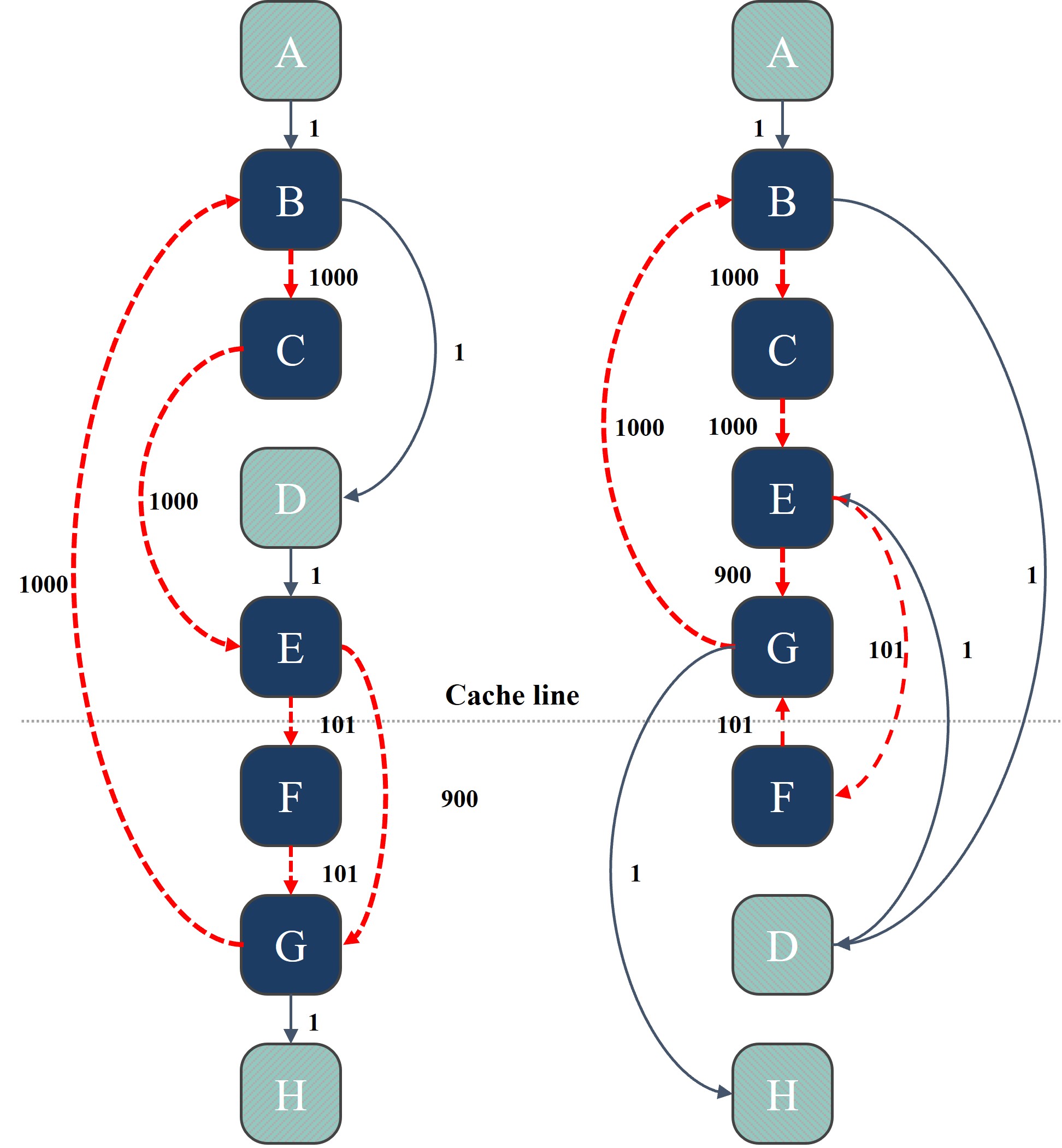}
    \label{fig:cache-opt-before-after}}
  \caption{PGO-driven block placement to improve I-cache performance. colored blocks and lines are hot. }
  \label{fig:cache-opt}
\end{figure}

Figure~\ref{fig:cache-opt} illustrate a tiny CFG before and after a PGO-driven reordering that dramatically cuts conflict misses in the L1 I-cache.  In the “simplified CFG” (Figure~\ref{fig:cfg4-original} ), hot edges (in red) loop between blocks B→C→E→G, but their physical layout forces the processor to repeatedly evict the instructions for G when falling back to block D.  After running profile-guided cache optimizer, shown in Figure~\ref{fig:cache-opt-before-after}, we reorder and group hot blocks so that B,C,E,G reside contiguously in a cache line.

\subsubsection{Value Profiling and Specialization}
In GCC, instrumentation-based FDO can record value profiles for certain expressions. For example, the values of indirect call targets or the values of division operands can be profiled at runtime via combining \texttt{-fprofile-arcs} with \texttt{-fprofile-values}. With \texttt{-fbranch-probabilities} (profile use), GCC will use these recorded values to drive optimizations. One such optimization is specialization of division operations: if a divisor is almost always a certain constant value, GCC can replace the division with a faster sequence when that value is encountered, and only fall back to a general case otherwise . This is controlled by the value profile transform pass ( -fvpt ) .

\subsubsection{Function-Level Optimization}
At the function level, optimizations use strategy to place frequently calling functions physically adjacent, reducing long-jump and page-switch overhead. For example, the Pettis-Hansen (PH) algorithm \cite{10.1145/93548.93550} merges high-frequency call-graph nodes via weighted clustering to minimize instruction cache conflicts. Propeller extends this by leveraging distributed build systems: only hot object files are relinked, while cold code reuses cached artifacts, dramatically lowering optimization time \cite{10.1145/3575693.3575727}.
% \paragraph{过程级优化：}针对过程（函数）间的调用关系，采用“邻近优先”策略，将频繁互调的过程物理相邻，降低长跳转指令和页面切换开销。例如，Pettis-Hansen\(PH\)算法 \cite{10.1145/93548.93550} 通过加权调用图合并高频过程节点，减少指令缓存冲突；Propeller 进一步结合分布式构建系统，仅对热点对象文件重生成，冷代码复用缓存，显著降低优化开销 \cite{10.1145/3575693.3575727} 。  

\paragraph{Function Inlining}
Profile feedback heavily influences the inliner. Hot call sites are prioritized for inlining, even if they would normally be considered too large. In GCC Interprocedural Inline pass, call graph nodes and edges carry profile metadata. If a call is hot (high frequency), GCC raises inlining thresholds to inline it aggressively. With AutoFDO, GCC performs an early inline pass that uses the sample profile to decide inlines even if it increases code size. The early inline phase in AutoFDO is iterative and can also promote indirect calls to direct calls when the profile suggests a likely target, which is a form of indirect call promotion, then inline those as well in subsequent iterations. In instrumentation FDO, a similar effect occurs: indirect call value profiling and promotion are enabled by \texttt{-fprofile-values} or \texttt{-fvpt}, described below, and the regular inliner will also inline indirect call targets that became direct through profile feedback.

LLVM also uses profile data to guide inlining decisions. Calls that are "hot" (high execution count) get a bonus in the inlining cost model, making them more likely to be inlined, whereas cold calls are disfavored. The InlineAdvisor or legacy inliner will query the profile  via ProfileSummaryInfo pass to see if a callee or callsite is hot. Conversely, functions with zero or very low counts may be marked with a “cold” attribute or treated as cold, which drastically lowers the inlining threshold for callers. There is also support for context-sensitive inlining in sample PGO: the SampleProfileLoader pass can direct early inlining of certain calls before the main inlining pass. Essentially, profile data ensures that inlining decisions align with observed execution: hot call chains are expanded for speed, while cold code is kept out-of-line to save space.

\paragraph{Indirect Call Promotion}
Indirect call promotion is enabled by value profiling: if the profile shows that an indirect function call (e.g. via function pointer or virtual call) almost always calls a specific target function, GCC will introduce a direct call check for that target (with guard) to avoid the indirect call overhead. In AutoFDO, the sample profile format also retains information about likely indirect call targets and their frequencies. This achieves a similar result to instrumentation-based indirect call promotion, but using sampled data.
Similarly, LLVM PGO use pass can turn indirect calls into direct calls for hot targets. Additionally, there is a late pass in code generation that might insert conditional jumps to likely targets. Much of this is handled in the IR PGO use pass itself by adding direct-call fast paths. The result is that by the time codegen happens, many indirect call sites have been optimized based on profile. This can yield significant speedups in C++ programs with virtual calls or function pointer calls.

\subsubsection{Loop Optimizations}
Loop optimizations are a key area where PGO shines. Profile data allows the compiler to make informed decisions about loop unrolling, peeling, unswitching, and vectorization. Optimizer uses the profiled trip counts to decide if a loop executes often enough (or has a small constant bound) to justify unrolling, peeling, or vectorizing. Many of these loop optimizations are enabled by profile-use flags. For example, \texttt{-funroll-loops}, \texttt{-fpeel-loops}, and \texttt{-funswitch-loops} are automatically enabled under FDO or AutoFDO . This means that with profile feedback, GCC will perform aggressive loop unrolling and peeling (normally done at -O3 ) even at -O2 optimization levels in GCC. The profile-estimated loop iteration count also helps both GCC and LLVM avoid bloating code for cold loops: loops that rarely run may not be unrolled at all, whereas hot loops with predictable trip counts can be fully unrolled or peeled. These decisions occur in passes like the GIMPLE loop optimizers (\texttt{tree-loop-*} passes) and RTL loop optimizer in GCC, which query basic block frequencies. Likewise, loop unswitching (moving invariant conditionals out of loops) uses profile info to focus on profitable cases.

%% file: contents/sections/sec4-3_LTO.tex
\subsection{Link Time and Post-Link Time Optimization}
% \subsection{链接与后链接阶段优化}
In a typical build process, compilers perform optimizations during compilation using static or heuristic profile data, which can misalign with actual runtime behavior when mapping back to IR. In contrast, link time and post-link time optimizations work on fully assembled binaries, observing and manipulating exact branch targets, code layout, and machine-instruction patterns. The primary goals of this profile guided strategy are to refine code layout, reduce branch-prediction errors, improve ICache locality, and eliminate redundant or unreachable code, resulting in a more compact, efficient binary that not only runs faster but also scales better in compile time and resource usage.
% 在典型的构建过程中，大部分编译器在编译时基于静态或启发式程序描述数据来执行优化。然而，这些方法在将运行时行为映射到中间表示形式时往往存在不准确性。相比之下，链接时和后链接时的优化在完全组装好的二进制文件中工作，可以直接观察并操控诸如精确分支目标、二进制代码布局以及机器级模式等信息。这种反馈导向策略的主要目标是通过精细化代码布局、减少分支预测错误、提升指令缓存局部性以及最小化冗余或不可达代码，从而提高执行性能。预期的结果是生成更紧凑、更高效的二进制文件，这不仅能显著提升运行速度，还能在编译时间和资源使用方面表现出更好的可扩展性。

\subsubsection{Link Time Optimization}

Li et al.\ \cite{36355} were the first to combine cross-module optimization with profile guided techniques at link time, proposing a lightweight framework that integrates cross-module analysis and profile data. By avoiding traditional heavyweight cross-module link time phases, their approach reduces code-regeneration overhead and greatly improves scalability, making it practical for large-scale projects. This integrated design enables more aggressive function inlining and indirect-call promotion across module boundaries, overcoming limitations of per-source-file optimization. Propeller \cite{10.1145/3575693.3575727} introduces the “Basic Block Section” abstraction in the linker, enabling profile-driven block ordering at link time without costly disassembly.
% Li 等人\cite{36355} 率先做出了尝试，通过在链接时将跨模块优化与反馈导向技术相结合，他们提出了一种轻量级的框架。这一方法不再依赖于传统的链接时跨模块优化阶段，而是融合了跨模块分析和反馈数据的优势，从而在性能上相对于基于反馈导向优化的基线实现了显著提升。该方法不仅减少了代码再生成的开销，还显著提高了扩展性，使得该技术在大规模项目中更具实用性。这种集成设计使得优化器能够有效地进行函数内联和间接调用推广，突破了以往由源文件边界所造成的优化限制。 Propeller则引入“基本块节”（Basic Block Sections）链接器抽象，允许链接阶段按剖面信息灵活调整块顺序，避免反汇编开销 \cite{10.1145/3575693.3575727} 。
% Lightweight Feedback-Directed Cross-Module Optimization
% Propeller: A Profile Guided, Relinking Optimizer for Warehouse-Scale Applications

\paragraph{Function Reordering}

Without profile data, GCC enables \texttt{-freorder-functions}, which works by segregating frequently executed functions into a .text.hot section and infrequently used ones into .text.unlikely. The linker then groups hot functions together for better locality. Additionally, GCC introduced \texttt{-fprofile-reorder-functions}, which uses instrumentation data to order functions by time of first execution. In effect, functions that execute early (during program startup) are placed first, which can improve startup performance by clustering initialization code. Both techniques rely on the collected profile: without real profile data or manual hot/cold annotations, function reordering does nothing.
% GCC 12 LTO
% When the recompilation reaches the link stage, profile-guided function reordering refines code layout at a coarser granularity. Functions are assigned to hot or unlikely sections based on aggregated call frequencies gathered during profile collection. The linker then positions hot functions contiguously in memory—often in a dedicated .text.hot segment—while relegating cold or infrequently used functions to a separate .text.unlikely area. This organization improves instruction-cache performance across function boundaries and enhances page-fault behavior by clustering frequently executed code pages.
% GCC 12 LTO end

\paragraph{Link Time Inlining}
Link time inlining is a powerful optimization that allows the compiler to inline functions across module boundaries. In LLVM, when LTO or ThinLTO is enabled, the profile data is applied during the LTO link stage. In Full LTO, all IR is merged and the PGO use passes run on the whole program IR, so cross-module profile information is naturally utilized. In ThinLTO, each module is optimized separately, it uses a pre-link phase to decide which functions to import or inline based on profile. LLD coordinates ThinLTO by orchestrating the parallel backends, passing the same profile data to each. The profile is indexed by function GUIDs (Global Unique IDs), so each ThinLTO backend can pick out the profile for the functions it’s responsible for. In practice, the PGO use passes run in each ThinLTO backend on its portion of the program, but they consult the global profile summary for thresholds. If a function was not present (because it was in another module and not imported), its profile data might be ignored or, in CSSPGO, used to trigger an import. Thus, LLD ensures that any necessary cross-module profile info is carried over via the combined index.

With LTO, GCC also has the opportunity to inline or clone functions across what were original file boundaries. Profile data informs these decisions globally. Profile-directed function cloning (e.g. cloning a function with a constant argument or for a hot caller) can also operate across modules. Flags like \texttt{- fipa-cp-clone} are enabled under FDO , and during LTO the IPA constant propagation pass texttt{-fipa-cp} uses profile information to decide which instances to clone. For example, if under profile it’s observed that 90\% of calls to function foo(int mode) always pass mode=1 , the IPA CP pass might clone foo into a specialized version for mode=1 . This is more effective with whole-program visibility and profile data.

% LLVM 15
% \paragraph{Link-Time and Cross-Module Enhancements}
% PGO’s impact extends beyond single-module IR optimizations into link-time whole-program reasoning. LLVM’s ThinLTO mechanism merges summaries from all modules into a global index that includes profile-derived hotness data. During the ThinLTO back-end phase, each module consults this index to decide which functions to import for cross-module inlining. Only those functions with significant execution frequency or hot call edges are brought into local contexts, enabling aggressive inlining where it yields substantial speedups while avoiding unnecessary code duplication of cold routines. This lightweight, summary-driven approach achieves much of Full LTO’s optimization power without the quadratic memory and CPU costs associated with merging all IR at once.

% In addition to driving inlining, link-time PGO information can guide symbol ordering. Hot functions—determined by top percentile execution counts—can be emitted into designated “.text.hot” sections or ordered via linker scripts to reside contiguously in the final binary. Cold functions, by contrast, may be grouped separately, minimizing their interference with hot-code instruction-cache locality. While LLVM 15 does not automatically reorder functions solely based on PGO data, it provides the necessary hooks—through section prefixes and linker options—for developers to elicit these layout improvements.

\subsubsection{Post-Link Time Optimization}
Post-link time profile guided optimizations leverage precise profile data to reorder and prune code at the binary level after the compilation and link phases. These techniques postpone optimization until most code generation and linking are complete, allowing them to operate directly on the final binary, where mappings between profile data and executed instructions are most accurate.
% 链接时和后链接时的反馈导向优化技术利用精确的剖面数据直接在二进制级别分析并重新排序、内联和精简代码，这一过程是在传统编译完成之后进行的。这种方法与传统的基于源代码或编译时反馈导向优化不同，因为它将优化阶段推迟到大部分代码生成和链接完成后，从而允许优化过程直接处理最终的二进制文件，在该阶段程序描述信息中的数据与实际执行指令的映射更加准确。

Panchenko et al.\ \cite{10.5555/3314872.3314876} introduced BOLT, which reorders and restructures code in the final executable based on sampling data. Working directly on the binary, BOLT accurately identifies hot functions and basic blocks and rearranges them to boost instruction cache performance and reduce branch mispredictions.  
%BOLT delivers double-digit speedups over binaries optimized by standard PGO and link time optimizations, providing substantial improvements even on extensively pre-optimized code.
% 在后链接时优化方面，Panchenko 等人 \cite{10.5555/3314872.3314876} 着手在二进制文件生成后重新排序和重构代码以优化性能。这种方法利用基于采样的数据推动二进制级别的代码布局优化。通过直接在最终可执行文件上工作，BOLT 可以精确判断哪些部分的代码被频繁执行，并据此重新排列函数和基本块。这一方法极大地提升了指令缓存性能并减少了分支预测失误，对于那些典型的数据中心应用中大而复杂的二进制文件尤为关键。其预期结果是相比经过标准反馈导向优化和链接时优化生成的二进制文件，能实现双位数的性能提升，从而对即便经过广泛编译时处理的代码也能带来显著改进。
% BOLT: a practical binary optimizer for data centers and beyond
Building on BOLT, Lightning BOLT \cite{10.1145/3446804.3446843} adds parallel processing and selective optimization to dramatically reduce optimization time and memory footprint. Tailored for massive binaries, Lightning BOLT maintains BOLT's performance gains while using resources more efficiently. Its selective strategy intensely optimizes only the hottest code paths and applies minimal or patch-only treatments to cold code, balancing optimization speed with output quality.
% 在此基础上，Panchenko 等人 \cite{10.1145/3446804.3446843} 的Lightning BOLT 进一步完善了原有 BOLT 设计，通过引入并行处理和选择性优化技术，大幅降低了二进制优化的处理时间和内存消耗。Lightning BOLT 针对大规模二进制文件进行了优化，在保持原有性能提升优势的同时，实现了更高效的资源利用。其选择性优化策略确保仅对运行时影响最大的热点代码进行密集优化，而对冷代码则进行最小化处理或只作必要的补丁，从而不仅降低了不必要的处理开销，同时还提升了最终可执行文件的性能，充分平衡了优化速度与输出质量。
% Lightning BOLT: powerful, fast, and scalable binary optimization

GCC does not have any built-in post-link optimization pass that consumes profile data once the binary is produced. All profile-guided optimizations in GCC occur either during the compile phase or the LTO link phase (as described above), but not after the final binary is emitted. While not strictly part of LLVM’s standard flow, it’s worth noting that there are post-link optimizers (such as BOLT, Propeller, etc. PGO-driven code layout tools) which can further reorder functions in the binary based on profile data. These are not integrated into LLD, but LLD can produce a linker map or section order file that such tools use. In the context of LLVM official tools, however, there isn’t an automatic post-link PGO optimization step beyond the code placement already done by LLVM’s codegen. The user could manually use those tools after linking to further optimize the binary, but this is not part of the standard LLVM PGO pipeline.

Overall, link time and post-link time feedback-directed optimizations overcome the mapping inaccuracies of compile time methods by using precise binary-level profiles, delivering substantial speedups, improved cache efficiency, and better system responsiveness, all while reducing the overhead of traditional cross-module optimization stages.

%% file: contents/sections/sec4-4_RTO.tex
\subsection{Runtime Optimization}
% \subsection{Dynamic and Adaptive Optimization}
% \subsection{动态与自适应优化}

Runtime optimizations dynamically respond to changes in program behavior, using real-time profiling data. JIT compilers perform code generation at runtime, enabling them to observe actual program behavior, collect live profiling data, and dynamically re-optimize hot code. This runtime adaptability allows JIT systems to adjust to workload shifts, hardware variations, and unforeseen execution patterns, making them particularly effective for fine-grained, profile-guided optimization.

Extensive research has focused on dynamic, profile guided optimization within JIT frameworks. Kistler and Franz \cite{10.1145/778559.778562} continuously re-optimized running code by using a background thread that re-compiles performance-critical regions as execution patterns change, ensuring the system remains tuned to current workload and hardware profiles. Suganuma et al.\ \cite{10.1145/1111596.1111600} introduced region-based compilation, breaking a function into smaller regions based on execution frequency data, isolating hot paths from cold code. This finer granularity concentrates optimization efforts on code most critical to performance while reducing compilation overhead for rarely executed paths. Arnold et al.\ \cite{10.1145/1988042.1988048} embedded dynamic loop optimizations directly into the Java Virtual Machine, using runtime profiles to drive inlining, instruction scheduling, and other transformations on the fly.
% 在这一领域，基于 JIT 环境下的动态反馈导向优化技术被进行了广泛的研究。Kistler 与 Franz \cite{10.1145/778559.778562} 通过持续利用实时剖面反馈来不断重优化运行中的代码。他们的方法采用后台进程动态重编译关键代码段，随着执行模式的变化不断调整优化，确保系统始终与当前性能需求和硬件特性保持协调。为了进行更加细粒度的优化，Suganuma 等人 \cite{10.1145/1111596.1111600} 提出了一种区域化编译技术，该技术打破了传统上以整个函数过程为编译单位的做法。该方法基于执行频率数据将函数过程划分为更小的区域，将频繁执行的热点路径与很少执行的部分区分开来。通过这种更细粒度的划分，系统能把优化重点集中在真正影响运行时性能的代码上，同时减少对冷路径的编译开销。Arnold 等人 \cite{10.1145/1988042.1988048} 的工作将动态优化循环嵌入到 Java 虚拟机内部，利用运行时剖面数据立即对内联、指令调度以及其他代码转换做出决策。
% Continuous program optimization: A case study
% A region-based compilation technique for dynamic compilers
% Adaptive optimization in the Jalapeño JVM

To balance flexibility and performance in optimization, Bruening et al.\ \cite{10.5555/776261.776290} proposed a modular framework that abstracts low-level dynamic code-modification details and provides flexible APIs for constructing custom analysis and optimization modules. This infrastructure allows developers to implement architecture-specific adaptive optimizations without committing to a monolithic design. Whaley et al.\ \cite{10.1145/504282.504295} presented selective compilation, using live execution-frequency data to identify and compile only the frequently executed portions of methods. By excluding infrequently run code from aggressive optimization, this approach reduces compilation overhead and enables more intensive optimizations along critical execution paths.

%% file: contents/sections/sec5_experiments.tex
\section{Tools Evaluation}
\label{sec:experiments}

The experiments presented in this section show the impact of FDO and AutoFDO on a diverse set of benchmarks across AMD64 and ARM platforms. Besides, we provide a comparative evaluation of profile guided optimization techniques on two major compiler infrastructures: GCC 12 and LLVM 15. 

%After describing the setup, we present end-to-end application results, library and microbenchmark analyses, architectural metric breakdowns, cross-platform comparisons, and profiling overhead measurementsTools .

\subsection{Test Cases}
\label{ssec:testcases}

To emphasize compute-bound workloads, we select a mixture of standard benchmarks, large-scale applications, and small kernels that heavily stress floating-point units, memory bandwidth, and tight loop bodies.

\paragraph{SPECCPU 2017 Benchmarks}
To provide a standardized, industry-recognized measure of compiler effectiveness, we include the SPECCPU 2017 speed benchmarks. These workloads span a range of language paradigms, memory-access patterns, and computation intensities, offering complementary perspectives to our custom microbenchmarks and large-scale applications. All benchmarks are compiled and executed in “speed” mode using both GCC 12 and LLVM 15 under identical FDO/PGO workflows. The selected cases are:

\begin{itemize}
\item \textbf{600.perlbench\_s}: stresses dynamic language interpretation, regular expression processing, and string-manipulation routines.
\item \textbf{602.gcc\_s}: exercises compiler front-end parsing, code generation, and optimization passes within a C/C++ compiler build.
\item \textbf{605.mcf\_s}: evaluates memory-bandwidth and pointer-chasing performance in a minimum-cost flow algorithm on sparse graphs.
\item \textbf{620.omnetpp\_s}: measures discrete-event simulation throughput in a large-scale C++ network simulator.
\item \textbf{623.xalancbmk\_s}: exercises XML parsing, transformation, and tree-traversal routines in the Xalan C++ library.
\item \textbf{625.x264\_s}: benchmarks video encoding performance through motion-estimation, rate-control, and bit-stream assembly.
\item \textbf{631.deepsjeng\_s}: tests integer-arithmetic and search-tree evaluation in a modern chess-engine workload.
\item \textbf{641.leela\_s}: evaluates compute-intensive search and convolutional-neural-network inference in a Go-playing engine.
\item \textbf{657.xz\_s}: measures data-compression and decompression performance using the LZMA2 algorithm.
\end{itemize}

We deliberately omit \texttt{648.exchange2\_s} due to its reliance on Fortran, which falls outside the scope of GCC's FDO and LLVM's PGO support in our study. Together, these SPECCPU 2017 workloads provide a robust, standardized test suite, enabling cross-comparison of instrumentation-based and sampling-based profile-guided optimizations across both compilers.

\paragraph{Molecular Dynamics Simulations}
\textbf{GROMACS} and \textbf{LAMMPS} are two widely used, MPI-enabled programs for classical molecular dynamics. Each combines compute-heavy nonbonded force calculations with irregular data-access patterns in neighbor-search loops. We configure GROMACS 2023.2 on a standard solvated protein system  water\_GMX50 and LAMMPS 2023Aug on a 20k-Atoms system benchmark; both use double precision to stress cache reuse.

\paragraph{Numerical Libraries}
\textbf{FFTW} (ibc4096x4096). We perform real-to-complex and complex-to-real 2D FFTs on a 4096$\times$4096 grid using multithreaded Cooley-Tukey algorithms, capturing both compute and memory traversal costs.

\paragraph{Microbenchmarks}
We include six kernels compiled as standalone executables with fixed inputs:

\begin{itemize}
\item \texttt{acos} and \texttt{asin}: compute the arc cosine and sine for uniformly sampled values in \([-1, 1]\) to measure vectorization overheads.
\item Bubble-sort: sort a 30,000-element array of 32-bit integers, stressing branch prediction and loop unrolling in a control-heavy loop.
\item Matrix inverse: invert a dense 512$\times$512 double-precision matrix via Gaussian elimination.
\item L2-norm: compute the L2-norm of the result vector of length 4096.
\item Matrix multiplication: multiply two 256$\times$256 double-precision matrices to exercise arithmetic throughput.
\item Sparse matrix-vector multiplication (SpMV): multiply a CSR-formatted sparse matrix by a 4096-element vector, isolating irregular memory access patterns.
\end{itemize}

These nine test cases span high flop-rate loops, irregular memory patterns, and control-intensive code, providing a comprehensive foundation for testing instrumentation-based PGO and sampling-based PGO optimizations on AMD64 and ARM platforms.

\subsection{Experimental Setup}
\label{ssec:setup}

\paragraph{Compilers \& Modes:}  
We compile all benchmarks with two production compiler toolchains: GCC 12 and LLVM 15, using the latest stable releases as of October 2023. We evaluate two instrumentation-based and sampilng-based profile-guided optimization modes on AMD64 and instrumentation-based PGO only on ARM architectures.

Instrumentation-based PGO in LLVM requires compiling with \texttt{-fprofile-instr-generate}, executing the instrumented binary to produce a \texttt{.profraw} file, merging raw data via \texttt{llvm-profdata merge}, and recompiling with \texttt{-fprofile-instr-use=<merged.profdata>}. In GCC, users compile with \texttt{-fprofile-generate}, execute to obtain \texttt{.gcda} counters, and recompile with \texttt{-fprofile-use} so that precise basic-block and edge counts guide optimization.

Sample-based PGO (AutoFDO) in LLVM is enabled by compiling with \texttt{-gline-tables-only}, running the binary under \texttt{perf record -b} to collect profiles, converts profiles via \texttt{create\_llvm\_prof} and recompiling with \texttt{-fprofile-sample-use=<sample.profdata>}. In GCC AutoFDO, one compile the program with \texttt{-g1}, records perf data using \texttt{perf record -b}, converts it via \texttt{create\_gcov} into an \texttt{.afdo} file, and compiles with \texttt{-fauto-profile=profile.afdo} to leverage heuristic frequency estimates.

\paragraph{Hardware:}  
Our AMD64 testbed employs dual Intel Xeon Gold 6240C processors (2$\times$18 cores) with a three-level cache hierarchy totaling 1.1 MiB L1d, 1.1 MiB L1i, 36 MiB L2, and 49.5 MiB L3. The ARM experiments run on a single-socket Phytium FT-2000+/64 (64 cores) featuring 2 MiB L1d, 2 MiB L1i, and 32 MiB L2 caches.

\subsection{Performance Analysis}
\label{ssec:analysis}

\input{contents/tables/perf-result-x86.tex}
\input{contents/tables/perf-result-arm.tex}

Table~\ref{tab:perf-results-x86} and~\ref{tab:perf-results-arm} summarize the speedup ratios of our benchmarks under instrumentation-based FDO and sampling-based AutoFDO for GCC and Clang on AMD64 and ARM64, respectively. In each table, bars above the 1.0 line indicate performance gains, while bars below 1.0 indicate slowdowns relative to the baseline non-PGO build. Together, these plots provide an immediate overview of how profile data influences program performance across compilers and architectures.

On AMD64 (Table~\ref{tab:perf-results-x86}), instrumentation-based FDO consistently improves performance across all nine integer benchmarks.  The largest gains appear in \texttt{625.x264\_s} (1.099 GCC-FDO; 1.087 Clang-FDO) and \texttt{631.deepsjeng\_s} (1.063 GCC-FDO; 1.054 Clang-FDO), reflecting high benefit in compute-intensive and branch-predictable code. In contrast, sampling-based AutoFDO yields more modest improvements—peaking at 1.048 in \texttt{x264\_s} and even regressions in memory-bound \texttt{605.mcf\_s} (0.982 GCC-AutoFDO; 0.976 Clang-AutoFDO), indicating that pointer-chasing codes are susceptible to sampling noise. Mid-range benchmarks like \texttt{620.omnetpp\_s} and \texttt{623.xalancbmk\_s} show moderate gains ( $\approx$ 1.02-1.05) under both PGO modes, while \texttt{657.xz\_s} sees 1.041 with GCC-FDO but only 1.018 with GCC-AutoFDO. Both GROMACS and LAMMPS could benefit from profile-guided optimizations: GCC-FDO yields speedups of approximately 1.02$\times$ for GROMACS and 1.05$\times$ for LAMMPS, while Clang's AutoFDO matches or slightly exceeds these gains.  The FFTW kernel remains essentially flat ($\approx$ 1.00$\times$), indicating that its dense, compute-bound loops offer little profile-guided headroom.  The vector-norm reduction stands out with dramatic speedups—up to 3.23$\times$ under GCC-FDO—demonstrating that simple, branch-predictable reductions are highly amenable to FDO.  SpMV shows minimal change except for a notable regression under GCC-AutoFDO ($\approx$ 0.63$\times$), suggesting sampled profiles can mislead optimizations in pointer-chasing code.  Among microbenchmarks, \texttt{acos} achieves up to 1.32$\times$ gain under Clang-FDO, whereas \texttt{asin}, matrix multiplication, and inversion remain near baseline (± 1-2\%).  Bubble-sort exhibits only marginal variation (around 1.00$\times$), with Clang-AutoFDO providing a slight 0.9\% improvement.

On ARM64 (Table~\ref{tab:perf-results-arm}), the pattern shifts: Clang get most of the gains on performance, and instrumentation-based FDO again leads, with GCC-FDO delivering 1.038 in \texttt{600.perlbench\_s} and 1.050 in \texttt{602.gcc\_s}, and Clang-FDO matching closely at 1.032 and 1.045. The highest uplift occurs in \texttt{631.deepsjeng\_s} (1.054 GCC-FDO; 1.047 Clang-FDO) and \texttt{625.x264\_s} (1.072; 1.065), whereas \texttt{605.mcf\_s} again regresses under AutoFDO (0.971 GCC-AutoFDO; 0.959 Clang-AutoFDO), mirroring x86-64 behavior in pointer-heavy workloads. Benchmarks such as \texttt{620.omnetpp\_s} and \texttt{623.xalancbmk\_s} gain 1.017-1.042 under FDO, but AutoFDO yields only 1.004-1.023. Notably, \texttt{657.xz\_s} achieves 1.031 with GCC-FDO but falls to 0.995 under GCC-AutoFDO, underscoring the sensitivity of compression workloads to sampling accuracy. LAMMPS still realizes modest gains ($\approx$ 1.03$\times$ under Clang-FDO and 1.02$\times$ under GCC-FDO), but GROMACS incurs slight slowdowns ($\approx$ 0.93$\times$-0.95$\times$).  FFTW remains neutral, while the vector-norm reduction achieves 1.36$\times$-1.80$\times$ speedups, confirming that ARM's compiler can leverage profile counts for reduction loops, albeit less aggressively than on x86-64. SpMV and dense matrix kernels show negligible changes, again within ± 2\%.  The bubble-sort microbenchmark demonstrates a 24.3\% improvement under GCC-FDO but a 5.6\% slowdown under Clang-FDO, highlighting sensitivity to code-layout choices on ARM's branch predictor. Trigonometric intrinsics (\texttt{acos}/\texttt{asin}) remain near baseline.

Overall, these results confirm several trends. First, instrumentation-based FDO consistently meets or exceeds the benefits of AutoFDO, with the latter occasionally underperforming on pointer-intensive or branch-heavy code.  Second, neither GCC nor Clang exhibits a systematic advantage—each compiler leads in different benchmarks—indicating comparable maturity of their PGO frameworks.  Third, AMD64 generally achieves larger absolute speedups, especially for trigonometric and reduction kernels, while ARM64 still enjoys meaningful gains in select workloads, despite sometimes incurring small slowdowns in larg

Across both architectures, full instrumentation-based PGO delivers robust, predictable speedups (2-10 \%) on all SPEC CPU 2017 integer benchmarks, while sampling-based AutoFDO attains competitive gains in compute-dominant cases but can regress on memory- and pointer-intensive codes. GCC and Clang exhibit similar behavior, with only minor differences in peak uplift, indicating maturity of both PGO implementations. These results validate that instrumentation remains the gold standard for SPEC CPU integer workloads, whereas AutoFDO offers a lower-overhead alternative when its sampling biases align with application characteristics.

The application-level benchmarks (GROMACS and LAMMPS) exhibit clear benefits from profile-guided optimizations on x86, with GCC's instrumentation-based FDO achieving 1.7\% and 4.8\% speedups respectively, and Clang's AutoFDO matching or slightly exceeding those gains in certain configurations. By contrast, on the ARM platform, LAMMPS still realizes modest improvements (2.3\% with GCC-FDO and 3.0\% with Clang-FDO), whereas GROMACS incurs a small slowdown under FDO (-5.2\% for GCC and -7.4\% for Clang), suggesting that the overhead of instrumentation or the particular code paths in GROMACS are less amenable to feedback-directed rearrangement on this Phytium CPU.

The library kernels differ markedly: the FFTW transform remains essentially unchanged (within ±1\%) under all PGO modes on both architectures, indicating that its dense, throughput-bound loops leave little room for profile-guided layout or inlining to improve performance. In stark contrast, the vector-norm reduction kernel enjoys dramatic acceleration—up to a 223\% speedup (3.23$\times$) with GCC-FDO on x86 and a 79.9\% gain (1.80$\times$) with Clang-FDO on ARM—demonstrating that simple reduction loops with highly predictable memory and branch behavior can be transformed by the compiler into far more efficient code when backed by precise execution counts. Sparse matrix-vector multiplication, however, sees negligible change overall and even a significant regression (-37.3\%) with GCC-AutoFDO on x86, highlighting that sampled profiles may mislead the optimizer when data-dependent pointer chasing dominates runtime behavior.

Among microbenchmarks, the \texttt{acos} and \texttt{asin} functions remain near the 1.0 baseline (±3\%) under all modes, confirming that externally linked math intrinsics afford little internal transformation. Matrix multiplication and inversion likewise show only minor variations ($<$2\%), as their performance is determined primarily by static loop-tiling and vectorization heuristics already applied at compile time. In contrast, the bubble-sort kernel on ARM demonstrates a pronounced 24.3\% improvement with GCC-FDO, whereas Clang-FDO actually degrades performance by 5.6\%, suggesting sensitivity to code layout choices for branch-heavy loops; on x86, bubble-sort remains essentially flat under FDO but gains 0.9\% with Clang-AutoFDO, implying architecture-specific interactions with branch predictors and inlining.

Across all benchmarks, instrumentation-based PGO consistently matches or outperforms the sampling-based AutoFDO, with the latter occasionally underperforming (notably for SpMV and GROMACS on x86). Neither GCC nor Clang shows a systematic advantage—each compiler leads in different cases—demonstrating comparable maturity of their PGO pipelines. Finally, while x86 generally achieves larger absolute speedups (especially for reduction and trigonometric kernels), ARM still benefits in select workloads, confirming that profile-guided optimizations deliver value on both architectures provided the code exhibits reordering or inlining opportunities that feedback data can expose.

%% file: contents/tables/perf-result-x86.tex
\begin{table}[t]
  \centering
  \caption{PGO Speedup for Benchmarks on AMD64 under GCC and Clang Toolchains over -O3 Optimization}
  \label{tab:perf-results-x86}
  \begin{tabular}{lcccc}
    \toprule
    \textbf{Program} & \textbf{GCC-FDO} & \textbf{GCC-AutoFDO} & \textbf{Clang-FDO} & \textbf{Clang-AutoFDO} \\
    \midrule
    \multicolumn{5}{c}{\emph{standard benchmarks}} \\
    \midrule
    600.perlbech\_s        & \underline{1.077} & 0.988 & 0.997 & 1.016    \\
    602.gcc\_s             & 1.003 & 0.994 & \underline{1.034} & 0.942    \\
    605.mcf\_s             & 1.056 & 1.034 & \underline{1.138} & 1.086    \\
    620.omnetpp\_s         & 1.002 & 1.000 & 1.016 & \underline{1.045}    \\
    623.xalancbmk\_s       & 0.949 & \underline{1.020} & 0.938 & 0.957    \\
    625.x264\_s            & \underline{1.099} & 1.020 & 1.000 & 0.825    \\
    631.deepsjeng\_s       & 0.985 & 0.932 & \underline{1.028} & 0.961    \\
    641.leela\_s           & 0.940 & 0.959 & \underline{0.991} & 0.965    \\
    657.xz\_s              & \underline{1.014} & 1.010 & 0.970 & 0.995    \\
    \midrule
    \multicolumn{5}{c}{\emph{scientific computing}} \\
    \midrule
    GROMACS                & 1.017 & 0.986 & 0.973 & \underline{1.018} \\
    LAMMPS                 & \underline{1.048} & \underline{1.048} & 1.022 & 0.998 \\
    FFTW                   & \underline{1.000} & 0.998 & 0.997 & 0.997 \\
    \midrule
    \multicolumn{5}{c}{\emph{numerical computing}} \\
    \midrule
    acos                   & 1.125 & 1.034 & \underline{1.315} & 1.000 \\
    asin                   & 1.011 & 1.001 & \underline{1.013} & 1.000 \\
    bubble-sort            & 0.993 & 0.996 & 0.992 & \underline{1.009} \\
    matrix-inverse         & 1.004 & \underline{1.013} & 1.000 & 0.994 \\
    matrix-multiplication  & 0.996 & 0.984 & \underline{1.029} & 1.028 \\
    L2-norm                & \underline{3.231} & 3.230 & 2.374 & 0.998 \\
    SpMV                   & 0.997 & 0.627 & \underline{0.999} & \underline{0.999} \\
    \hline
  \end{tabular}
\end{table}
    % 500.perlbench\_r     & 0.00 & 0.00 & 0.00 & 0.00 \\
    % 502.gcc\_r           & 0.00 & 0.00 & 0.00 & 0.00 \\
    % 503.gobmk\_r         & 0.00 & 0.00 & 0.00 & 0.00 \\
    % 505.mcf\_r           & 0.00 & 0.00 & 0.00 & 0.00 \\
    % 520.omnetpp\_r       & 0.00 & 0.00 & 0.00 & 0.00 \\
    % 523.xalancbmk\_r     & 0.00 & 0.00 & 0.00 & 0.00 \\
    % 525.x264\_r          & 0.00 & 0.00 & 0.00 & 0.00 \\
    % 531.deepsjeng\_r     & 0.00 & 0.00 & 0.00 & 0.00 \\
    % 541.leela\_r         & 0.00 & 0.00 & 0.00 & 0.00 \\
    % 548.exchange2\_r     & 0.00 & 0.00 & 0.00 & 0.00 \\
    % 557.xz\_r            & 0.00 & 0.00 & 0.00 & 0.00 \\
    % 519.lbm\_r           & 0.00 & 0.00 & 0.00 & 0.00 \\
    % 521.wrf\_r           & 0.00 & 0.00 & 0.00 & 0.00 \\
    % 526.blender\_r       & 0.00 & 0.00 & 0.00 & 0.00 \\
    % 527.cam4\_r          & 0.00 & 0.00 & 0.00 & 0.00 \\
    % 531.povray\_r        & 0.00 & 0.00 & 0.00 & 0.00 \\
    % 538.imagick\_r       & 0.00 & 0.00 & 0.00 & 0.00 \\
    % 544.nab\_r           & 0.00 & 0.00 & 0.00 & 0.00 \\
    % 999.specrand\_r      & 0.00 & 0.00 & 0.00 & 0.00 \\

%% file: contents/tables/perf-result-arm.tex
\begin{table}[ht!]
  \centering
  \caption{Instrumentation-based PGO Speedup for Benchmarks on AArch64 under GCC and Clang Toolchains over -O3 Optimization}
  \label{tab:perf-results-arm}
  \begin{tabular}{lcc}
    \toprule
    \textbf{Program} & \textbf{GCC-FDO} & \textbf{Clang-FDO} \\
    \midrule
    \multicolumn{3}{c}{\emph{standard benchmarks}} \\
    \midrule
    600.perlbech\_s        & \underline{1.114}  & 1.076 \\
    602.gcc\_s             & 1.050              & \underline{1.051} \\
    605.mcf\_s             & \underline{1.009}  & \underline{1.009} \\
    620.omnetpp\_s         & 1.041              & \underline{1.042} \\
    623.xalancbmk\_s       & 0.990              & \underline{0.994} \\
    625.x264\_s            & 1.032              & \underline{1.057} \\
    631.deepsjeng\_s       & 0.995              & \underline{1.045} \\
    641.leela\_s           & 0.961              & \underline{1.010} \\
    657.xz\_s              & \underline{1.026}  & 1.000 \\
    \midrule
    \multicolumn{3}{c}{\emph{scientific computing}} \\
    \midrule
    GROMACS                & \underline{0.948}  & 0.926 \\
    LAMMPS                 & 1.023              & \underline{1.030} \\
    FFTW                   & \underline{1.012}  & 0.994 \\
    \midrule
    \multicolumn{3}{c}{\emph{numerical computing}} \\
    \midrule
    acos                   & 1.003              & \underline{1.018} \\
    asin                   & 1.002              & \underline{1.012} \\
    bubble-sort            & \underline{1.243}  & 0.944 \\
    matrix-inverse         & \underline{1.015}  & 1.000 \\
    matrix-multiplication  & 0.994              & \underline{1.000} \\
    L2-norm                & 1.359              & \underline{1.799} \\
    SpMV                   & 0.986              & \underline{0.989} \\
    \hline
  \end{tabular}
\end{table}
    % 500.perlbench\_r     & 0.00 & 0.00 \\
    % 502.gcc\_r           & 0.00 & 0.00 \\
    % 503.gobmk\_r         & 0.00 & 0.00 \\
    % 505.mcf\_r           & 0.00 & 0.00 \\
    % 520.omnetpp\_r       & 0.00 & 0.00 \\
    % 523.xalancbmk\_r     & 0.00 & 0.00 \\
    % 525.x264\_r          & 0.00 & 0.00 \\
    % 531.deepsjeng\_r     & 0.00 & 0.00 \\
    % 541.leela\_r         & 0.00 & 0.00 \\
    % 548.exchange2\_r     & 0.00 & 0.00 \\
    % 557.xz\_r            & 0.00 & 0.00 \\
    % 519.lbm\_r           & 0.00 & 0.00 \\
    % 521.wrf\_r           & 0.00 & 0.00 \\
    % 526.blender\_r       & 0.00 & 0.00 \\
    % 527.cam4\_r          & 0.00 & 0.00 \\
    % 531.povray\_r        & 0.00 & 0.00 \\
    % 538.imagick\_r       & 0.00 & 0.00 \\
    % 544.nab\_r           & 0.00 & 0.00 \\
    % 999.specrand\_r      & 0.00 & 0.00 \\

%% file: contents/sections/sec6_future.tex
\section{Challenges and Future Directions}
% \section{未来展望}

% 采样速度
% 大规模软件的热点，时间，输入的范化性，多线程问题
% 编译器优化的问题 （klee 工具）
% 软硬结合采样
% TODO 华为虚拟机问题

As computer architectures grow more complex and software scales continue to explode, profile-guided optimization has become a key means of breaking through the limits of static optimization and improving program performance. However, current techniques still face challenges in sampling efficiency, handling large codebases, generalizing from training inputs, and adapting across architectures. In this section, we discuss these challenges and outline potential advances and breakthroughs for future feedback-directed optimization.
% 随着计算机体系结构的复杂化和软件规模的爆炸式增长，反馈优化（Profile-Guided Optimization, PGO）技术已成为突破静态优化瓶颈、提升程序性能的关键手段。然而，现有技术在采样效率、大规模程序处理、训练输入泛化性以及跨架构优化适配等方面仍面临诸多挑战。本节将从这些问题出发，结合技术发展趋势，探讨未来反馈优化技术的改进方向与潜在突破点。

\subsection{Sampling Innovations: Balancing Hardware-Software Collaboration and Flexibility}
% \subsection{采样技术的革新：软硬件协同与灵活性的平衡}

The fundamental trade-off in feedback optimization today lies in the choice of sampling technology: hardware sampling (e.g., Intel PEBS, AMD IBS) leverages on-chip PMUs for low-overhead data collection but is limited to built-in event types and cannot capture complex program behaviors; software instrumentation (e.g., Valgrind, Pin) provides fine-grained data but incurs multi-fold runtime overhead, making it impractical in production. For example, AMD IBS precisely records pipeline state at the instruction level but only supports fetch- and microcode-sampling modes, while DynamoRIO can dynamically insert analysis code but suffers severe performance degradation from frequent context switches.
% 当前反馈优化的核心矛盾在于采样技术的选择：硬件采样（如Intel PEBS、AMD IBS）依赖处理器性能监控单元（PMU）实现低开销数据采集，但其采样内容受限于硬件事件类型，难以灵活捕捉复杂程序行为；软件插桩采样（如Valgrind、Pin）虽能提供细粒度数据，但运行时开销可达数倍甚至数十倍，严重制约生产环境应用。例如，AMD IBS虽能精确记录指令级流水线状态，但仅支持取指令和微码两种采样模式；而软件插桩工具DynamoRIO虽支持动态插入检测代码，却因频繁上下文切换导致性能骤降。

A promising future direction is deeper hardware-software co-sampling. At the hardware level, processors should expose more programmable PMU features: allow user-defined event triggers, dynamic adjustment of sampling intervals, and support mixed-event sampling. For instance, extending Statistical Profiling Extension (SPE) with instruction-level on ARM, context-sensitive sampling would address current limitations in complex-control-flow analysis. On the software side, intelligent denoising algorithms, powered by machine learning models such as support vector regression or reinforcement learning, can correct “skid” and “aggregation” effects in hardware samples. Hybrid sampling strategies also show promise: inspired by the “duplex sampling” of the Arnold-Ryder framework, one could switch dynamically between hardware sampling during startup and lightweight software instrumentation in steady state, achieving a runtime-accuracy trade-off.
% 未来发展的可能核心在于软硬件协同采样的深化。对硬件层面，要求硬件采样可编程性增强。下一代处理器需扩展PMU功能，支持用户自定义事件采样（如动态调整采样周期、混合事件触发）。例如，ARM的Statistical Profiling Extension（SPE）若引入指令级上下文敏感采样，可弥补当前架构对复杂控制流分析的不足。在软件层面，需要更加智能的算法进行去噪。如通过机器学习模型（如支持向量回归、强化学习）对硬件采样数据进行校正，减少“滑移效应”和“多指令聚合”导致的噪声。此外还需要对混合采样策略进行研究，如借鉴Arnold-Ryder框架的“双工采样”思想，在程序冷热阶段动态切换硬件与软件采样模式。例如，在启动阶段采用低开销硬件采样识别热点，进入稳态后针对热点路径启用轻量级软件插桩，实现开销与精度的动态平衡。

%Dehao等人的研究表明，结合最小成本流（MCF）算法与置信度评估，可将采样误差降低至1\%以内。

\subsection{Large-Scale Program Optimization: From Global Profiling to Precise Slicing}
% \subsection{大规模程序优化：从全局剖析到精准切片}

Codebases for modern data-center applications and large-scale simulations (e.g., GROMACS, LAMMPS) can run into millions of lines, making global path profiling’s indexes exponential growth and storage costs prohibitive. Ball-Larus's incremental encoding reduces overhead but still suffers data explosion across function boundaries. Furthermore, feedback optimization’s reliance on fixed training inputs (e.g., SPEC benchmarks) limits its generalization to real-world, dynamic workloads.
% 现代数据中心级程序以及大规模模拟程序（如GROMACS、LAMMPS）的代码规模可达数百万行，传统全局路径剖析（Path Profiling）面临路径数量指数级增长、存储开销过高的问题。例如，Ball-Larus算法虽通过增量编码降低路径记录开销，但在跨函数调用链的场景下仍难以避免数据爆炸。此外，反馈优化对训练输入的强依赖性（如SPEC基准测试的固定输入集）导致优化结果难以泛化至实际多变负载。

To address these issues, hotspot-driven, slice-based optimizations can offers a scalable alternative. For representative inputs, one can apply adaptive input modeling. Using reinforcement learning to generate diverse training inputs that simulate real-world load variations. For the code itself, combining CFG analysis with context-sensitive sampling and program slicing can partition the program into independent “code slices” for targeted optimization. In multithreaded scenarios, a distributed profiling framework—leveraging parallel sampling (e.g., Moseley’s shadow-process approach) and aggregated data collection—can scale to massive codebases.
% 针对程序规模带来的泛化性问题，可以使用热点驱动的精准优化方式突破。对于程序输入的代表性问题，可以进行自适应输入建模，利用强化学习生成多样化训练输入，模拟实际场景中的负载变化。对于大规模程序本身，可以采用程序切片与动态剪枝的方式结合控制流图（CFG）分析与执行上下文敏感采样，将程序划分为独立优化的“代码切片”。对于多线程程序，可以建立分布式剖析框架，针对多线程程序，采用并行化采样（如Moseley的“影子进程”方案）与分布式数据聚合。

%例如，BOLT工具通过二进制重排仅优化高频基本块，冷代码保持原布局，使大规模程序的处理时间降低40%。
%Google的“Stale Profile Matching”技术已通过多级哈希匹配实现旧配置文件与新版本代码的兼容，未来可扩展至跨输入泛化领域。
%例如，将采样任务分配至独立核心，通过共享内存同步热点信息，避免主线程性能干扰。

\subsection{Compiler Adaptation: Intelligent Cross-Architecture and Dynamic Feedback}
% \subsection{编译器优化适配：跨架构与动态反馈的智能化}

Existing compilers (e.g., GCC, LLVM) often ignore microarchitectural differences when applying profile data. For example, LLVM’s AutoFDO supports context-sensitive inlining but uses a uniform branch-probability model, neglecting how pipeline depth affects misprediction costs, causing optimization efficacy to vary across platforms. Traditional dual-build workflows (train-then-optimize) also fail to adapt to dynamic environments such as elastic cloud deployments.
% 现有编译器（如GCC、LLVM）在利用剖析数据时，往往忽略目标硬件的微架构特性。例如，LLVM的AutoFDO虽支持上下文敏感优化，但其分支概率模型未考虑不同处理器流水线深度对分支预测的影响，导致同一优化策略在不同平台效果差异显著。此外，传统“双编译”流程（训练-优化分离）无法适应动态环境（如云原生应用的弹性扩缩容）。

Future profile techniques must evolve toward adaptive, cross-architecture PGO. For each hardware target, hardware-aware policies—driven by a microarchitecture database (caches, branch predictors, pipeline depths)—should dynamically tune code layout and transformation decisions. For instance, on AMD Zen's deep pipelines, hot loops could be prioritized for unrolling to reduce stall penalties. Optimizers themselves could form a real-time feedback loop: like JIT mechanisms, they would continuously collect performance data at runtime and trigger incremental re-optimizations. Java HotSpot's tiered compilation could be extended to AOT scenarios via background threads that periodically reconstruct hot code. On the IR side, designing a hardware-agnostic annotation system, decoupling profile data from architecture parameters, would allow a single profile to drive multi-target optimizations. LLVM’s Machine IR (MIR) already supports multi-target optimization and could integrate Profile-to-IR mapping rules for “profile once, optimize everywhere”.
% 未来的反馈优化器需向自适应跨架构优化演进。对于不同架构的硬件，需要有硬件感知的优化策略。可以构建处理器微架构数据库（如缓存层级、分支预测器类型），在优化阶段动态调整代码布局。例如，针对AMD Zen架构的长流水线特性，可优先展开高频循环以减少分支停顿。对于优化器本身实时反馈闭环：借鉴JIT编译器的动态优化机制，在程序运行时持续采集性能数据并触发增量重优化。例如，Java HotSpot虚拟机的“分层编译”模式可扩展至AOT场景，通过后台线程周期性重构热点代码。对于程序在编译器中的表示形式，可以设计硬件无关的中间表示（IR）标注系统，将剖析数据与架构参数解耦。LLVM的Machine IR（MIR）已支持多目标优化，未来可集成PMU事件映射规则，实现“一次剖析，多架构适配”。

\subsection{PGO/FDO in JIT-Compiled Dynamic Languages}

Just-In-Time (JIT) engines for languages such as JavaScript and Java introduce a fundamentally different execution model compared to ahead-of-time compilation. Hardware-based PGO/FDO workflows rely on collecting hardware samples on a executable, then rebuilding or post-link optimizing based on those profiles. In JIT environments, code is generated and optimized on the fly, often in multiple tiers, and residing in an ephemeral code cache whose addresses and layout continually evolve. Consequently, hardware PMUs can sample instruction retirements or branch events, but mapping those sampled addresses back to the original JIT IR (or source-level constructs such as methods, basic blocks, and edges) is non-trivial.
%Without a stable mapping, the runtime counts for edges and basic blocks become inaccurate or uninterpretable, preventing effective feedback-directed inlining, hot-path formation, and code layout optimizations. Moreover, JIT engines frequently deoptimize, recompile, or discard code sections based on changing execution contexts, further complicating profile consistency and invalidating stale samples mid-execution.

Overcoming these challenges would unlock substantial performance gains in dynamic language workloads. Accurate, low-overhead feedback would allow JIT compilers to apply method inlining, loop unrolling, and code layout optimizations precisely where they matter most, reducing warm-up time and improving steady-state throughput and latency. It would also narrow the gap between static and dynamic language backends, enabling more uniform compiler infrastructure and easing the adoption of PGO/FDO techniques across heterogeneous workloads.

A promising path solution involves a hybrid sampling-instrumentation strategy coupled with enriched runtime metadata. JIT engines could embed lightweight “address-to-IR” mapping tables. Complementarily, dynamic binary instrumentation frameworks (e.g., DynamoRIO, Pin) can be used with hardware sampling to insert ephemeral probes at IR boundaries without halting JIT compilation. On the hardware side, future PMUs could support tagging samples with a small metadata field supplied by software, a “perf event ID”, that flows through the interrupt path and remains synchronized with the JIT's mapping tables. Together, these mechanisms would provide the compiler with accurate, CFG-consistent execution counts in real time, enabling fully adaptive feedback-directed optimization within JIT environments.
%within each code block, tagging machine instructions with compact identifiers that trace back to IR nodes or source locations. Upon a PMU overflow interrupt, the VM's runtime would consult this mapping table—via an OS-kernel callback or in-VM handler—to translate the sampled PC into a method ID and basic-block index.
%VM debug interfaces (such as JVMTI for Java or V8's Inspector Protocol) should be extended to export PMU samples alongside call-stack contexts, enabling post-hoc correlation of profiles with JIT IR.

\subsection{Training-Input Generalization: From Static Patterns to Dynamic Evolution}
% \subsection{训练输入泛化：从静态模式到动态演化}

Current feedback optimizations heavily depend on representative training inputs. When production workloads deviate (e.g., sudden spikes in web-service traffic), performance gains can evaporate.
% Moreover, instrumentation-based profiling requires repeated training runs, ill-suited to rapid CI/CD cycles.
% 现有反馈优化严重依赖训练输入的代表性，若实际运行场景偏离训练集（如Web服务突发流量），优化效果可能显著下降。此外，传统插桩流程需反复执行训练阶段，难以适应持续集成（CI/CD）的快速迭代需求。

% 。例如，SPEC CPU测试中基于固定输入的优化在阿里云真实负载下性能提升不足5%

A key future direction is dynamic evolution and robustness. One approach is online, incremental training. After deployment, lightweight hardware sampling continuously updates profile data. Facebook’s HHVM runtime combines LBR sampling with flow-sensitive analysis for real-time JIT tuning. Transfer learning, using pre-trained models of CFG embeddings, could predict hotspots for never-seen inputs. Automated input-generation methods, blending symbolic execution and fuzz testing (e.g., extending KLEE \cite{10.5555/1855741.1855756}), could generate high-coverage training sets for PGO.
% 针对这一问题，未来的核心在于动态演化与鲁棒性增强。对于训练集的建设，可以采用在线增量训练的方式，即在程序部署后，通过轻量级硬件采样持续更新配置文件。例如，Facebook的HHVM运行时结合LBR采样与流敏感分析，实现JIT代码的实时调优。并且可以利用机器学习中的学习方式解决泛化问题，如利用迁移学习进行泛化，即利用预训练模型（如程序控制流图嵌入）预测未见过输入的潜在热点。最后合成输入生成也可以被用来解决输入集合的问题，结合符号执行与模糊测试（Fuzzing），自动生成覆盖多路径的输入集。例如，KLEE工具可扩展至反馈优化领域，为编译器提供高覆盖率的训练数据。
%微软的“PGO for .NET”项目已通过图神经网络（GNN）实现跨输入优化泛化，加速比波动降低30%。

% 总结
PGO stands at the cusp of moving from “static tuning” to “dynamic intelligence”. At the sampling layer, hardware-software co-sampling and intelligent denoising will break the efficiency-precision trade-off. For large codebases, precise slicing and distributed frameworks will tame data explosion. Cross-architecture adaptability and real-time feedback loops will propel intelligent optimization decisions. Dynamic, evolving training inputs will endow PGO with robust performance under changing loads. Looking further ahead, as emergent hardware like quantum accelerators and computation-in-memory platforms materialize, profile-guided optimizations will increasingly fuse with architectural innovations. This convergence will form a “sense-decide-optimize” autonomous loop that braces up efficient operation at extreme scale. The co-evolution of algorithms, compilers, and hardware will be the master key to unlocking the next generation of computational potential.  

%% file: contents/sections/sec7_conclusion.tex
\section{Conclusion}
% \section{结论}

This article has demonstrated how profile-guided optimization techniques collect and generate runtime profile data and use that data to guide optimizations. Instrumentation-based PGO yields the most accurate profile data but incurs significant runtime overhead. Thanks to the synergy of hardware sampling and profile-mapping algorithms, sampling-based feedback optimizations overcome the high cost of data collection and the mapping inaccuracies inherent in compile-time methods, achieving high performance gains comparable to instrumentation-based approaches while reducing cross-module optimization overhead. Empirical results show that instrumentation-based PGO can deliver higher peak performance but at the expense of costly profile-collection phases; in contrast, Sampling-based PGO achieves consistent speedups with much lower deployment and runtime costs. In terms of compiler support, LLVM offers more mature implementations of feedback optimizations, while GCC retains advantages in certain specialized scenarios. Looking forward, realizing a true “sense-decide-optimize” closed loop will depend on hardware-software co-design for near-zero-overhead sampling, online dynamic evolution of training inputs, and robust cross-architecture adaptation.
% 本文展示了反馈优技术如何进行运行时程序描述信息的采集、生成以及利用反馈信息进行优化。传统插桩型反馈优化技术可以提供最准确的程序描述信息，但运行时开销不可忽略。得益于硬件采样与数据映射算法的协同，采样型反馈优化技术克服了运行时采集数据的巨大开销以及编译时方法固有的映射不准确性，实现了近似于插桩的性能提升，同时降低了传统跨模块优化所带来的额外开销 。实验证明，插桩型 FDO 虽能带来更高的性能峰值，但需付出更大的配置文件收集代价；相比之下，AutoFDO 在保持一致性能提升的同时，实现了更低的部署及运行时开销；在编译器实现上，反馈优化的各项技术在 LLVM 编译器上的支持更为成熟，而 GCC 编译器在某些特定场景中仍具优势。展望未来，实现真正的“感知-决策-优化”闭环需依赖软硬件协同设计以获得接近零开销的采样能力、训练输入的在线动态演化，以及跨架构的鲁棒优化适配，为下一代超大规模系统的高效运行提供坚实支撑 。

% 根据前文所述，不难发现，工作者已将采集到的性能信息应用在各个阶段，包括编译阶段、链接前阶段、以及链接后阶段。针对生成程序的各个阶段使用性能数据的尝试已经基本穷尽。但是对于反馈优化的各个阶段依旧有着可以尝试的方向。在采集信息阶段，当前技术主要根据中央处理的性能监控单元中的分支事件报告进行优化，而硬件厂商在性能监控单元中提供的监控事件如数据缓存未命中等与程序性能息息相关的事件仍然没有被充分利用。

% 宏观上对不同程序的优化方向方面，工作重心主要着重于对复杂逻辑代码的性能优化，如在大型数据中心的业务逻辑程序进行优化。但是，对于着重于数值计算的大型科学计算程序还缺乏有针对性的优化，如数据的局部性，减少数据缓存的未命中率。

% 微观上，目前的主要工作仅仅着眼于对分支指令的优化，包括对分支预测的优化，以及如何改变程序的布局以减少分支语句的运行等方面。本质上，目前的优化程序认为各个指令的运行时间基本一致。实际上，一些特殊的指令如 I/O 相关指令或 SIMD 指令在执行事件上可能与普通指令有着本质上的差距，如何根据运行时的反馈信息对不同指令进行有针对性的优化措施也是未来的一种可行的优化方向。